\documentclass[useAMS,usenatbib]{mn2e}
\usepackage{graphicx}
\usepackage{amssymb}

\def\jref@jnl#1{{\rm#1}}

\def\aj{\jref@jnl{AJ}}                   
\def\araa{\jref@jnl{ARA\&A}}             
\def\apj{\jref@jnl{ApJ}}                 
\def\apjl{\jref@jnl{ApJ}}                
\def\apjs{\jref@jnl{ApJS}}               
\def\ao{\jref@jnl{Appl.~Opt.}}           
\def\apss{\jref@jnl{Ap\&SS}}             
\def\aap{\jref@jnl{A\&A}}                
\def\aapr{\jref@jnl{A\&A~Rev.}}          
\def\aaps{\jref@jnl{A\&AS}}              
\def\azh{\jref@jnl{AZh}}                 
\def\baas{\jref@jnl{BAAS}}               
\def\jrasc{\jref@jnl{JRASC}}             
\def\memras{\jref@jnl{MmRAS}}            
\def\mnras{\jref@jnl{MNRAS}}             
\def\pra{\jref@jnl{Phys.~Rev.~A}}        
\def\prb{\jref@jnl{Phys.~Rev.~B}}        
\def\prc{\jref@jnl{Phys.~Rev.~C}}        
\def\prd{\jref@jnl{Phys.~Rev.~D}}        
\def\pre{\jref@jnl{Phys.~Rev.~E}}        
\def\prl{\jref@jnl{Phys.~Rev.~Lett.}}    
\def\pasp{\jref@jnl{PASP}}               
\def\pasj{\jref@jnl{PASJ}}               
\def\qjras{\jref@jnl{QJRAS}}             
\def\skytel{\jref@jnl{S\&T}}             
\def\solphys{\jref@jnl{Sol.~Phys.}}      
\def\sovast{\jref@jnl{Soviet~Ast.}}      
\def\ssr{\jref@jnl{Space~Sci.~Rev.}}     
\def\zap{\jref@jnl{ZAp}}                 
\def\nat{\jref@jnl{Nature}}              
\def\iaucirc{\jref@jnl{IAU~Circ.}}       
\def\aplett{\jref@jnl{Astrophys.~Lett.}} 
\def\apspr{\jref@jnl{Astrophys.~Space~Phys.~Res.}}
\def\bain{\jref@jnl{Bull.~Astron.~Inst.~Netherlands}} 
\def\fcp{\jref@jnl{Fund.~Cosmic~Phys.}}  
\def\gca{\jref@jnl{Geochim.~Cosmochim.~Acta}}   
\def\grl{\jref@jnl{Geophys.~Res.~Lett.}} 
\def\jcp{\jref@jnl{J.~Chem.~Phys.}}      
\def\jgr{\jref@jnl{J.~Geophys.~Res.}}    
\def\jqsrt{\jref@jnl{J.~Quant.~Spec.~Radiat.~Transf.}}
\def\memsai{\jref@jnl{Mem.~Soc.~Astron.~Italiana}}
\def\nphysa{\jref@jnl{Nucl.~Phys.~A}}   
\def\physrep{\jref@jnl{Phys.~Rep.}}   
\def\physscr{\jref@jnl{Phys.~Scr}}   
\def\planss{\jref@jnl{Planet.~Space~Sci.}}   
\def\procspie{\jref@jnl{Proc.~SPIE}}   

\title[Covariant Compton Scattering Kernel]{Covariant Compton Scattering Kernel in General Relativistic Radiative Transfer}
\author[Z. Younsi and K. Wu]{Ziri Younsi$^{1}$\thanks{E-mail: zy2@mssl.ucl.ac.uk (ZY); kw@mssl.ucl.ac.uk (KW)} and Kinwah Wu$^{1}$
\\
$^{1}$Mullard Space Science Laboratory, University College London, Holmbury St Mary, Dorking, Surrey, RH5 6NT, UK\\
}

\begin{document}
\onecolumn

\date{Accepted ***. Received *** in original form ****}

\pagerange{\pageref{firstpage}--\pageref{lastpage}} \pubyear{2012}

\maketitle

\label{firstpage}

\begin{abstract}  
A covariant scattering kernel is a core component in any self-consistent general relativistic radiative transfer formulation in scattering media.  
An explicit closed-form expression for a covariant Compton scattering kernel 
  with a good dynamical energy range has unfortunately not been available thus far. 
Such an expression is essential to obtain numerical solutions to the general relativistic radiative transfer equations 
  in complicated astrophysical settings where strong scattering effects are coupled with highly relativistic flows and steep gravitational gradients. 
Moreover, this must be performed in an efficient manner.
With a self-consistent covariant approach, 
   we have derived a closed-form expression for the Compton scattering kernel for arbitrary energy range. 
The scattering kernel and its angular moments are expressed in terms of hypergeometric functions, 
  and their derivations are shown explicitly in this paper.   
We also evaluate the kernel and its moments numerically,  
  assessing various techniques for their calculation. 
Finally, we demonstrate that our closed-form expression 
 produces the same results as previous calculations,   
 which employ fully numerical computation methods and are applicable only in more restrictive settings.

\end{abstract}


\section{Introduction} 

Compton scattering of photons by relativistic electrons is an efficient process 
  to produce high-energy cosmic X-rays and $\gamma$-rays. 
It plays an important role in determining spectral formation and in regulating energy transport  
  in a variety of astrophysical systems, e.g.\ 
   accretion disks of black hole systems 
     \citep{Sunyaev1985, Dermer1989, Haardt1993, Poutanen1993, Titarchuk1994, Hua1995, Stern1995}, 
   relativistic AGN jets   
     \citep{Begelman1987, McNamara2009, Krawczynski2012}, 
   neutron-star X-ray bursts  
    \citep{Titarchuk1988, Madej1991, Titarchuk1994, Madej2004}, 
  and in some accreting white dwarfs
    \citep{Kylafis1982, Matt2004, McNamara2008A, McNamara2008B, Titarchuk2009}.  
Compton scattering of cosmic microwave background photons by hot gases trapped inside the potential wells 
  of large gravitating systems, such as galaxy clusters, 
  also leads to SunyaevÐ-Zel'dovich effects \citep{Sunyaev1980, Rephaeli1995, Dolgov2001, Colafrancesco2003}, 
  through which various aspects of cosmology and the evolution of large-scale structures in the Universe may be investigated.  


Compton scattering in astrophysical plasmas is often investigated using Monte-Carlo simulations 
  \citep[e.g.][]{Pozdnyakov1983, Hua1995}.  
The Monte-Carlo approach is an approximation scheme to proper radiative transfer calculations, 
  where the radiative transfer equation is derived from the laws of conservation \citep[see][]{Rybicki_Lightman, Chandrasekhar1960, Peraiah2001}.  
It has the advantage of being able to handle complicated system geometries, 
  as well as the flexibility to incorporate relevant additional physics, such as absorption and pair production,  into the system.    
However, 
  it is not straightforward to implement the usual Monte-Carlo method in certain extreme astrophysical environments,  
  such as systems with steep density gradients or fractal-like inhomogeneities, 
  and ultra-relativistic flows near the event-horizon of a black hole. 
In the latter, relativistic and space-time curvature effects are important, 
  and radiative transfer in these systems requires a covariant formulation \citep{Lindquist1966, Baschek1997, Fuerst2004, Younsi2012}. 
In the absence of scattering, 
  the covariant radiative transfer can be solved along the null geodesic \citep[see][]{Viergutz1993, Reynolds1999, Dexter2009} 
  using a ray-tracing technique \citep[e.g.][]{Fuerst2004, Vincent2011}. 
In the presence of scattering,  
  the covariant transfer equation is much more complicated, 
 and the transfer equation is no longer a differential equation but an integro-differential equation.  
A key ingredient in the radiative transfer formulation is the scattering kernel, 
  which describes how photons interact with electrons. 
The moment expansion \citep{Thorne1981, Turolla1988, Rezzolla1994, Challinor2000, Fuerst2006, Wu2008, Shibata2011} of this kernel is essential 
  in deriving a practical (numerical) scheme to solve the integro-differential radiative transfer equation \citep[see][]{Fuerst2006, Farris2008, Zanotti2011}.

This article shows explicitly the derivation of the invariant scattering kernel  
  for Compton scattering in a general relativistic setting 
  and finds a closed-form expression in terms of hypergeometric functions. The method is not limited by energy range and is valid both for Compton and inverse Compton scattering.
The article is organised as follows. 
\S2 introduces the covariant radiative transfer equation in the presence of scattering and discusses methods for its solution. \S3 derives the covariant Klein-Nishina cross-section for relativistic Compton scattering. \S4 derives from first principles the relativistic electron distribution function, which must later be convolved with the Klein-Nishina cross-section. \S5 presents an outline of the derivation of the integral from of the covariant Compton scattering Kernel. \S6 outlines a method to simplify the calculation of successive angular moments of the scattering kernel through changing the order of integration. \S7 derives algebraic expressions for the first three angular moments of the scattering kernel. \S8 outlines a method for deriving the angular moments of the scattering kernel, through employing recursion identities. \S9 demonstrates how these moment integrals may be expressed in closed-form, for arbitrary order $n$, in terms of Gauss hypergeometric functions. This yields an analytic result for moments of the Klein-Nishina cross-section, $\mathcal{M}_{n}$. \S10 performs the integration of the convolution of the moments of the Klein-Nishina cross-section with the relativistic electron distribution function, using the methods outlined in the previous chapters. \S11 is devoted to the discussion and \S12 the summary.

\section{Radiative Transfer With Scattering}
In Newtonian space-time the radiative transfer equation in a medium reads 
\begin{equation}  
 \left(\frac{1}{c}\frac{\partial }{\partial t} + {\hat \mathbf{\Omega}}\cdot \nabla \right) I_{\nu} ({\hat \mathbf{\Omega}}) 
  =  j_{\nu}({\hat \mathbf{\Omega}}) - \kappa_{\nu} I_{\nu}({\hat \mathbf{\Omega}}) 
     + \int\!\!\!\!\int \mathrm{d}\Omega \ \mathrm{d}\nu' \ \sigma(\nu,   {\hat \mathbf{\Omega}};   \nu', {\hat \mathbf{\Omega}}') I_{\nu'}({\hat \mathbf{\Omega}}')  \ ,  
\label{nrt0}
\end{equation}      
\citep[see][]{Mihalas1984, Peraiah2001}
  where $I_{\nu} ({\hat \mathbf{\Omega}})$ is the intensity of the radiation at a frequency $\nu$ propagating in the $\hat \mathbf{\Omega}$-direction, 
  $j_{\nu}$ and $\kappa_{\nu}$ are the emission and absorption coefficient respectively, 
  and $\sigma(\nu,   {\hat \mathbf{\Omega}};   \nu', {\hat \mathbf{\Omega}}')$ is the scattering kernel 
  which determines the amount of radiation intensity at a frequency $\nu'$ in a direction ${\hat \mathbf{\Omega}}'$ 
  being scattered into the intensity $I_{\nu'}(\hat \mathbf{\Omega}')$. 
For instance, in the photon-electron scattering process, 
  the scattering kernel is determined by the momentum distribution of the electrons 
  and the differential scattering cross-section, the Klein-Nishina \citep[][]{Klein1929} differential cross-section
\begin{equation} 
 \left( \frac{\mathrm{d}\sigma}{\mathrm{d}\Omega}\right)_{\rm KN} = \left(\frac{e^{2}}{m_{\rm e}c^{2}} \right)^{2} \left(\frac{k_{f}}{k_{i}} \right)^{2}  
   f(k_{f}, {\hat \bepsilon}_{f}; k_{i}, {\hat \bepsilon}_{i}) 
  =\frac{3 \sigma_{\mathrm{T}}  }{8\pi} \left(\frac{k_{f}}{k_{i}} \right)^{2}  
   f(k_{f}, {\hat \bepsilon}_{f}; k_{i}, {\hat \bepsilon}_{i})     \ , 
\end{equation} 
 where $e$ is the electron charge, $m_{\mathrm{e}}$ is the electron mass, $\sigma_{\mathrm{T}}$ is the Thomson cross-section 
 \citep[][]{Thomson1906, Compton1923}, 
   $k_{i}$ and $k_{f}$ are the wave numbers of the photon before and after the scattering respectively, and 
   ${\hat \bepsilon}_{i}$ and ${\hat \bepsilon}_{f}$ are the corresponding polarisation vectors of the photon.  
The function  $f(k_{f}, {\hat \bepsilon}_{f}; k_{i}, {\hat \bepsilon}_{i})$ is given by
\begin{equation} 
f(k_{f}, {\hat \bepsilon}_{f}; k_{i}, {\hat \bepsilon}_{i})  
  = \left|{\hat \bepsilon}_{f}^{*}\cdot{\hat \bepsilon}_{i}  \right|^{2} 
    + \frac{\left(k_{f}-k_{i}\right)^{2}}{4~k_{f}k_{i}}\left[ 1+   \left({\hat \bepsilon}_{f}^{*}\times {\hat \bepsilon}_{f}  \right)   
       \cdot   \left({\hat \bepsilon}_{i}\times {\hat \bepsilon}_{i}^{*}  \right)   \right],
\end{equation}   
\citep[see][]{Jackson75}. 

In the absence of scattering, the covariant form of the radiative transfer equation may be written as
\begin{equation}   
\frac{\mathrm{d}{\cal I} }{\mathrm{d}\xi}  = k_{\alpha} \frac{\partial {\cal I}}{\partial x^{\alpha}}
  - \Gamma^{\alpha}_{\beta \gamma}k^{\beta}k^{\gamma}  \frac{\partial {\cal I}}{\partial k^{\alpha}}  
   = -k^{\alpha}u_{\alpha} \big|_{\xi} \left(~   \eta_{0} - \chi_{0} {\cal I} ~ \right),     
\label{grrt0}
\end{equation} 
\citep{Baschek1997, Fuerst2004, Wu2008, Younsi2012}, 
 where ${\cal I}$ is the invariant intensity of the radiation,  $x^{\alpha}$ is a position 4-vector, 
  $\eta_{0}$ and $\chi_{0}$ are the invariant emission and absorption coefficients respectively (evaluated in a local inertial frame), 
  $\xi$ is the affine parameter, $k^{\alpha}$ is the propagation (wave number) 4-vector of the radiation, 
  and $u^{\alpha}$ is the 4-velocity of the medium interacting with the radiation. 
Equation ({\ref{grrt0}) is similar in form to equation (\ref{nrt0}) without the scattering term.  
The term $k^{\alpha}u_{\alpha} |_{\xi}$ is a correction factor for the aberration and energy shift 
   in the transformation between reference frames. 
For covariant transfer of radiation in the presence of scattering, 
 the radiative transfer equation is of the form
\begin{equation}  
\frac{{\mathrm d}{\cal I}(x^{\beta},k^{\beta})}{{\mathrm d}\xi}  
   = -k^{\alpha}u_{\alpha} \big|_{\xi} 
    \left[     \eta_{0}(x^{\beta},k^{\beta}) 
        - \chi_{0}(x^{\beta},k^{\beta}) {\cal I}(x^{\beta},k^{\beta})     
     +\int {\mathrm d}^{4}k^{\beta}\   \sigma (x^{\beta}; k^{\beta}, k'^{\beta})   {\cal I} ( x^{\beta},  k'^{\beta})      \right]  \ ,
\label{grrt1}
\end{equation} 
 analogous to equation (\ref{nrt0}).  
Several methods have been proposed to solve the above equation or to obtain an approximate solution. 
For instance, one could transform the integro-differential radiative transfer equation 
  into a set of differential equations using a moment expansion 
  \citep{Thorne1980, Thorne1981, Fuerst2006, Wu2008, Shibata2011}.  
Nevertheless, one needs to specify the properties of the medium spanning the space-time. 
In addition to the global flow dynamics, 
  one also needs to know how the radiation interacts with the medium
  (via the emission coefficient, absorption coefficient and the scattering kernel), 
  at least in the local inertial frame. 
The invariant emission and absorption coefficients can be easily derived from the conventional emission and absorption coefficients 
 \citep[see][]{Fuerst2004, Fuerst2007}.   
The derivation of the scattering kernel is more complicated.  
Some attempts have been made \citep[e.g.][]{Shestakov1988}, but only numerical results were obtained due to the complexity of the underlying mathematics. 
To date a closed-form expression for the corresponding scattering kernel is not available. 
The lack of a closed-form scattering kernel hinders  
  the development of fast and accurate numerical algorithms to solve the covariant radiative transfer equation, 
  which itself can be numerically intensive. 
  
\section{Covariant Compton Scattering}

Here and hereafter this article adopts the geometrical unit convention (with $G=c=h=1$) and employs the $(-,+,+,+)$ metric signature. 
Energy-momentum conservation implies that 
\begin{equation}
k^{\alpha}+p^{\alpha} = k'^{\alpha}+p'^{\alpha} \ ,
\end{equation}  
 in a photon-electron scattering process. 
Here unprimed and primed variables denote, respectively, variables evaluated before and after scattering.   
The 4-momentum of a photon $k^{\alpha}$ and the 4-momentum of an electron $p^{\alpha}$  
  satisfy $k^{\alpha}k_{\alpha} = k'^{\alpha}k'_{\alpha} = 0$ 
  and $p^{\alpha}p_{\alpha} = p'^{\alpha}p'_{\alpha} = -m_{\mathrm{e}}^{2}$, respectively.
Energy-momentum conservation also leads to the invariance relation 
\begin{equation}
k^{\alpha}p_{\alpha}  =   k'^{\alpha}p'_{\alpha} \ ,  
\end{equation} 
  and a covariant generalised energy-shift formula for the scattered photon,
\begin{equation}  
k'^{\alpha}(k_{\alpha}+p_{\alpha})  =  k^{\alpha}p_{\alpha} \  . 
\label{cs-1}
\end{equation} 
As the scattering process occurs in a relativistic fluid, 
  the derivation of the scattering opacity due to ensembles of photons and electrons requires expressing the scattering variables of the particles in the local reference rest-frame (co-moving with the fluid 4-velocity), 
  as well as specifying the transformation between the fluid rest-frame and the observer's frame.    
The fluid 4-velocity, in the fluid rest frame, is denoted as $u^{\alpha}$. 
  The electron 4-velocity is $v^{\alpha}$.
Clearly $u^{\alpha}u_{\alpha}=-1$ and $v^{\alpha}v_{\alpha}\equiv v < 1$. 
The directional unit 4-vector of the photon in the fluid rest frame may be specified as $n^{\alpha}$, 
   which is given by 
\begin{equation}
n^{\alpha}=\frac{P^{\alpha \beta}k_{\beta}}{|| P^{\alpha \beta}k_{\beta} ||} \ ,
\end{equation}
   where the tensor $P^{\alpha \beta}=g^{\alpha \beta}+u^{\alpha}u^{\beta}$ 
   projects onto the 3-surface orthogonal to $k_{\beta}$. 
A variable
\begin{equation}
\gamma \equiv -\frac{k^{\alpha}u_{\alpha}}{m_{\mathrm{e}}}\ ,
\end{equation}
  may be constructed, from which $n^{\alpha}$ may be expressed as 
\begin{equation}
n^{\alpha}=\frac{k^{\alpha}}{m_{\mathrm{e}}\gamma}-u^{\alpha} \ . 
\end{equation}
Hence, it follows the photon 4-momentum may be expressed as
\begin{equation}
k^{\alpha} = m_{\mathrm{e}}\gamma(n^{\alpha}+u^{\alpha})\ .
\end{equation}
Similarly, for the electrons, 
\begin{equation}
\lambda \equiv -\frac{p^{\alpha}u_{\alpha}}{m_{\mathrm{e}}} \ .  
\end{equation} 
Clearly $\lambda=1/\sqrt{1-v^{2}}$, which is simply the Lorentz factor of the electron. 
The directional 4-velocity of the electron in the fluid frame is therefore 
\begin{equation}
\hat{v}^{\alpha} = \frac{P^{\alpha \beta}p_{\beta}}{|| P^{\alpha \beta}p_{\beta} ||}  
                            =  \frac{p^{\alpha}-m_{\mathrm{e}}\lambda u^{\alpha}}{m_{\mathrm{e}}\lambda v} \ . 
\label{hat-v}
\end{equation} 
It therefore follows that  
\begin{equation} 
            v^{\alpha}  =  \frac{p^{\alpha}}{m_{\mathrm{e}}\lambda}-u^{\alpha}  \ ,
\end{equation}
  and 
\begin{equation}
  p^{\alpha} = m_{\mathrm{e}}\lambda (v^{\alpha}+u^{\alpha}) \ .
\end{equation} 
Note that the photon 4-momentum after the scattering event is 
\begin{equation}
k'^{\alpha} = m_{\mathrm{e}}\gamma'(n'^{\alpha}+u^{\alpha})\ .
\end{equation}
Thus, the following expressions are obtained:
\begin{eqnarray}
k^{\alpha}k'_{\alpha}&=&m_{\mathrm{e}}^{2}\gamma \gamma'(\zeta -1)\ ,  \label{e-1} \\
p^{\alpha}k_{\alpha}&=&m_{\mathrm{e}}^{2}\lambda\gamma(v^{\alpha}n_{\alpha}-1)\ , \label{e-2} \\
p^{\alpha}k'_{\alpha}&=&m_{\mathrm{e}}^{2}\lambda\gamma'(v^{\alpha}n'_{\alpha}-1) \ , \label{e-3} 
\end{eqnarray} 
  where $\zeta=n^{\alpha}n'_{\alpha}$ is the direction cosine of the angle between the incident and scattered photon. 
  Hence energy-momentum conservation, equation (\ref{cs-1}), may be expressed as:
\begin{equation}
m_{\mathrm{e}}^{2}\gamma\gamma' 
   \left[\zeta-1+\lambda\left(\frac{1-v^{\alpha}n_{\alpha}}{\gamma'}-\frac{1-v^{\alpha}n'_{\alpha}}{\gamma} \right) \right]=0 \ . 
\label{cs-2}
\end{equation}
The cross-section for scattering of a photon by an electron is given in \cite{Kershaw1986} as:
\begin{equation}
\sigma(\gamma\rightarrow\gamma', \hat{\mathbf{\Omega}}\rightarrow \hat{\mathbf{\Omega}'},\mathbf{v})
    =\frac{3\sigma_{\mathrm{T}}}{16\pi \gamma \nu \lambda}\left[1+\left(1-\frac{1-\zeta}{\lambda^{2}D D'}\right)^{2}
   +  \frac{(1-\zeta)^{2}\gamma \gamma'}{\lambda^{2}D D'}\right] \ 
    \delta \left[\zeta-1+\lambda\left(\frac{D}{\gamma'}-\frac{D'}{\gamma} \right)\right] \ ,  
\label{cross-section-A}
\end{equation} 
where $D\equiv 1-\hat{\mathbf{\Omega}}\cdot\mathbf{v}/c = 1-v^{\alpha}n_{\alpha}$, and similarly for $D'$.   
Using equations (\ref{e-1})--(\ref{e-3}), the photon-electron scattering cross-section, 
  equation (\ref{cross-section-A}), may be expressed in the following covariant form:
\begin{equation}
  \sigma(\gamma\rightarrow\gamma', n^{\alpha}\rightarrow n'^{\alpha},v^{\alpha})
    =\frac{3\sigma_{\mathrm{T}}}{16\pi \gamma \nu \lambda}\left[1+\left(1+\frac{m_{\mathrm{e}}^{2}\mathcal{T}}{k^{\alpha}k'_{\alpha}} \right)^{2}+\mathcal{T} \right] \ 
      \delta\left(\frac{\mathcal{P}}{m_{\mathrm{e}}^{2}\gamma\gamma'} \right) \ , 
\label{cross-section-B}
\end{equation}
where $\delta$ denotes the Dirac delta function, and $\mathcal{T}$, $\mathcal{P}$ are defined respectively as  
\begin{eqnarray}
 \mathcal{T} &=& \frac{(k^{\alpha}k'_{\alpha})^{2}}{(p^{\alpha}k_{\alpha})(p^{\beta}p'_{\beta})} \ , \\
 \mathcal{P} &=& k^{\alpha}k'_{\alpha}+p^{\alpha}k'_{\alpha}-p^{\alpha}k_{\alpha} \ . 
\end{eqnarray} 
It follows that $\mathcal{P}$ represents energy and momentum conservation of the scattering process. The delta function enforces the conservation of energy and momentum in the scattering process, by weighting the scattering cross-section such that it is zero if energy and momentum are not conserved. Integrating this cross-section, equation (\ref{cross-section-B}), over a relativistic electron distribution function yields the kernel for Compton scattering.

\section{Electron Distribution Function}
In order to calculate the Compton scattering kernel the relativistic electron distribution function, $f(\lambda)$, must be determined. This may be derived as follows.
The energy of an electron is $E=\lambda \ \! m_{\mathrm{e}}c^{2}$, and its linear momentum is given by $p=\lambda \ \! m_{\mathrm{e}}v$, from which it follows that
\begin{equation}
\frac{\mathrm{d}p}{\mathrm{d}v} =   m_{\mathrm{e}}\frac{\mathrm{d}}{\mathrm{d}v}(\lambda v) = m_{\mathrm{e}}\lambda^{3} \ .
\end{equation} 
As an example, consider an ensemble of relativistic electrons with isotropic momenta for which the distribution function is given by the pseudo-Maxwellian
\begin{equation}
\Psi(\mathbf{p})=\mathrm{C} \ \! e^{-E(\mathbf{p}) /k_{\mathrm{B}}T_{\mathrm{e}}}\ ,
\end{equation}
   where $E$ is the electron energy, $T_{\mathrm{e}}$ the electron temperature, $k_{\mathrm{B}}$ the Boltzmann constant 
   and $\mathrm{C}$ is a normalisation constant. 
Note that the distributions of electrons in momentum space and in velocity space are related via
\begin{equation} 
f(\mathbf{v})v^{2}\mathrm{d}v=\Psi(\mathbf{p})p^{2}\mathrm{d}p \ ,  
\end{equation} 
which may be expressed as
\begin{equation}
f(\mathbf{v})=\frac{p^{2}}{v^{2}}\frac{\mathrm{d}p}{\mathrm{d}v}\Psi(\mathbf{p}) \ .
\end{equation} 
It immediately follows that
\begin{equation}
f(\mathbf{v})=\mathrm{C}'\lambda(v)^{5}e^{-\lambda(v)/\tau} \ ,
\end{equation}
   where $\mathrm{C}'=m_{\mathrm{e}}^{3}\mathrm{C}$ is a constant and $\tau={k_{\mathrm{B}}T_{\mathrm{e}}}/{m_{\mathrm{e}}}$. 
The normalisation of the distribution function $f(\mathbf{v})$ to unity, i.e. 
\begin{equation}
 \int \mathrm{d}\mathbf{v}f(\mathbf{v}) =  4\pi \int_{0}^{1}\mathrm{d}v \ \! v^{2}f(\mathbf{v}) =1 \ , 
\end{equation}
  yields the familiar relativistic Maxwellian form, 
\begin{equation}
  f(\lambda) = \frac{\lambda^{5}e^{-\lambda/\tau}}{4\pi\tau K_{2}(1/\tau)} \ ,
\end{equation}
where $K_{2}$ denotes the modified Bessel function of the second kind.

\section{Compton Scattering Kernel}

The Compton scattering kernel, as seen in equation (\ref{grrt1}), is essential in solving the radiative transfer equation. It is determined by the convolution of the photon-electron scattering cross-section 
 with the electron velocity distribution, i.e.\ 
\begin{equation}
   \sigma_{\mathrm{s}}(\gamma \rightarrow \gamma', \zeta, \tau)
    =\frac{3\rho \sigma_{\mathrm{T}}}{16\pi \gamma \nu}\int \mathrm{d}\mathbf{v} \ 
     \frac{f(\lambda)}{\lambda}\left[1+\left(1+\frac{m_{\mathrm{e}}^{2}\mathcal{T}}{k^{\alpha}k'_{\alpha}} \right)^{2}+\mathcal{T} \right] \ 
     \delta\left(\frac{\mathcal{P}}{m_{\mathrm{e}}^{2}\gamma\gamma'}\right) \  ,
\label{scat_k}
\end{equation}
where $\rho$ is the electron density. To evaluate the above integral, first consider the argument of the delta function
\begin{equation}
y = \frac{\mathcal{P}}{m_{\mathrm{e}}^{2}\gamma \gamma'} \ . \label{y-1}
\end{equation}
Rewriting (\ref{y-1}) in terms of a linear combination of a scalar and an inner product of two unit vectors is a succinct way of expressing the energy-momentum conservation. More importantly, aside from the more compact notation, the inner product of two unit vectors (the magnitude of which never exceeds unity) provides constraints on the electron energy. This makes the subsequent integrals easier to solve, and is the most natural way of proceeding with the problem. Substituting equations (\ref{e-1})--(\ref{e-3}) into (\ref{y-1}) yields
\begin{eqnarray}
y&=& \left[\zeta-1+\lambda\left(\gamma'^{-1}-\gamma^{-1}\right)+\frac{\lambda}{\gamma\gamma'}v^{\alpha}\left(\gamma'n'_{\alpha}-\gamma n_{\alpha}\right) \right] \ , \nonumber \\
  &=& \Gamma+\hat{v}^{\alpha}w_{\alpha} \ , \label{y-2}
\end{eqnarray}
where
\begin{eqnarray}
\Gamma &=& \zeta-1+\lambda\left(\gamma'^{-1}-\gamma^{-1}\right) \ , \\
w_{\alpha} &=& \frac{\lambda v}{\gamma\gamma'}\left(\gamma'n'_{\alpha}-\gamma n_{\alpha}\right) \ ,
\end{eqnarray}
and hence (\ref{y-2}) is split into a scalar and vector component. It immediately follows that $y$ may be rewritten as
\begin{equation}
y = w\left(\frac{\Gamma}{w}+\hat{v}^{\alpha}\hat{w}_{\alpha} \right) \ ,
\end{equation}
where
\begin{eqnarray}
\hat{w}_{\alpha} &=& \frac{\gamma'n'_{\alpha}-\gamma n_{\alpha}}{q} \ , \\
w &=& \frac{\lambda v}{\gamma\gamma'}q \ ,
\end{eqnarray}
and $q$, akin to the resultant photon energy along the direction of photon momentum transfer, is defined as
\begin{equation}
q = \sqrt{\gamma^{2}+\gamma'^{2}-2\gamma\gamma' \zeta} \ . \label{q}
\end{equation}
Therefore $\hat{w}_{\alpha}$ represents a unit vector along the direction of photon momentum transfer and $\hat{v}^{\alpha}\hat{w}_{\alpha}$ is simply the projection of the electron velocity onto this preferred direction. Under integration, the delta function can be rewritten as $\delta\left(\Gamma/w+\hat{v}^{\alpha}\hat{w}_{\alpha} \right)/w$, and the energy-momentum conservation may be rewritten as
\begin{equation}
\hat{v}^{\alpha}\hat{w}_{\alpha} = -\frac{\Gamma}{w} \ .
\end{equation}
 From this it immediately follows $|| -{\Gamma}/{w} || \le 1$ and therefore
\begin{equation}
(1-\zeta)+\lambda(\gamma^{-1}-\gamma'^{-1})\le\frac{\lambda v q}{\gamma \gamma'} \  , \label{ineq}
\end{equation}
 which is akin to solving the quadratic equation $A\lambda^{2}-B\lambda-C=0$, 
 with coefficients $A$, $B$ and $C$ given by:
\begin{eqnarray}
A&=&2\gamma \gamma'(\zeta-1) \ , \\
B&=& (\gamma'-\gamma)A \ , \\
C&=&q^{2}+\frac{A^{2}}{4} \ .
\end{eqnarray}
Taking the positive solution to (\ref{ineq}) yields, upon employing the identity $q^{2}=(\gamma'-\gamma)^{2}-A$,
\begin{equation}
\lambda_{+}=\left(\frac{\gamma'-\gamma}{2} \right)+\frac{q}{2}\sqrt{1+\frac{2}{\gamma\gamma'(1-\zeta)}}\ ,
\end{equation}
which is essentially the minimum electron energy in the Compton scattering process.  
The form of $\lambda$ as a function of $\zeta$ is crucial in later calculations involving integrations over $\lambda$ and $\zeta$.
The integral in equation (\ref{scat_k}) may now be rewritten as
\begin{equation}
\int \mathrm{d}\mathbf{v}=\int_{0}^{1}  \mathrm{d}v \ v^{2}   \int_{-1}^{1}\mathrm{d}(\hat{v}^{\alpha}\hat{w}_{\alpha})\int_{0}^{2\pi}\mathrm{d}\phi\ . 
\label{int}
\end{equation}
Hence it follows that the delta function fixes this preferred direction naturally \citep[][]{Prasad1986,Beason1991}, and this is clearly the most straightforward approach.
Note, as in \cite{Kershaw1986}, the angular addition formula:
\begin{equation}
\hat{v}^{\alpha}\hat{m}_{\alpha}=(\hat{n}^{\alpha}\hat{m}_{\alpha})(\hat{v}^{\alpha}\hat{n}_{\alpha})+\sqrt{1-(\hat{n}^{\alpha}\hat{m}_{\alpha})^{2}}\sqrt{1-(\hat{v}^{\alpha}\hat{n}_{\alpha})^{2}}\cos \phi,
\end{equation}
where $\hat{m}_{\alpha}$ is equal to $\hat{w}_{\alpha}$ or $\hat{w}'_{\alpha}$, the unit vector of the photon velocity before or after collision respectively. It is easily verified that
\begin{eqnarray}
n^{\alpha}\hat{w}_{\alpha}&=&\frac{\gamma'\zeta-\gamma}{q}\ , \\
n'^{\alpha}\hat{w}_{\alpha}&=&\frac{\gamma'-\gamma\zeta}{q}\ , \\
\hat{v}^{\alpha}\hat{w}_{\alpha}&=&-\frac{\gamma\gamma'\Gamma}{q\lambda v} \ .
\end{eqnarray} 
As such, in equation (\ref{int}) only the $\phi$ integral need be evaluated explicitly. 
The square-bracketed term in the kernel may be rewritten \citep[e.g.][]{Kershaw1986} as
\begin{eqnarray}
\left[1+\left(1+\frac{m_{\mathrm{e}}^{2}\mathcal{T}}{k^{\alpha}k'_{\alpha}} \right)^{2}+\mathcal{T} \right]
  = 2+\left[\gamma\gamma'(1-\zeta)-2-\frac{2}{\gamma\gamma'(1-\zeta)} \right]  
  \left[\left(\lambda\gamma' D'\right)^{-1}-\left(\lambda\gamma D\right)^{-1} \right]+ \left(\lambda\gamma' D'\right)^{-2} + \left(\lambda\gamma D\right)^{-2} \  , 
\label{curvy}
\end{eqnarray}
  which must be integrated term-by-term over $\phi$. 
The integrals to solve have the forms: 
\begin{eqnarray}
I_{1}&=&\int_{0}^{2\pi}\frac{\mathrm{d}\phi}{\alpha+\beta\cos\phi}\ , \\
I_{2}&=&\int_{0}^{2\pi}\frac{\mathrm{d}\phi}{(\alpha+\beta\cos\phi)^{2}}\ ,
\end{eqnarray}
where
\begin{eqnarray}
\alpha &=& 1-v(\hat{w}^{\alpha}n_{\alpha})(\hat{w}^{\alpha}\hat{v}_{\alpha})\  , \\
\beta   &=& -v\sqrt{1-(\hat{w}^{\alpha}n_{\alpha})^{2}}\sqrt{1-(\hat{w}^{\alpha}\hat{v}_{\alpha})^{2}}\  .
\end{eqnarray}
Clearly, the two integrals are related, via $I_{2}=-\frac{\mathrm{d}I_{1}}{\mathrm{d}\alpha}$ and therefore only $I_{1}$ need be evaluated, yielding
\begin{eqnarray}
I_{1}&=&\frac{2\pi}{(\alpha^{2}-\beta^{2})^{1/2}}\ , \\
I_{2}&=&\frac{2\pi\alpha}{(\alpha^{2}-\beta^{2})^{3/2}}\ ,
\end{eqnarray}
where the coefficients $\alpha\equiv\alpha(x)$, $\beta$ and $\alpha^{2}-\beta^{2}$ are given by 
\begin{eqnarray}
\alpha &=& \frac{\gamma'}{\lambda q^{2}}\left[x\left(\gamma^{-1}+\gamma'^{-1}\right)-\left(1+\zeta\right) \gamma\gamma' \right]\ , \\
\alpha' &=&\frac{\gamma}{\gamma'}\alpha \ , \\
\beta &=& \frac{\gamma'\omega(\zeta-1)}{\lambda q^{2}}\sqrt{A\lambda^{2}-B\lambda-C}\ , \\
\beta' &=& \frac{\gamma}{\gamma'} \ \! \beta\ , \\
\alpha^{2}-\beta^{2} &=& \frac{\gamma'^{2}\left(1-\zeta\right)^{2}\left(x^{2}+\omega^{2} \right)}{\lambda^{2}q^{2}}\ , \\
\alpha'^{2}-\beta'^{2} &=& \left(\frac{\gamma}{\gamma'}\right)^{2}\left(\alpha^{2}-\beta^{2}\right)\ ,
\end{eqnarray}
wherein the notation $x\equiv\gamma+\lambda$ prior to collision and $x\equiv\gamma'-\lambda$ after collision is adopted. Additionally, $\omega^{2} = (1+\zeta)/(1-\zeta)$. 
The $\phi$-integrals immediately follow, yielding 
\begin{eqnarray}
\int_{0}^{2\pi} \mathrm{d}\phi \   D^{-1} &=& \frac{2\pi \lambda q}{\gamma'}
   \frac{(1-\zeta)^{-1}}{\left(x^{2}+\omega^{2} \right)^{1/2}}\ , \\
\int_{0}^{2\pi} \mathrm{d}\phi \  D'^{-1} &=& \frac{\gamma'}{\gamma}\int_{0}^{2\pi} \mathrm{d}\phi \   D^{-1} \ , \\
\int_{0}^{2\pi} \mathrm{d}\phi \  D^{-2}  &=& \frac{2\pi\gamma\lambda^{2}q}{\gamma'\left(1-\zeta\right)^{2}}\frac{\left[x\left(\gamma^{-1}+ 
   \gamma'^{-1}\right)-\left(1+\zeta\right) \right]}{\left(x^{2}+\omega^{2} \right)^{3/2}}\ , \\
\int_{0}^{2\pi} \mathrm{d}\phi \   D'^{-2} &=& \frac{\gamma'}{\gamma}\int_{0}^{2\pi} \mathrm{d}\phi \  D^{-2} \ .
\end{eqnarray}
The Compton scattering kernel in equation (\ref{scat_k}) may now be rewritten as  
\begin{equation}
\sigma_{\mathrm{s}}\left(\gamma\rightarrow\gamma',\zeta,\tau\right)=\frac{3\rho \sigma_{\mathrm{T}}}{8\gamma\nu}\int_{\lambda_{+}}^{\infty} \mathrm{d}\lambda \ 
\frac{f(\lambda)}{\lambda^{5}}\left[\frac{2\gamma\gamma'}{q}+R\left(\gamma+\lambda\right)-R\left(\gamma'-\lambda\right) \right]   \ , \label{Kernel_Symmetric}
\end{equation}
where the function $R(x)$ is defined as
\begin{equation}
R(x) = \frac{w-\zeta}{(1-\zeta)^{2}(x^{2}+\omega^{2})^{3/2}}  
 + \left[-\gamma\gamma'+\frac{2}{1-\zeta}+\frac{2}{\gamma\gamma'(1-\zeta)^{2}} \right]\frac{1}{(x^{2}+\omega^{2})^{1/2}} \ , \label{R(x)}
\end{equation} 
where $w\equiv w(x)$, with $w(x)=\left[x \left(\gamma^{-1}+\gamma'^{-1}\right)-1 \right]$. The scattering kernel, as it is written in equation (\ref{Kernel_Symmetric}), is highly symmetric and essentially the sum of three terms: the resultant photon energy along the direction of momentum transfer, a pre-collisional photon-electron interaction term, and less a post-collisional photon-electron interaction term, with the interaction term defined in equation (\ref{R(x)}).

\section{Angular Moments of the Compton Cross Section} 

In solving the full radiative transfer equation with Compton scattering, a generalised Eddington approximation \citep{Eddington1926,Rybicki_Lightman} to compute successive angular moment integrals of $\sigma_{s}$ may be employed \citep{Thorne1981,Fuerst2006,Wu2008}. 
In this section, angular moments of the form $\zeta^{n}$ \citep[e.g.][and references therein]{Shestakov1988} are used to define the moment expansion of the Compton scattering kernel. This requires solving integrals of the form
\begin{eqnarray}
\int \mathrm{d}\zeta \  \zeta^{n}\sigma_{\mathrm{s}}\left(\gamma\rightarrow\gamma',\zeta,\tau\right) = 
\frac{3\rho\sigma_{\mathrm{T}}}{8\gamma\nu}
 \int_{-1}^{1}\mathrm{d}\zeta \ \!  \zeta^{n}\int_{\lambda_{+}}^{\infty}  \mathrm{d}\lambda \ 
 \frac{f(\lambda)}{\lambda^{5}}\left[\frac{2\gamma\gamma'}{q}+R\left(\gamma+\lambda\right) -R\left(\gamma'-\lambda\right) \right]   \ . 
\label{int_zeta}
\end{eqnarray}
However, as equation (\ref{int_zeta}) stands, integrating over $f(\lambda)$ is analytically impossible. Rather than perform the $\lambda$ integration first, it is more straightforward to switch the order of integration. Not only does this enable the derivation of analytic results, performing the $\lambda$ integration after the $\zeta$ integration affords the method greater generality, since the $\zeta$ integral is independent of the assumed electron distribution function (in the isotropic case). To change the order of integration, first consider $\lambda_{+}(\zeta)$ (which must be inverted, i.e. $\zeta(\lambda_{+})$ found), with the left boundary $\lambda_{+}(-1)$ found as
\begin{equation}
\lambda_{+}(-1)\equiv\lambda_{\mathrm{L}}=\frac{\gamma'-\gamma}{2}+\frac{\gamma'+\gamma}{2}\sqrt{1+\frac{1}{\gamma\gamma'}}  \ ,
\end{equation}
whereas
\begin{equation}
\lim_{\zeta\rightarrow1}\lambda_{+}=+\infty    \ , 
\end{equation}
is the right boundary. The minimum value of $\lambda_{+}$, i.e. the value of $\zeta$ such that $\lambda_{+}$ is minimised, is found as
\begin{equation}
\zeta_{1,2}=1\pm\left(\gamma^{-1}-\gamma'^{-1}\right)\ ,
\end{equation}
and hence
\begin{equation}
\lambda_{\mathrm{min}}=1+\frac{1}{2}\left[\left(\gamma'-\gamma\right)+\left|\gamma'-\gamma\right| \right] \ .
\end{equation}
Normally $\lambda_{\mathrm{min}}<\lambda_{\mathrm{L}}$ by definition. However, $\lambda_{\mathrm{L}}\le \lambda_{\mathrm{min}}$ if the following condition is satisfied:
\begin{equation}
\left|\gamma^{-1}-\gamma'^{-1}\right|\ge 2. \label{Condition}
\end{equation}
Rearranging $\lambda_{+}$ to find $\zeta$ as a function of $\lambda$ yields
\begin{equation}
\zeta_{\pm}=\frac{1}{\gamma\gamma'}\left[1+\left(\gamma+\lambda\right)\left(\gamma'-\lambda\right)\pm\sqrt{\lambda^{2}-1}\sqrt{\left(\lambda+\gamma-\gamma'\right)^{2}-1} \right]\ . \label{zeta_pm}
\end{equation}
It immediately follows that the order of integration may be reversed as
\begin{equation}
\int_{-1}^{1}\mathrm{d}\zeta\int_{\lambda_{+}}^{\infty}\mathrm{d}\lambda=\int_{\lambda_{\mathrm{L}}}^{\infty}\mathrm{d}\lambda\int_{-1}^{\zeta_{+}}\mathrm{d}\zeta+\int_{\lambda_{\mathrm{min}}}^{\lambda_{\mathrm{L}}}\mathrm{d}\lambda\int_{\zeta_{-}}^{\zeta_{+}}\mathrm{d}\zeta\ , \label{integration_limits}
\label{int_many}
\end{equation}
at the expense of evaluating two different integrals. However, if $\lambda_{\mathrm{L}}\le \lambda_{\mathrm{min}}$ then $\lambda_{\mathrm{min}}=\lambda_{\mathrm{L}}$ and the second term in equation (\ref{int_many}) vanishes, necessitating evaluation of the first double integral only (see Fig. 1).

\section{Performing the Angular moment integrals} 

In evaluating equation (\ref{int_zeta}) with equation (\ref{int_many}), three different types of moment integral arise, namely
\begin{eqnarray}
Q_{n}&=&\int \mathrm{d}\zeta \ \frac{\zeta^{n}}{q}\ , \label{Qn} \\
R_{n}&=&\int\frac{\mathrm{d}\zeta \ \zeta^{n}}{(1-\zeta)^{2}\left(x^{2}+\frac{1+\zeta}{1-\zeta}\right)^{3/2}}\ , \label{Rn} \\
S_{n,m}&=&\int  \frac{\mathrm{d}\zeta\ \zeta^{n}}{(1-\zeta)^{m}\left(x^{2}+\frac{1+\zeta}{1-\zeta}\right)^{1/2}}\ ,  
 \hspace*{0.5cm} m=0, \ 1, \ 2 \ . \label{Snm} 
\end{eqnarray}
Note the identity $\frac{\mathrm{d}S_{n,2}}{\mathrm{d}x}\equiv-xR_{n}$. With the aforementioned definitions the angular moment function of order $n$, $\mathcal{M}_{n}$, may be written as
\begin{eqnarray}
\mathcal{M}_{n}&=&\int \mathrm{d}\zeta\ 
   \zeta^{n}\left[\frac{2\gamma\gamma'}{q}+R\left(\gamma+\lambda\right)-R\left(\gamma'-\lambda\right) \right] \\
&=& A_{n}+B_{n}\left(\gamma+\lambda\right)-B_{n}\left(\gamma'-\lambda\right)\ , \label{Mn}
\end{eqnarray}
where
\begin{eqnarray}
A_{n}&=&2\gamma\gamma'Q_{n}, \label{An} \\
B_{n}&=&\left(w \ \! R_{n}-R_{n+1}\right)+\left(-\gamma\gamma'S_{n,0}+2S_{n,1}+\frac{2}{\gamma\gamma'}S_{n,2} \right)\ . \label{Bn}
\end{eqnarray}
In equations (\ref{Rn}) and (\ref{Snm}), the integrals have an $x$-dependence which is crucial to their evaluation. As noted earlier, $x\equiv(\gamma+\lambda)$ or $x\equiv(\gamma'-\lambda)$, depending on whether the integral is pre-collisional or post-collisional. The evaluation of these integrals yields different results depending on whether $x^{2}<1$, $x^{2}=1$ or $x^{2}>1$.

The moment integrals may be integrated analytically, although the resultant expressions are algebraically cumbersome. The $n=0,1,2$ moments for $A_{0}$ and $B_{0}$ are as follows:
\begin{eqnarray}
A_{0}&=&-2q\ , \label{A0_1} \\
A_{1}&=&-\frac{2q}{3\gamma\gamma'}\left(q^{2}+3\gamma\gamma'\zeta\right)\ , \\
A_{2}&=&-\frac{2q}{15\gamma^{2}\gamma'^{2}}\left[2\left(\gamma^{2}+\gamma'^{2}\right)\left(q^{2}+3\gamma\gamma'\zeta\right)+3\gamma^{2}\gamma'^{2}\zeta^{2}\right]\  , \label{A2_1}
\end{eqnarray}
\\ 
\underline{For $x^{2}\ne 1$:}
\begin{eqnarray}
B_{0}&=&b\left[2-x\left(\gamma^{-1}+\gamma'^{-1}\right)+\frac{2\left(1+\gamma\gamma'\right)}{x^{2}-1}+ \gamma\gamma'\left(1-\zeta\right)+  
\frac{2}{\gamma\gamma'}\left(x^{2}+\omega^{2}\right) \right]+\frac{2\left(2x^{2}+\gamma\gamma'-1\right)}{1-x^{2}}\texttt{C}\left(x\right)\ , \\
\nonumber \\
B_{1}&=&\frac{b}{1-x^{2}}\left\{x\left(1+x^{2}\right)\left(\gamma^{-1}+\gamma'^{-1}\right)-\left(7+\zeta\right)+2x^{2}\left(\zeta-2\right)+ 
\left[\left(1+x^{2}\right)+\left(1-x^{2}\right)\zeta\right]\left[\frac{4\left(1-x^{2}\right)-\gamma^{2}\gamma'^{2}\left(1-\zeta^{2}\right)}{2\gamma\gamma'\left(1-\zeta\right)} \right] \right\} \nonumber  \\ &&   
+\left[\frac{2\left(w-2x^{2}\right)}{x^{2}-1}+\frac{\left(2-\gamma\gamma'\right)\left(2x^{2}+1\right)}{\left(x^{2}-1\right)^{2}}+\frac{4}{\gamma\gamma'} \right]\texttt{C}\left(x\right)\ , \\
B_{2}&=&\frac{b}{(1-x^{2})^{2}}\bigg\{-w\left[2+\zeta+x^{2}\left(3-\zeta\right)+x^{4}\right]+\left[1+\zeta+x^{2}\left(1-\zeta\right)\right]\left[x^{2}\left(3+\zeta\right)-\zeta\right]+  \nonumber \\ &&
                 \frac{6}{x^{2}-1}\Big\{\left(4+9x+2x^{4}\right)\left(3+3x^{2}+\gamma\gamma'\right)+\left(x^{2}-1\right)\left[3\left(\gamma\gamma'-1\right)+2x^{2}\left(\gamma\gamma'-6\right)\right]\zeta+ \nonumber \\ &&
                  \left(x^{2}-1\right)^{2}\left(2\gamma\gamma'-3\right)\zeta^{2}\Big\}+\frac{2\left(1-x^{2}\right)\left[\left(1+\zeta\right)+x^{2}\left(1-\zeta\right)\right]\left(2-x^{2}-\zeta\right)}{\gamma\gamma'\left(1-\zeta\right)}\bigg\}+ \nonumber \\ &&
                  \frac{1}{1-x^{2}}\bigg\{\frac{2\left[w\left(2x^{2}+1\right)-\left(2x^{4}+1\right)\right]}{1-x^{2}}+\frac{3+\gamma\gamma'\left[5+6\left(x^{2}-1\right)+2\left(x^{2}-1\right)^{2}\right]}{\left(1-x^{2}\right)^{2}}+\frac{4\left(1-2x^{2}\right)}{\gamma\gamma'} \bigg\}\texttt{C}\left( x\right) \  ,
\end{eqnarray} 
where
\begin{equation}
b=\frac{\sqrt{1-\zeta}}{\sqrt{\left(1+x^{2}\right)+\left(1-x^{2}\right)\zeta}}\ ,
\end{equation}
and the function $\texttt{C}\left(x\right)$ is defined as
\begin{eqnarray}
\texttt{C}(x) = \left\{
  \begin{array}{ c c }
     \frac{1}{\sqrt{1-x^{2}}}~\textrm{Arctan}\left(\frac{\sqrt{1-x^{2}}\sqrt{1-\zeta}}{\sqrt{\left(1+x^{2}\right)+\zeta\left(1-x^{2}\right)}} \right) \ ,  & \textrm{if} \ x^{2}<1 \ ; \\
  \! \! \!  \! \! \! \! \! \! \! \! \frac{1}{\sqrt{1-x^{2}}}~\textrm{Arcsinh}\left(\frac{\sqrt{x^{2}-1}\sqrt{1-\zeta}}{\sqrt{2}} \right) \ , & \textrm{if} \ x^{2}>1 \ .
  \end{array} \right.
\end{eqnarray}
\\ 
\underline{For $x^{2}=1$:}
\begin{equation}
B_{0} = \sqrt{\frac{1-\zeta}{2}}\left[\frac{4}{\gamma\gamma'(1-\zeta)}-4-w+\frac{2\gamma\gamma'(1-\zeta)}{3}+\frac{2+\zeta}{3} \right]\ ,
\end{equation}
\begin{eqnarray}
B_{1} =  \sqrt{\frac{1-\zeta}{2}}\left[\frac{4(2-\zeta)}{\gamma\gamma'}(1-\zeta)-\frac{4}{3}(2+\zeta)-\frac{w(2+\zeta)}{3}+  
\frac{2\gamma\gamma'(1-\zeta)(2+3\zeta)}{15}+\frac{8+4\zeta+3\zeta^{2}}{15} \right]\ ,
\end{eqnarray}
\begin{eqnarray}
B_{2} =  \sqrt{\frac{1-\zeta}{2}}\left[\frac{4(8-4\zeta-\zeta^{2})}{3\gamma\gamma'(1-\zeta)}-\frac{(4+w)(8+4\zeta+3\zeta^{2})}{15}+  
\frac{2\gamma\gamma'(1-\zeta)(8+12\zeta+15\zeta^{2})}{105}+\frac{16+8\zeta+6\zeta^{2}+5\zeta^{3}}{35} \right]\ . 
\label{B2_b}
\end{eqnarray}
In principle equations (\ref{Qn})--(\ref{Snm}) may be integrated for arbitrary $n$, 
but, as seen in equations (\ref{A0_1})--(\ref{B2_b}), the resultant algebraic expressions become extremely cumbersome. Moreover, the expressions for $A_{n}$ must be evaluated either two or four times per scattering event, and $B_{n}$ either four or eight times per scattering event. Given the inherent algebraic complexity, and the number of calls required per scattering event, this will lead to significant loss of precision, in particular between cancellations of terms of similar value or of particular smallness \citep[e.g.][]{Poutanen2010}.

Using equations (\ref{Mn})--(\ref{Bn}) the Compton scattering kernel may be written more compactly as
\begin{equation}
\sigma_{\mathrm{s}n}\left(\gamma\rightarrow\gamma',\tau\right)=\frac{3\rho\sigma_{\mathrm{T}}}{8\gamma\nu}
  \left( \int_{\lambda_{\mathrm{L}}}^{\infty}  \mathrm{d}\lambda \   \frac{f(\lambda)}{\lambda^{5}}\mathcal{M}_{n}\big|_{-1}^{\zeta_{+}} 
   +\int_{\lambda_{\mathrm{min}}}^{\lambda_{\mathrm{L}}} \mathrm{d}\lambda\   \frac{f(\lambda)}{\lambda^{5}}\mathcal{M}_{n}\big|_{\zeta_{-}}^{\zeta_{+}}   \right)\ ,
\label{scat2}
\end{equation}
where, as noted before, the second term in square brackets in the above equation vanishes when $\lambda_{\mathrm{L}}\le\lambda_{\mathrm{min}}$, saving significant computational expense. In the case of $x^{2}=1$, the moment integrals simplify significantly. This is as far as it proves possible to proceed analytically. Integrations over $\lambda$ would have to be performed with an appropriate numerical scheme.

Naturally, the question arises as to whether the integrals in equation (\ref{scat2}) can be performed analytically. As it stands, the method presented thus far would require arbitrary precision arithmetic to evaluate, and therefore be computationally expensive and time consuming. In the following section, the evaluation of integrals (\ref{Qn})--(\ref{Snm}) is demonstrated analytically and in closed form, for arbitrary moment order.

\section{Evaluating the Moment Integrals for Arbitrary order}
The previous section derived analytic expressions for the first three moments of the Compton scattering kernel. As the order of the moments increases, the algebraic complexity of the resultant expression grows rapidly. Clearly the method, as it stands, does not lend itself readily to the evaluation of higher-order moments. These are necessary for more accurate evaluation of radiation transport problems. A much faster method is to evaluate equations (76)--(78) recursively.
Firstly, consider equation (\ref{Qn}) for $Q_{n}$.
By employing the identity
\begin{equation}
\frac{\mathrm{d}q}{\mathrm{d}\zeta}=-\frac{\gamma\gamma'}{\zeta} \ ,
\end{equation}
upon integrating $Q_{n}$ by parts, the following recurrence relation immediately follows
\begin{equation}
\gamma\gamma'\left(2n+1\right)Q_{n}=\left(\gamma^{2}+\gamma'^{2}\right)n \ \! Q_{n-1}-q \ \! \zeta^{n} \ . \label{Qn_it}
\end{equation}
With the seed $Q_{0}=\left(\sqrt{\gamma^{2}+\gamma'^{2}}-q\right)/\gamma\gamma'$, $Q_{n}$ may be evaluated for arbitrary $n$. Next consider equation (\ref{Rn}) in the form
\begin{equation}
R_{n}=-\frac{1}{2\sqrt{2}}\int \mathrm{d}u \ \! u^{-1/2}\left(1-u\right)^{n}\left(1- c \ \! u\right)^{-3/2} \ , \label{Rn_integral}
\end{equation}
where the substitution $u=1-\zeta$ has been employed, and $c\equiv (1-x^{2})/2$. By expanding in series the term $\left(1-u\right)^{n}$, equation (\ref{Rn_integral}) may be written as
\begin{equation}
R_{n}=\sum_{k=0}^{n}\frac{\left(-1\right)^{k+1}{{n}\choose{k}}}{2\sqrt{2}}\int du\frac{u^{k-1/2}}{\left(1- c \ \! u\right)^{3/2}} \ . 
\end{equation}
Defining the integral
\begin{equation}
I_{_{R}}\left(k\right)=\int \mathrm{d}u\frac{u^{k-1/2}}{\left(1- c \ \! u\right)^{3/2}} \ , \label{Ir_kp1}
\end{equation}
a recursion relation for equation (\ref{Ir_kp1}) may be found by integrating by parts
\begin{equation}
2\left(k-1\right)c \ \! I_{_{R}}(k)=\left(2k-1\right) I_{_{R}}\left(k-1\right)-\frac{2u^{k-1/2}}{\sqrt{1-c \ \! u}}\ .
\end{equation}
The value $ I_{_{R}}(0)$ immediately follows, but to perform recursively the seed value $ I_{_{R}}(1)$ is also needed
\begin{equation}
 I_{_{R}}(1)=\frac{2\sqrt{u}}{c\sqrt{1-c \ \! u}}-\frac{2}{c^{3/2}}\mathrm{arcsin}\left(\sqrt{c \ \! u}\right) \ .
\end{equation}
Therefore $R_{n}$ may now be defined as
\begin{equation}
R_{n}=\sum_{k=0}^{n}\frac{(-1)^{k+1}{{n}\choose{k}}}{2\sqrt{2}}I_{_{R}}(k) \ , \label{Rn_it}
\end{equation}
which can be solved for arbitrary $n$.
Similarly, for $S_{n,m}$
\begin{equation}
S_{n,m}=\sum_{k=0}^{n}\frac{(-1)^{k+1}{{n}\choose{k}}}{\sqrt{2}}I_{_{S}}\left(k,m\right) \ , \label{Snm_it}
\end{equation}
where
\begin{equation}
I_{_{S}}(k,m)=\int \mathrm{d}u \frac{u^{k-m+1/2}}{\sqrt{1-c\ \! u}} \ .
\end{equation}
After some working, the recursion relation for $I_{_{S}}(k,m)$ is obtained as
\begin{equation}
\left(k-m+1\right)c \ \! I_{_{S}}\left(k,m\right)=\left(k-m+1/2\right)I_{_{S}}\left(k-1,m\right)-u^{k-m+1/2}\sqrt{1-c \ \! u} \ .
\end{equation}
This identity requires four different seed values for the cases $m=0$, $1$ and $2$:
\begin{eqnarray}
I_{_{S}}(0,0)&=&\frac{u}{2c}I_{_{S}}(0,2)+\frac{1}{2c}I_{_{S}}(0,1) \ , \\
I_{_{S}}(0,1)&=&\frac{2 \ \! \textrm{arcsin}\left(\sqrt{c \ \! u} \right)}{\sqrt{c}} \ , \\
I_{_{S}}(0,2)&=&-\frac{2\sqrt{1-c \ \! u}}{\sqrt{u}} \ , \\
I_{_{S}}(1,2)&=& I_{_{S}}(0,1) \ .
\end{eqnarray}
The numerical evaluation of these recursion relations in \textsc{Fortran95} is shown in Fig. 2 for $Q_{n}$ and $S_{n,2}$. For $Q_{n}$ it is clear the method is inaccurate for $n>20$, regardless of the cosine of the scattering angle, $\zeta$. For $R_{n}$ the method is numerically unstable for $n>30$ for $\zeta=-1$, as well as slowly convergent, regardless of the value of $x$. However, for $\zeta>-1$ the method appears both numerically stable and rapidly convergent, even for $n=50$. Similar results are obtained for $S_{n,m}$ as for $R_{n}$, with the exception that for lower energies, $S_{n,0}$ is numerically unstable both for extreme backward scattering and extreme forward scattering beyond $n=30$. More accurate evaluation would require the implementation of arithmetic precision beyond that of standard double precision.

Thus equations (\ref{Qn_it}), (\ref{Rn_it}) and (\ref{Snm_it}) enable (\ref{Mn}) to be solved iteratively. In computing angular moments of the Klein-Nishina cross-section this will greatly reduce the computational time and resources required. Each moment integral can be computed recursively using the stored numerical value of the previous moment. Unfortunately, as the order increases, there will inevitably be loss of precision through differences of terms in the recursion relations. Further, it is impossible to perform the final integral over the electron distribution function without either an algebraic expression for each moment, or an appropriate closed-form expression for each moment in terms of more generalised functions. The following sections detail such a method based on the latter.

\section{Evaluating Moment Integrals - Hypergeometric Function Method}
In this section the moment integrals in equations (\ref{Qn})--(\ref{Snm}) are evaluated in terms of ordinary hypergeometric functions \citep{Bateman1955}. In terms of this function, the problem of relativistic Compton scattering is greatly simplified \citep{Aharonian1981}. Hypergeometric functions are a very general class of functions which contain many of the known mathematical functions as special or limiting cases \citep{Luke1969, Abramowitz1972}.

The ordinary hypergeometric function of one variable, or Gauss hypergeometric function \citep{Gauss1866}, is defined by the series
\begin{equation}
_{2}F_{1}\left(a,b;c;z\right)=\sum_{n=0}^{\infty}\frac{\left(a\right)_{n}\left(b\right)_{n}}{\left(c\right)_{n}\ \! n!}z^{n} \ ,
\end{equation}
where the notation
\begin{equation}
(a)_{n}\equiv\frac{\Gamma(a+n)}{\Gamma(a)} \ ,
\end{equation}
is the rising factorial or Pochhammer symbol \citep{Bateman1955}. The series is absolutely convergent for $|z|<1$, and terminates after a finite number of terms if either $a$ or $b$ is a negative integer. The case $|z|\ge 1$ may be solved by analytic continuation \citep{Zhang1996}. Although $z$ may take complex values, in this paper $z$ is always real. With this definition the integrals $Q_{n}$, $R_{n}$ and $S_{n,m}$ may be solved. Having written $R_{n}$ and $S_{n,m}$ in summation form in equations (\ref{Rn_it}) and (\ref{Snm_it}) simplifies things considerably. Using the series expansion $(1-u)^{n}=\sum_{k=0}^{n}(-1)^{k}{{n}\choose{k}}u^{k}$, the following expressions for (\ref{Qn})--(\ref{Snm}) are found
\begin{eqnarray}
Q_{n} &=& \frac{\zeta^{n+1}}{(n+1)\sqrt{\gamma^{2}+\gamma'^{2}}}{}_{2}\textrm{F}_{1}\left[\frac{1}{2},n+1;n+2;\frac{2\gamma\gamma'}{\gamma^{2}+\gamma'^{2}}\zeta \right]\ , \label{Qn_Gauss} \\
R_{n} &=& -\frac{(1-\zeta)^{1/2}}{\sqrt{2}}\sum_{k=0}^{n}\frac{{{n}\choose{k}}(\zeta-1)^{k}}{2k+1}{}_{2}\textrm{F}_{1}\left[\frac{3}{2},k+\frac{1}{2};k+\frac{3}{2};\frac{1}{2}\left(1-x^{2}\right)\left(1-\zeta\right) \right] \ , \label{Rn_Gauss} \\
S_{n,m} &=& -\sqrt{2}(1-\zeta)^{\frac{3}{2}-m}\sum_{k=0}^{n}\frac{{{n}\choose{k}}(\zeta-1)^{k}}{2k-2m+3}\ \! {}_{2}\textrm{F}_{1}\left[\frac{1}{2},k-m+\frac{3}{2};k-m+\frac{5}{2}; \frac{1}{2}\left(1-x^{2}\right)\left(1-\zeta\right) \right] \ . \label{Snm_Gauss}
\end{eqnarray}
Here are a few notes about the continuity of expressions (\ref{Qn_Gauss})--(\ref{Snm_Gauss}). $Q_{n}$ is always within the convergence region, and only lies on the boundary in the case of a perfectly elastic collision i.e. Thomson scattering \citep{Thomson1906}. Equations (\ref{Rn_Gauss}) and (\ref{Snm_Gauss}) can be divided into two cases: those which lie within the convergence region ($|z|<1$) and those that lie on the boundary or outside it ($z\le -1$). The case $z\le -1$, i.e. $\zeta\ge(x^{2}+1)/(x^{2}-1)$, may be solved by analytic extension with the following expression
\begin{equation}
{}_{2}\mathrm{F}_{1}\left[a,b;b+1;z\right]=(1-z)^{-a}{}_{2}\mathrm{F}_{1}\left[a,1;b+1;\frac{z}{z-1}\right] \ ,
\end{equation}
which brings $R_{n}$ and $S_{n,m}$ into the convergence region. The Gauss hypergeometric function is well documented in the literature and there exist several codes in \textsc{Fortran} which can evaluate it both accurately and rapidly \citep[e.g.][]{Forrey1997,Zhang1996}, in addition to handling all cases of differences of parameters and values which can give rise to numerical problems \citep[e.g.][]{Zhang1996}.

In the special case $x^{2}=1$ the expressions for $R_{n}$ and $S_{n,m}$ reduce to
\begin{eqnarray}
R_{n}&=&-\frac{\left(1-\zeta\right)^{1/2}}{\sqrt{2}}{}_{2}\mathrm{F}_{1}\left(\frac{1}{2},-n;\frac{3}{2};1-\zeta\right) \ , \label{Rn_xe1} \\
S_{n,m}&=& -\frac{\sqrt{2}\left(1-\zeta\right)^{\frac{3}{2}-m}}{3-2m}{}_{2}\mathrm{F}_{1}\left(\frac{3}{2}-m,-n;\frac{5}{2}-m;1-\zeta \right) \ , \label{Snm_xe1}
\end{eqnarray}
which are detailed in Appendix \ref{AppendixA}.

Thus the moment integrals for all values of $x$ have been defined in closed-form. Results of the direct numerical evaluation of the moment integrals $Q_{n}$ and $S_{n,2}$ are presented in Fig. 3. For $Q_{n}$ the direct hypergeometric function method is a significant improvement. This is obvious since, in closed-form, $Q_{n}$ only ever requires one function evaluation, irrespective of the moment order. However, for $R_{n}$ and $S_{n,m}$ this method fares no better, and is in fact worse for larger scattering angles than the recursive method. This is due to oscillating sums in the corresponding expressions. However, the closed-form nature of these expressions is necessary to define the scattering kernel analytically. Plots of the numerical evaluation of the moment integral $\mathcal{M}_{n}$ as a function of $n$, evaluated in \textsc{Python} to high numerical precision, are shown in Fig. 4. For very low scattering angles the angular moments are oscillatory, as can be seen in the $\zeta=-1$ case. However, this is not a numerical issue, but rather an intrinsic physical issue with the form of the Compton scattering kernel itself. Recall equation (\ref{q}), which was derived in taking the direction of photon momentum transfer as the $\mathit{z}$-axis of integration. In doing this, $q$ is uniquely defined by equation (\ref{q}) and so the method is inherently somewhat oscillatory for $\zeta$ close to $-1$, i.e. scattering angles close to $0$.

In Fig. 5, $\mathcal{M}_{n}$ is plotted as a function of $\zeta$ for low order and high order, odd and even moments $n$. Odd and even moments are plotted separately to emphasise the change in shape and decrease in size of $\mathcal{M}_{n}$ as the order increases. Odd and even moments have a distinct shape which flattens and decreases in magnitude as the order increases. Clearly as the moment order increases, $\mathcal{M}_{n}$ becomes less sensitive to moderate scattering angles and remains unchanged over an increasingly large range of $\zeta$. The effect of increasing electron velocity is to shift the maximum of $\mathcal{M}_{n}$ towards $\zeta=1$, i.e. back scattering, as well as reducing the absolute magnitude of $\mathcal{M}_{n}$.

The remainder of the paper proceeds with the hypergeometric function method, with the aforementioned numerical considerations in mind.
The final step in computing the Compton scattering cross-section is integrating over the relativistic electron distribution function, which is detailed in the next section.

\section{Integrating over the electron distribution function}
In the general case, in all of the literature at present, only integration over $\zeta$ or $\lambda$ has been performed analytically --- generally a choice must be made between performing integrals of the angular moments or integrating over the electron distribution function. The sixth and final integration over photon energy can be performed numerically during the radiative transfer calculations at each point along a ray. Regardless, with the methods at present, one is left with at best two further sets of integrals to evaluate. Further, the problem as formulated in the current literature \citep{Prasad1986,Nagirner1993,Poutanen2010} is algebraically cumbersome. It is common to resort to Monte-Carlo methods to solve the multi-dimensional integrals. To have a closed-form solution to the first five integrals, including the electron distribution function, would eliminate the need for evaluating multi-dimensional integrals and entail solving only the photon frequency integral along the ray, as is common in ray-tracing \citep[see e.g.][]{Vincent2011,Younsi2012}.
\\
\subsection{Integrating over the electron distribution function for constant $\zeta$}
Convolving the moment integrals with the electron distribution function necessitates solving integrals of the form
\begin{equation}
T = -\frac{\tau}{2}\mathrm{e}^{\pm \gamma^{(')}/\tau}\int \mathrm{d}y \frac{\mathrm{e}^{-\sqrt{1-y}}}{\sqrt{1-y}}  {}_{2}\mathrm{F}_{1}\left(a,b;c;\alpha+\beta y \right) \ , \label{T}
\end{equation}
where the change of variable for pre-collision (post-collision) as $\tilde{x}=\gamma+\lambda$ ($\tilde{x}=\lambda-\gamma'$), followed by $y=1-\tilde{x}^{2}/\tau^{2}$ has been introduced. The $\pm$ sign indicates pre/post-collision and $\alpha=u(1-\tau^{2})/2$, $\beta=u\tau^{2}/2$.

Consider the Taylor expansion
\begin{equation}
 \frac{\mathrm{e}^{-\sqrt{1-y}}}{\sqrt{1-y}}=\frac{1}{\mathrm{e}}\sum_{p=0}^{\infty}\frac{\mathrm{s}_{p}}{2^{p}p!}y^{p} \ ,
\end{equation}
where the recursion $\mathrm{s}_{p}$ is defined as
\begin{eqnarray}
\mathrm{s}_{p}&=&(2p-1)\mathrm{s}_{p-1}+\mathrm{s}_{p-2} \ , \\
\mathrm{s}_{0}&=& 1 \ , \\
\mathrm{s}_{1}&=& 2 \ .
\end{eqnarray}
Next consider the integral
\begin{equation}
J_{p}=\int \mathrm{d}y \ \! y^{p} \ \! {}_{2}\mathrm{F}_{1}\left(a,b;c;\alpha+\beta y\right) \ . \label{J_p}
\end{equation}
The integral (\ref{J_p}) may be easily solved recursively:
\begin{equation}
J_{p}\left(a,b,c\right)=y^{p}J_{0}\left(a,b,c\right) -\frac{p\left(c-1\right)}{\left(a-1\right)\left(b-1\right)\beta}J_{p-1}\left(a-1,b-1,c-1\right) \ ,
\end{equation}
where
\begin{equation}
J_{0}\left(a,b,c\right)=\frac{\left(c-1\right)}{\left(a-1\right)\left(b-1\right)\beta}{}_{2}\mathrm{F}_{1}\left(a-1,b-1;c-1;\alpha+\beta y \right) \ .
\end{equation}
Hence equation (\ref{T}) may now be written as
\begin{equation}
T=-\frac{\tau}{2}\mathrm{e}^{-1\pm \gamma^{(')}/\tau}\sum_{p=0}^{\infty}\frac{s^{p}}{2^{p}p!}J_{p} \ .
\end{equation}
For fixed values of $\zeta$ this method works well. However, in the case of evaluating the full Compton scattering kernel, from the limits of integration in equation (\ref{integration_limits}) it is clear that $\zeta_{\pm}$ is a function of $\lambda$. Consequently, the integral in equation (\ref{J_p}) is no longer trivial and cannot be expressed in closed-form. The resultant integrations over $\lambda$ must the be performed numerically. An algorithm to perform the integration is presented in the following subsection.

\subsection{Integrating over the electron distribution function in general}
To evaluate the full Compton scattering kernel, in full generality, there are two expressions of importance, namely
\begin{eqnarray}
A_{n}&=&2\gamma\gamma'Q_{n} \ , \\
B_{n}&=&x(\gamma^{-1}+\gamma'^{-1})R_{n}-(R_{n}+R_{n+1})-\gamma\gamma'S_{n,0}+2\ \! S_{n,1}+\frac{2}{\gamma \gamma'}S_{n,2} \ .
\end{eqnarray}
Computing the Compton scattering kernel involves evaluating the following expression:
\begin{eqnarray}
T &=& T_{1}+T_{2}\nonumber \\ 
&=&\int_{\lambda_{\mathrm{L}}}^{\infty}\mathrm{d}\lambda \ \! \mathrm{e}^{-\lambda/\tau}\left[A_{n}+B_{n}(\gamma+\lambda)-B_{n}(\gamma'-\lambda) \right]\big\vert_{-1}^{\zeta_{+}}+\int_{\lambda_{\mathrm{min}}}^{\lambda_{\mathrm{L}}}\mathrm{d}\lambda \ \! \mathrm{e}^{-\lambda/\tau}\left[A_{n}+B_{n}(\gamma+\lambda)-B_{n}(\gamma'-\lambda) \right]\big\vert_{\zeta_{-}}^{\zeta_{+}} \ . \label{T_full}
\end{eqnarray}
The second term in equation (\ref{T_full}), $T_{2}$, vanishes if the condition given by equation (\ref{Condition}) is satisfied, as noted previously. Solving equation (\ref{T_full}) necessitates the definition of the following seven integrals
\begin{eqnarray}
I_{1}(\zeta,\lambda_{1},\lambda_{2}) &=& 2\gamma\gamma' \int_{\lambda_{1}}^{\lambda_{2}}\mathrm{d}\lambda \ \! \mathrm{e}^{-\lambda/\tau}\ \! Q_{n}(\zeta) \nonumber \\
&=& \frac{2\gamma\gamma'}{(n+1)\sqrt{\gamma^{2}+\gamma'^{2}}}\int_{\lambda_{1}}^{\lambda_{2}}\mathrm{d}\lambda \ \! \mathrm{e}^{-\lambda/\tau}\ \! \zeta^{n+1} \ \! {}_{2}\mathrm{F}_{1}\left[\frac{1}{2},n+1;n+2;\frac{2\gamma\gamma'}{\gamma^{2}+\gamma'^{2}}\zeta \right] \ , \\
I_{2}(\zeta,x,\lambda_{1},\lambda_{2}) &=& \int_{\lambda_{1}}^{\lambda_{2}}\mathrm{d}\lambda \ \! \mathrm{e}^{-\lambda/\tau}\ \! R_{n}(\zeta,x) \nonumber \\
&=& \frac{1}{\sqrt{2}} \sum_{k=0}^{n}\frac{(-1)^{k+1}{{n}\choose{k}}}{(2k+1)}\int_{\lambda_{1}}^{\lambda_{2}}\mathrm{d}\lambda \ \! \mathrm{e}^{-\lambda/\tau}\ \! (1-\zeta)^{k+\frac{1}{2}}\ \! {}_{2}\mathrm{F}_{1}\left[\frac{3}{2},k+\frac{1}{2};k+\frac{3}{2};z \right] \ , \\ 
I_{3}(\zeta,x,\lambda_{1},\lambda_{2}) &=& (\gamma^{-1}+\gamma'^{-1})\int_{\lambda_{1}}^{\lambda_{2}}\mathrm{d}\lambda \ \! \mathrm{e}^{-\lambda/\tau}\ \! \lambda \ \! R_{n}(\zeta,x) \nonumber \\
&=& \frac{(\gamma^{-1}+\gamma'^{-1})}{\sqrt{2}}\sum_{k=0}^{n}\frac{(-1)^{k+1}{{n}\choose{k}}}{(2k+1)}\int_{\lambda_{1}}^{\lambda_{2}}\mathrm{d}\lambda \ \! \mathrm{e}^{-\lambda/\tau}\ \! \lambda \ \! (1-\zeta)^{k+\frac{1}{2}}\ \! {}_{2}\mathrm{F}_{1}\left[\frac{3}{2},k+\frac{1}{2};k+\frac{3}{2};z \right] \ , \\
I_{4}(\zeta,x,\lambda_{1},\lambda_{2}) &=& \int_{\lambda_{1}}^{\lambda_{2}}\mathrm{d}\lambda \ \! \mathrm{e}^{-\lambda/\tau}\ \! R_{n+1}(\zeta,x) \nonumber \\
&=& \frac{1}{\sqrt{2}}\sum_{k=0}^{n+1}\frac{(-1)^{k+1}{{n+1}\choose{k}}}{(2k+1)}\int_{\lambda_{1}}^{\lambda_{2}}\mathrm{d}\lambda \ \! \mathrm{e}^{-\lambda/\tau}\ \! (1-\zeta)^{k+\frac{1}{2}}\ \! {}_{2}\mathrm{F}_{1}\left[\frac{3}{2},k+\frac{1}{2};k+\frac{3}{2};z \right] \ , \\ 
I_{5}(\zeta,x,\lambda_{1},\lambda_{2}) &=& \gamma \gamma' \int_{\lambda_{1}}^{\lambda_{2}}\mathrm{d}\lambda \ \! \mathrm{e}^{-\lambda/\tau}\ \! S_{n,0}(\zeta,x) \nonumber \\
&=& \sqrt{2}\gamma\gamma'\sum_{k=0}^{n}\frac{(-1)^{k+1}{{n}\choose{k}}}{(2k+3)}\int_{\lambda_{1}}^{\lambda_{2}}\mathrm{d}\lambda \ \! \mathrm{e}^{-\lambda/\tau}\ \! (1-\zeta)^{k+\frac{3}{2}}\ \! {}_{2}\mathrm{F}_{1}\left[\frac{1}{2},k+\frac{3}{2};k+\frac{5}{2};z\right] \ , \\
I_{6}(\zeta,x,\lambda_{1},\lambda_{2}) &=& 2\int_{\lambda_{1}}^{\lambda_{2}}\mathrm{d}\lambda \ \! \mathrm{e}^{-\lambda/\tau}\ \! S_{n,1}(\zeta,x) \nonumber \\
&=& 2\sqrt{2} \sum_{k=0}^{n}\frac{(-1)^{k+1}{{n}\choose{k}}}{(2k+1)}\int_{\lambda_{1}}^{\lambda_{2}}\mathrm{d}\lambda \ \! \mathrm{e}^{-\lambda/\tau}\ \!(1-\zeta)^{k+\frac{1}{2}}\ \! {}_{2}\mathrm{F}_{1}\left[\frac{1}{2},k+\frac{1}{2};k+\frac{3}{2}; z\right] \ , \\
I_{7}(\zeta,x,\lambda_{1},\lambda_{2}) &=& \frac{2}{\gamma\gamma'}\int_{\lambda_{1}}^{\lambda_{2}}\mathrm{d}\lambda \ \! \mathrm{e}^{-\lambda/\tau}\ \! S_{n,2}(\zeta,x) \nonumber \\
&=& \frac{2\sqrt{2}}{\gamma\gamma'}\sum_{k=0}^{n}\frac{(-1)^{k+1}{{n}\choose{k}}}{(2k-1)}\int_{\lambda_{1}}^{\lambda_{2}}\mathrm{d}\lambda \ \! \mathrm{e}^{-\lambda/\tau}\ \! (1-\zeta)^{k-\frac{1}{2}}\ \! {}_{2}\mathrm{F}_{1}\left[\frac{1}{2},k-\frac{1}{2};k+\frac{1}{2}; z\right] \ ,
\end{eqnarray}
where $z\equiv \left(1-x^{2}\right)(1-\zeta)/2$ and the dependence of $\zeta$ and $x$ on $\lambda$ has been neglected, i.e. $\zeta\equiv\zeta(\lambda)$ and $x\equiv x(\lambda)$. Consider the functions
\begin{eqnarray}
f(\zeta,x,\lambda_{1},\lambda_{2})  &=& \frac{\gamma}{\gamma'}I_{2}+I_{3}-I_{4}-I_{5}+I_{6}+I_{7} \ , \label{f} \\
g(\zeta,x,\lambda_{1},\lambda_{2}) &=& \frac{\gamma'}{\gamma}I_{2}-I_{3}-I_{4}-I_{5}+I_{6}+I_{7} \ , \label{g}
\end{eqnarray}
where the dependence of $I$ on $\zeta$, $x$, $\lambda_{1}$ and $\lambda_{2}$ has been suppressed for the sake of brevity. The Compton scattering kernel may then be expressed as the composition of the following ten terms
\begin{eqnarray}
t_{1} &=& I_{1}(\zeta_{+},\lambda_{\mathrm{L}},\infty)-2\gamma\gamma'\tau \ \! Q_{n}(-1)\ \! \mathrm{e}^{-\lambda_{\mathrm{L}}/\tau} \ , \\
t_{2} &=& I_{1}(\zeta_{+},\lambda_{\mathrm{min}},\lambda_{\mathrm{L}})-I_{1}(\zeta_{-},\lambda_{\mathrm{min}},\lambda_{\mathrm{L}}) \ , \\
t_{3} &=& f(\zeta_{+},\gamma+\lambda,\lambda_{\mathrm{L}},\infty) \ , \\
t_{4} &=& f(-1,\gamma+\lambda,\lambda_{\mathrm{L}},\infty) \ , \\
t_{5} &=& g(\zeta_{+},\gamma'-\lambda,\lambda_{\mathrm{L}},\infty) \ , \\
t_{6} &=& g(-1,\gamma'-\lambda,\lambda_{\mathrm{L}},\infty) \ , \\
t_{7} &=& f(\zeta_{+},\gamma+\lambda,\lambda_{\mathrm{min}},\lambda_{\mathrm{L}}) \ , \\
t_{8} &=& f(\zeta_{-},\gamma+\lambda,\lambda_{\mathrm{min}},\lambda_{\mathrm{L}}) \ , \\
t_{9} &=& g(\zeta_{+},\gamma'-\lambda,\lambda_{\mathrm{min}},\lambda_{\mathrm{L}}) \ , \\
t_{10} &=& g(\zeta_{-},\gamma'-\lambda,\lambda_{\mathrm{min}},\lambda_{\mathrm{L}}) \ ,
\end{eqnarray}
where $Q_{n}(-1)$ is equivalent to $Q_{n}$ evaluated at $\zeta=-1$. Recall $\zeta_{\pm}\equiv\zeta_{\pm}(\lambda)$, as given in equation (\ref{zeta_pm}). Terms $t_{1}$ (pre-collisional) and $t_{2}$ (post-collisional) are independent of $x$. With the above ten terms $T_{1}$ and $T_{2}$ may now be written as
\begin{eqnarray}
T_{1} &=& t_{1}+t_{3}-t_{4}-t_{5}+t_{6} \ , \\
T_{2} &=& t_{2}+t_{7}-t_{8}-t_{9}+t_{10} \ ,
\end{eqnarray}
where, as noted before, $T_{2}$ vanishes if condition (\ref{Condition}) is satisfied. With $T_{1}$ and $T_{2}$ expressed, one may now evaluate equation (\ref{T_full}) numerically. It is easily shown that the number of numerical integrals scales linearly with the moment order $n$ and is given by $48n+51$ or $24n+25$, depending on whether $T_{2}$ need be evaluated. However, this is assuming the independent evaluation of each moment. In reality, in evaluating a moment $n$, all lower-order moments must also have been evaluated, and so the order of the method at each order $n$ is given by $(n+1)(24n+51)$ or $(n+1)(12n+25)$. 

The angular moments of the full Klein-Nishina Compton scattering kernel may now finally be written as:
\begin{eqnarray}
\sigma_{\mathrm{KN}}(\gamma\rightarrow\gamma',\tau) &=& \int \mathrm{d}\zeta \! \ \zeta^{n} \ \! \sigma_{\mathrm{S}}(\gamma\rightarrow\gamma',\zeta,\tau) \nonumber \\
&=&\frac{\mathcal{C}}{\gamma^{2}\ \! \tau \ \! \mathrm{K}_{2}(1/\tau)}T(\gamma,\gamma',\tau) \ , \label{sigmaKN}
\end{eqnarray}
where $T(\gamma,\gamma',\tau)\equiv T$, as given in equation (\ref{T_full}) and $\mathcal{C}=3\rho\sigma_{\mathrm{T}}/32\pi m_{\mathrm{e}}$. 

\subsection{Numerical implementation}
In implementing the formulation in the previous subsection numerically, several considerations and modifications of the formulae need to be considered. A prominent problem is the magnitude of the $1/\tau \ \! \mathrm{K}_{2}(1/\tau)$ term in the expression for the scattering kernel at electron temperatures below $10 \ \mathrm{keV}$. For an electron temperature of $10 \ \mathrm{keV}$ its value is $4.378\times 10^{24}$, at $1 \ \mathrm{keV}$ its value is $7.717\times 10^{225}$ and moving down to temperatures of  $1 \ \mathrm{meV}$, the lower-end of temperatures we will investigate numerically, the corresponding value is $1.649\times 10^{221924493}$. On this basis alone, any numerical computation of the scattering kernel would immediately require very high numerical precision indeed, particularly at temperatures below $1 \ \mathrm{keV}$. Accordingly, all of the numerical integrals in Equations (134)--(140), particularly in the case of nearly elastic collisions, will be of corresponding numerical smallness so as to cancel such large terms, since the value of the scattering kernel in this case is generally of the order of unity. Consequently, these numerical integrals will also require substantial numerical precision in memory storage alone.

Another issue is the need to define an efficient algorithm which computes the integrals and sums in Equations (134)--(140) with the minimum of computational overhead. Some integrals are repeated and consequently we introduce a new notation to make the formulation and its numerical implementation more transparent. Consider the following integral definition:
\begin{equation}
\mathcal{F}_{k}(a,b,\alpha)=\int_{\lambda_{1}}^{\lambda_{2}}\mathrm{d}\lambda \ \! \mathrm{e}^{-(\lambda-1)/\tau} \ \! \lambda^{\alpha} \ \! (1-\zeta)^{b}{}_{2}\mathrm{F}_{1}\left[a,b;b+1;z \right]. \label{Fk}
\end{equation}
We may rewrite Equations (134)--(140) as follows
\begin{eqnarray}
I_{1} &=& \frac{2\gamma\gamma'}{(n+1)\sqrt{\gamma^{2}+\gamma'^{2}}}\int_{\lambda_{1}}^{\lambda_{2}}\mathrm{d}\lambda \ \! \mathrm{e}^{-(\lambda-1)/\tau}\ \! \zeta^{n+1} \ \! {}_{2}\mathrm{F}_{1}\left[\frac{1}{2},n+1;n+2;\frac{2\gamma\gamma'}{\gamma^{2}+\gamma'^{2}}\zeta \right] \ \label{I1new}, \\
I_{2} &=& \frac{1}{\sqrt{2}}\sum_{k=0}^{n}\mathcal{D}(n,k,1)\ \! \mathcal{F}_{k}\left(\frac{3}{2},k+\frac{1}{2},0 \right) \ \label{I2}, \\
I_{3} &=& \frac{(\gamma^{-1}+\gamma'^{-1})}{\sqrt{2}}\sum_{k=0}^{n}\mathcal{D}(n,k,1)\ \! \mathcal{F}_{k}\left(\frac{3}{2},k+\frac{1}{2},1 \right) \ , \\
I_{4} &=& \frac{1}{\sqrt{2}}\sum_{k=0}^{n+1}\mathcal{D}(n+1,k,1)\ \! \mathcal{F}_{k}\left(\frac{3}{2},k+\frac{1}{2},0 \right) \ \label{I4}, \\
I_{5} &=& \sqrt{2}\gamma \gamma' \sum_{k=0}^{n}\mathcal{D}(n,k,2)\ \! \mathcal{F}_{k}\left(\frac{1}{2},k+\frac{3}{2},0 \right) \ , \\
I_{6} &=& 2\sqrt{2}\sum_{k=0}^{n}\mathcal{D}(n,k,1)\ \! \mathcal{F}_{k}\left(\frac{1}{2},k+\frac{1}{2},0 \right) \ , \\
I_{7} &=& \frac{2\sqrt{2}}{\gamma\gamma'}\sum_{k=0}^{n}\mathcal{D}(n,k,0)\ \! \mathcal{F}_{k}\left(\frac{1}{2},k-\frac{1}{2},0 \right) \ ,
\end{eqnarray}
where,
\begin{equation}
\mathcal{D}\left(n,k,l \right)=\frac{(-1)^{k+1}}{2k+2l-1}{{n}\choose{k}}.
\end{equation}
Note that the integrals in Equations (\ref{I2}) and (\ref{I4}) are identical, thus only the integral $\mathcal{F}_{n+1}\left(3/2,n+3/2,0\right)$ need be computed in $I_{4}$. With this, the scattering kernel may be written as
\begin{equation}
\sigma_{\mathrm{KN}}(\gamma\rightarrow\gamma',\tau) = \frac{\mathcal{C}\ \! \mathrm{e}^{-1/\tau}}{\gamma^{2}\ \! \tau \ \! \mathrm{K}_{2}(1/\tau)}T(\gamma,\gamma',\tau) \ , \label{sigmaKN2}
\end{equation}
which is far less expensive to compute numerically. Now, the modified term $\mathrm{e}^{-1/\tau}/\tau \ \! \mathrm{K}_{2}(1/\tau)$, at an electron temperature of $10 \ \mathrm{keV}$ has the value of $2.811\times 10^{2}$, at $1 \ \mathrm{keV}$ its value is $9.183\times 10^{3}$ and at $1 \ \mathrm{meV}$ its value is now $9.217\times10^{12}$. This method is readily parallelised, with each integral, or group of integrals, performed per CPU. Additionally, if the array $\mathcal{D}\left(n,k,l \right)$ is populated prior to runtime, and care is taken to handle positive and negative terms, performing one final subtraction at the end, then the method can be made very accurate. In the following subsection we detail a numerical investigation of a basic code we have written in \textsc{Python} to evaluate angular moments of the Compton scattering kernel.

\subsection{Numerical tests}
The computation of the angular moments of the Compton scattering kernel is based on the solution of many integrals of the form given in equations (\ref{Fk}) and (\ref{I1new}). We have written a code using the arbitrary-precision mathematics package \textsc{mpmath} in \textsc{Python} 2.7.3 from the Enthought \textsc{Python} Distribution 7.3-1 (64 bit). All calculations were performed on a Mid-2009 MacBook Pro with a 3.06GHz Intel Core 2 Duo CPU with 8GB of 1067 MHz DDR3 RAM - no computer-specific optimisations were performed. The code was designed and tested on Mac OSX 10.8.2, compatible with any OS with \textsc{Python} and \textsc{mpmath} installed.

To illustrate the functionality of the method, we show the relative errors, $\varepsilon$, for the first six angular moments of the Compton scattering kernel, for a broad range of photon energies, from $\gamma  =  1 \ \textrm{meV}$ to $\gamma =  1 \ \textrm{GeV}$. The relative error is defined with respect to an arbitrary precision code written in \textsc{Mathematica} with no less than 100 digits of accuracy. Values of $\sigma_{\mathrm{KN}}$ of magnitude less than $10^{-100}$ are neglected. The code is evaluated first with 53 bits of numerical precision (double precision - D). If the relative error is not less than $10^{-12}$ we then evaluate $\sigma_{KN}$ with 106 bit precision (double-double - DD), 159 bit precision (triple-double - DD) and, if necessary, with 212 bits of precision (quad-double - QD). We have chosen two electron temperatures for numerical testing, $1 \ \textrm{meV}$ and $1 \ \textrm{keV}$. We have chosen to iterate $\gamma'$ as $\gamma'=(1+\delta)\gamma$, with $\delta$ taking the values $10^{-6}$, $10^{-4}$, $10^{-2}$ and $1$ (for $\delta>1$, $\sigma_{\mathrm{KN}}$ is always negligibly small and so we omit those results).

To our knowledge there are no freely available codes in the literature which can compute successive angular moments of the Compton scattering kernel. Consequently, we have written a code in \textsc{Mathematica} 8 which computes the angular moments to arbitrary order. We then compare the results from \textsc{Mathematica} with those obtained from our \textsc{Python} code, evaluating the relative error $\varepsilon$ between the two.

In Table 1 the relative errors are computed for an electron temperature of $1 \ \textrm{meV}$. It is clear that at low incident photon energies, namely $1 \ \textrm{meV}$ and $1 \ \textrm{eV}$, double precision arithmetic is insufficient. Further, at $\gamma=1 \ \textrm{meV}$, even the errors at double-double precision are not sufficiently small, and so we display the result for triple-double precision. By photon energies of $\gamma = 1 \ \textrm{keV}$ double precision results become no worse than a few parts in $1000$.

In Table 2 the relative errors are computed for an electron temperature of $1 \ \textrm{keV}$. Again, at very low photon energies we have to resort to double-double, and even triple-double arithmetic precision. However, by photon energies of $1 \ \textrm{keV}$ double precision arithmetic is again sufficient. In those regions where $\gamma$ is large and the relative error at double precision is of the order of $10^{-3}$ or greater, the value of the scattering kernel is significantly less than unity, generally of the order of $10^{-50}$ or less. As the electron temperature increases still higher the results become even more accurate at double precision, and follow the same underlying trends, so we neglect them for the sake of brevity.

Clearly the method presented does not fare so well at low photon energies ($\gamma, \ \gamma' \ll 1$), as well as regions where $|\gamma-\gamma'|\ll 1$ and $\tau \ll 1$ and so we must resort to numerical precision greater than that of standard double precision. Regarding computation time, at double precision the numerical results can take from a few tenths of seconds to a few tens of seconds. Computation time increases drastically with increased numerical precision. We stress the system architecture these calculations were performed on was simply a laptop, and there is tremendous scope to improve the implementation of the underlying method. Since the method centrally revolves around solving specific definite integrals, it is easily parallelised and can be made significantly faster on that basis alone. Further, by careful consideration of positive and negative terms, only one subtraction need be performed per moment evaluation, greatly reducing round-off error (since the integral in equation (\ref{Fk}) is always positive). The terms $\mathcal{D}\left(n,k,l \right)$ may be tabulated prior to runtime and all values of $\mathcal{F}_{k}$ can be stored in an appropriate array. In addition, the integrals themselves could be pre-calculated on a standard grid of cases, with interpolation performed on this grid at run-time. In the regions where double precision accuracy is insufficient, asymptotic series expansions can be employed, particularly where $\tau \rightarrow 0$, $\gamma, \gamma' \rightarrow 0$, $\gamma/\gamma' \rightarrow 1$ and $\gamma$ fixed with $\gamma'\rightarrow 0$ (and vice-versa). However, in most regions of astrophysical interest, the electron and photon energies are of the order of $\mathrm{keV}$ energies or greater. The aforementioned refinements would make the Compton scattering code very robust across a much broader energy range, particularly at lower energies. 
\\
\\

\begin{table}
\caption{Relative errors for the first 6 moments of the Compton scattering kernel, evaluated at an electron temperature of $1 \ \mathrm{meV}$. Numbers between brackets denote multiplicative powers of 10. Hyphens indicate a relative error greater than unity.}      
\centering 
\begin{tabular}{l l l l l l l l l}               
\hline                         
$\gamma$ & $\delta$ & Precision & $\varepsilon_{n=0}$ & $\varepsilon_{n=1}$ & $\varepsilon_{n=2}$ & $\varepsilon_{n=3}$ & $\varepsilon_{n=4}$ & $\varepsilon_{n=5}$ \\ [0.5ex] 
\hline                                 
$1 \ \mathrm{meV}$ & $10^{-6}$ & D & -- & -- & --  & -- & -- & -- \\                 
& & DD & 1.11[-04] & 2.39[-04] & 3.95[-04]  & 8.94[-05] & 2.68[-04] & 3.55[-03] \\
& & TD & 1.80[-20] & 6.13[-20] & 5.64[-20] & 8.14[-20] & 5.33[-20] & 6.37[-21] \\
\\
& $10^{-4}$ & D & -- & -- & --  & -- & -- & -- \\
& & DD & 1.25[-06] & 3.22[-05] & 8.43[-05] & 8.13[-05] & 1.17[-04] & 3.38[-04] \\
& & TD & 2.10[-21] & 4.80[-21] & 4.77[-21] & 1.33[-20] & 1.52[-20] & 4.04[-19] \\
\hline
$1 \ \mathrm{eV}$ & $10^{-6}$ & D & -- & -- & --  & -- & -- & -- \\                 
& & DD & 7.35[-14] & 2.94[-14] & 2.06[-14]  & 7.74[-13] & 2.01[-13] & 7.24[-13] \\
\\
& $10^{-4}$ & D & -- & -- & --  & -- & -- & -- \\
& & DD & 6.63[-15] & 2.25[-14] & 8.62[-14] & 6.66[-14] & 1.30[-13] & 3.40[-13]  \\
\hline
$1 \ \mathrm{keV}$ & $10^{-6}$ & D & 4.97[-05] & 6.77[-05] & 4.27[-06]  & 4.71[-04] & 1.27[-03] & 3.09[-03] \\                 
& & DD & 4.90[-21] & 1.99[-21] & 7.07[-21] & 1.18[-21]  & 6.33[-21] & 1.27[-20]  \\
\\
& $10^{-4}$ & D & 2.76[-03] & 2.89[-03] & 3.01[-03]  & 3.12[-03] & 3.23[-03] & 3.32[-03] \\
& & DD & 2.76[-03] & 2.89[-03] & 3.01[-03]  & 3.12[-03] & 3.23[-03] & 3.32[-03] \\
& & TD & 2.76[-03] & 2.89[-03] & 3.01[-03]  & 3.12[-03] & 3.23[-03] & 3.32[-03] \\
& & QD & 8.52[-31] & 3.15[-18] & 1.01[-10] & 3.20[-10] & 6.78[-10] & 1.20[-09] \\ [1ex]          
\hline                                 
\end{tabular}
\label{table:tau1em6}                   
\end{table}

\begin{table}
\caption{As in Table 1, but now evaluated at an electron temperature of $1 \ \mathrm{keV}$}      
\centering 
\begin{tabular}{l l l l l l l l l}               
\hline                         
$\gamma$ & $\delta$ & Precision & $\varepsilon_{n=0}$ & $\varepsilon_{n=1}$ & $\varepsilon_{n=2}$ & $\varepsilon_{n=3}$ & $\varepsilon_{n=4}$ & $\varepsilon_{n=5}$ \\ [0.5ex]  
\hline                                 
$1 \ \mathrm{meV}$ & $10^{-6}$ & D & -- & -- & --  & -- & -- & -- \\                 
& & DD & 4.18[-02] & 2.42[-01] & 6.15[-02]  & 5.40[-01] & 9.72[-01] & 5.24[-01] \\
& & TD & 9.27[-18] & 5.83[-17] & 2.12[-17] & 1.97[-17] & 6.83[-18] & 1.46[-16] \\
\\
& $10^{-4}$ & D & -- & -- & --  & -- & -- & -- \\
& & DD & 1.47[-02] & 3.75[-02] & 2.48[-02] & 4.83[-02] & 3.08[-02] & 5.19[-02] \\
& & TD & 1.24[-18] & 3.22[-18] & 2.27[-18] & 3.25[-18] & 2.53[-18] & 3.85[-18] \\
\\
& $10^{-2}$ & D & -- & -- & --  & -- & -- & -- \\
& & DD & 3.53[-04] & 1.32[-03] & 7.55[-04] & 1.86[-03] & 1.13[-03] & 2.42[-03] \\
& & TD & 8.67[-21] & 1.31[-20] & 4.06[-21] & 4.40[-20] & 1.88[-20] & 5.94[-20] \\
\\
& $1$ & D & -- & -- & --  & -- & -- & -- \\
& & DD & 5.47[-06] & 1.12[-04] & 1.18[-04] & 5.39[-05] & 1.08[-04] & 9.49[-04] \\
& & TD & 6.25[-21] & 7.72[-21] & 3.13[-21] & 6.42[-21] & 9.38[-21] & 4.30[-20] \\
\hline
$1 \ \mathrm{eV}$ & $10^{-6}$ & D & -- & -- & --  & -- & -- & -- \\                 
& & DD & 9.24[-11] & 3.06[-10] & 7.12[-11]  & 4.13[-10] & 1.61[-10] & 7.85[-11] \\
\\
& $10^{-4}$ & D & -- & -- & --  & -- & -- & -- \\
& & DD & 2.35[-12] & 2.78[-12] & 1.27[-12] & 6.32[-12] & 2.89[-12] & 9.84[-12] \\
\\
& $10^{-2}$ & D & -- & -- & --  & -- & -- & -- \\
& & DD & 2.37[-14] & 2.05[-14] & 1.54[-14] & 3.02[-14] & 1.04[-13] & 4.93[-14] \\
\\
& $1$ & D & -- & -- & --  & -- & -- & -- \\
& & DD & 4.39[-13] & 1.23[-14] & 2.77[-14] & 4.07[-13] & 3.44[-13] & 3.88[-14] \\
\hline
$1 \ \mathrm{keV}$ & $10^{-6}$ & D & 1.75[-03] & 5.55[-03] & 2.98[-03]  & 1.80[-03] & 2.40[-03] & 7.01[-03] \\                 
& & DD & 7.19[-20] & 4.63[-19] & 5.24[-19]  & 7.15[-19] & 1.85[-19] & 7.73[-19] \\
\\
& $10^{-4}$ & D & 1.65[-05] & 2.15[-05] & 2.30[-05]  & 3.74[-05] & 1.07[-05] & 3.07[-05] \\
& & DD & 2.35[-12] & 2.78[-12] & 1.27[-12] & 6.32[-12] & 2.89[-12] & 9.84[-12] \\
\\
& $10^{-2}$ & D & 1.71[-07] & 4.90[-07] & 7.27[-08]  & 2.18[-07] & 6.10[-07] & 1.34[-07] \\
& & DD & 2.00[-24] & 1.31[-23] & 2.81[-23] & 1.29[-22] & 1.80[-23] & 1.17[-22] \\
\\
& $1$ & D & 5.28[-02] & -- & --  & -- & -- & -- \\
& & DD & 8.16[-25] & 3.83[-23] & 8.84[-23] & 1.92[-22] & 2.66[-22] & 6.29[-22] \\ 
\hline
$1 \ \mathrm{MeV}$ & $10^{-6}$ & D & 6.15[-11] & 1.17[-11] & 5.47[-11]  & 2.94[-11] & 7.44[-11] & 5.20[-11] \\                 
& & DD & 1.25[-27] & 4.58[-27]  & 6.01[-27]  & 9.13[-27]  & 4.43[-27]  & 9.32[-27]  \\
\\
& $10^{-4}$ & D & 3.47[-13] & 5.52[-13] & 3.74[-13] & 1.56[-13] & 1.94[-13] & 2.49[-13] \\
& & DD & 2.10[-28] & 2.01[-28]  & 2.44[-28]  & 2.11[-28]  & 1.97[-28]  & 2.59[-28]  \\
\\
& $10^{-2}$ & D & 4.24[-14] & 5.83[-14] & 5.84[-13] & 6.21[-12] & 2.84[-11] & 7.30[-11] \\
& & DD & 4.52[-30] & 3.25[-30] & 4.36[-30]  & 5.11[-30]  & 4.25[-30]  & 3.95[-30]  \\
\hline
$1 \ \mathrm{GeV}$ & $10^{-6}$ & D & 5.70[-09] & 5.70[-09] & 5.70[-09] & 5.70[-09] & 5.70[-09] & 5.70[-09] \\
& & DD & 1.22[-23] & 1.22[-23]  & 1.22[-23]  & 1.22[-23]  & 1.22[-23]  & 1.22[-23]  \\
\\
& $10^{-4}$ & D & 2.81[-01] & 2.81[-01] & 2.81[-01] & 2.81[-01] & 2.81[-01] & 2.81[-01] \\
& & DD & 2.81[-01] & 2.81[-01] & 2.81[-01] & 2.81[-01] & 2.81[-01] & 2.81[-01] \\
& & TD & 5.99[-08] & 5.99[-08] & 5.99[-08]  & 5.99[-08]  & 5.99[-08]  & 5.99[-08] \\
& & QD & 1.25[-30] & 1.02[-17] & 2.04[-17]  & 3.06[-17]  & 4.08[-17]  & 5.10[-17] \\ [1ex]          
\hline                                 
\\
\\
\end{tabular}
\label{table:tau1}                   
\end{table}

\section{Results and Discussion}  

We remark that this method can easily be generalised to include evaluation of moments of the cross-section in terms of more general functions of $\zeta$, 
  such as Legendre polynomials. 
This is shown in Appendix \ref{AppendixB}.   

We show in Figures 6--10 the computed moments of the Compton scattering kernel (in arbitrary units, i.e. $\mathcal{C}=1$)   
   obtained from the closed-form expression that we have derived as a function of scattered photon energy.  
Figures 6 and 7 illustrate the dependence of the zeroth moment of the scattering kernel on the electron temperature, for various incident photon energies.  
Fig. 8 shows the dependence of the 1st, 2nd, 3rd, 4th and 5th moments of the scattering kernel 
   for an incident photon energy of $40$ keV and an electron temperature of $1$ keV (top) and $20$ keV (bottom). 
Fig. 9 is similar to Fig. 8, except that the incident photon energy is $100$ keV. 
Fig. 10 is as in Fig. 9, except the incident photon energy is now $300$ keV. 

The parameters for the plots in these figures were chosen to enable comparison 
   with previous numerical calculations by \cite{Pomraning1972, Pomraning1973} 
   in which the angular moments are expanded in terms of Legendre polynomials $P_{n}(\zeta)$.  
Without a closed-form expression for the scattering kernel,    
   \cite{Pomraning1972, Pomraning1973} employed a fully numerical approach in his calculations.       
Although \cite{Pomraning1972, Pomraning1973} employed a Legendre polynomial moment expansion and we have considered different functions for the moment expansions,  
 in the classical limit, the zeroth order terms in both calculations are identical.  
Figures 6--7 indeed show that the zeroth order moments obtained by our derived closed-form expression 
   are the same as those obtained by the Legendre polynomial expansion of \cite{Pomraning1972}.   
We also note that the zeroth order moments of the kernel that we computed for various electron temperatures  
   are consistent with Monte-Carlo simulations of Compton scattering of monochromic emission lines 
   shown in \cite{Pozdnyakov1979} and \cite{Pozdnyakov1983}.  

In practical radiative transfer calculations, the full radiative transfer equation with scattering may in principle be decoupled, 
  in a truncated moment expansion, into a series of coupled ordinary differential equations \citep{Thorne1981, Fuerst2006, Wu2008}. 
 In solving the full radiative transfer equation in curved space times, 
   a covariant generalisation of the Eddington approximation \citep{Fuerst2006, Wu2008, Shibata2011} may be employed, 
   which, coupled with the aforementioned closed-form expressions for the angular moments, 
   yields a semi-analytic approach, necessitating the evaluation of two numerical integrals, 
   namely over $\lambda$ and $\gamma$ (or $\gamma'$, by detailed balance). 
The detailed procedures for such a decomposition are beyond the scope of this study, and we leave this to a future article.

\section{Summary}

We have derived a covariant expression for the relativistic Compton scattering kernel self-consistently.  

By specialising the $\mathit{z}$-axis of integration along the direction of photon momentum transfer, and re-arranging the order of integration, the problem of computing angular moments of the Klein-Nishina cross-section has been reduced to one of solving three types of moment integral. Further, in re-arranging the order of integration, our method is not restricted to the particular assumed electron distribution function, although for this work we assumed a relativistic Maxwellian distribution for the electrons. The analytical representation of these moment integrals in terms of hypergeometric functions enabled us to express the Klein-Nishina scattering kernel in the particularly elegant form given in equation (\ref{sigmaKN}). The problem of evaluating moments of the Klein-Nishina cross-section has been reduced to simply computing a series of one-dimensional integrals over the electron energy, $\lambda$, which are easily evaluated by quadrature methods. This is a significant improvement over current approaches.

We investigated the numerical stability of the evaluation of the angular moment integrals in \textsc{Fortran95}, both by recursive and direct evaluation of the hypergeometric functions. It was found that for $n>30$, numerical stability becomes an issue and double-precision arithmetic is no longer adequate. Further, as already described, the case of very low scattering angle ($\zeta\rightarrow -1$) is oscillatory, and slowly convergent, owing to the geometry of the problem. We also investigated the convergence of the angular moments of the Klein-Nishina scattering kernel ($\mathcal{M}_{n}$) and found the case of inverse-Compton scattering to be more slowly convergent than conventional Compton scattering, but also that the rate of convergence is strongly dependent on the electron velocity. We found that as the electron velocity increases, $\mathcal{M}_{n}$ converges much more rapidly as the moment order increases.

We carried out demonstrative calculations of the first six moments of the Klein-Nishina scattering kernel, 
   convolved with a relativistic Maxwellian distribution for electrons, for various incident photon energies and electron temperatures.  
The results we obtained were consistent with those obtained by fully numerical calculations 
  in which the moment expansion is performed in terms of Legendre polynomials \citep{Pomraning1972, Pomraning1973} 
  and by Monte-Carlo simulations of emission line broadening \citep{Pozdnyakov1979, Pozdnyakov1983}. 
We note that our closed-form expression enables us to perform covariant radiative transfer calculations efficiently 
  in astrophysical settings where general relativistic effects are important, 
  with the moment truncation carried out via an Eddington approximation scheme \citep[see][]{Fuerst2006, Wu2008}.

\section*{Acknowledgments}
We thank Curtis Saxton for comments, helpful suggestions and carefully proof reading the manuscript, as well as assistance with the presentation of some figures. We also thank the referee for pointing us towards some of the more recent works on this subject.

\bibliographystyle{mn2e}
\bibliography{references.bib}

\newpage

\begin{figure}
\begin{center} 
  \includegraphics[width=0.892\textwidth]{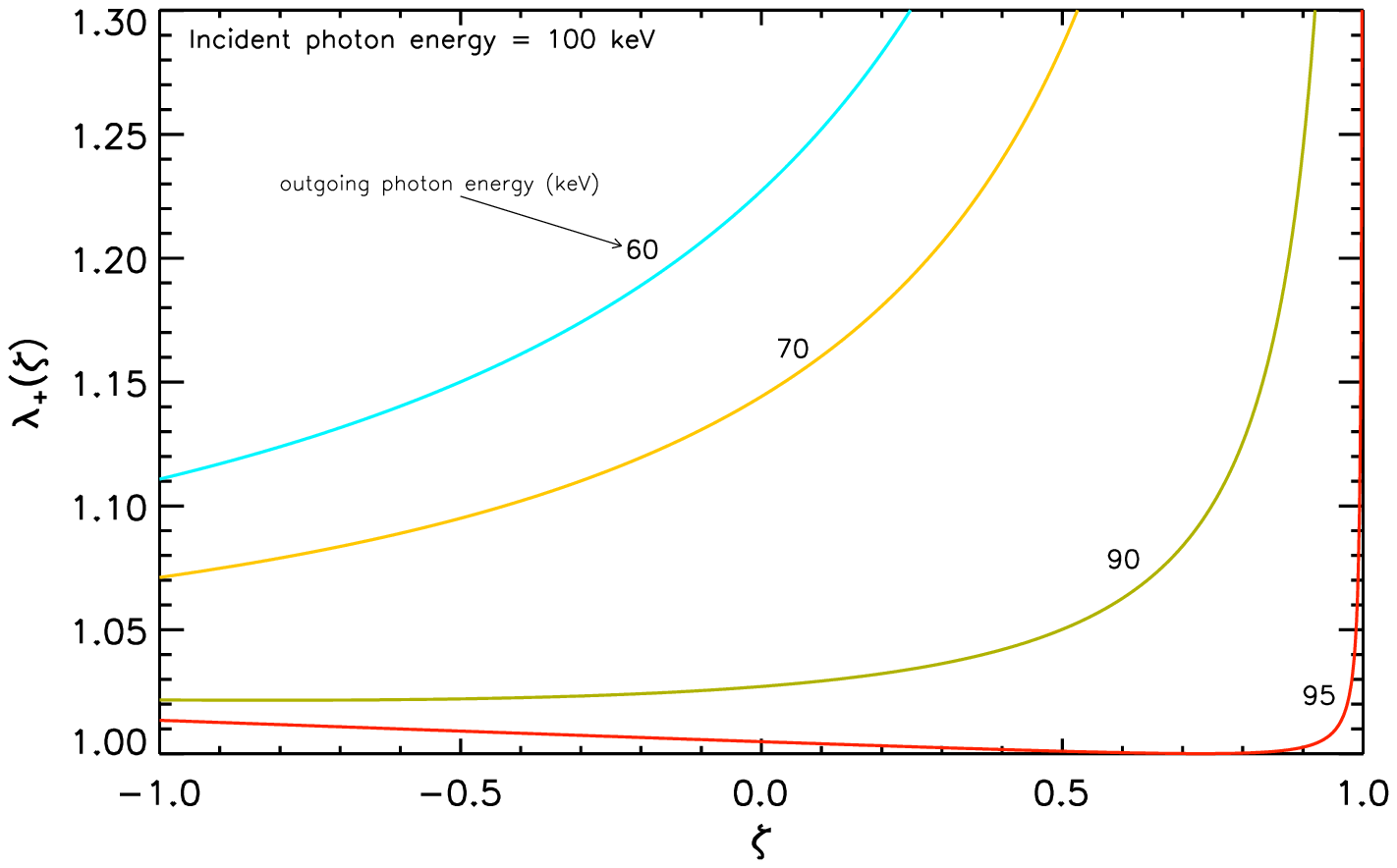}\vspace{-6mm}
\caption{Plot of $\lambda_{+}$ as a function of $\zeta$ for an incident photon of energy $100$ keV. For outgoing photon energies of $95$ keV and $90$ keV$, \lambda_{+}$ has a minimum and thus the integration over $\lambda$ must be divided into two regions. For outgoing photon energies of $60$ keV and $70$ keV, $\lambda_{+}$ does not have a minimum value between $\lambda_{+}(-1)$ and $\lambda_{+}(1)$, hence the integration over $\lambda$ is simply taken between $\lambda_{\mathrm{L}}$ and infinity.} 
\end{center}
\label{fig-1}
\end{figure}

\begin{figure}
\begin{center} 
  \includegraphics[width=0.49\textwidth]{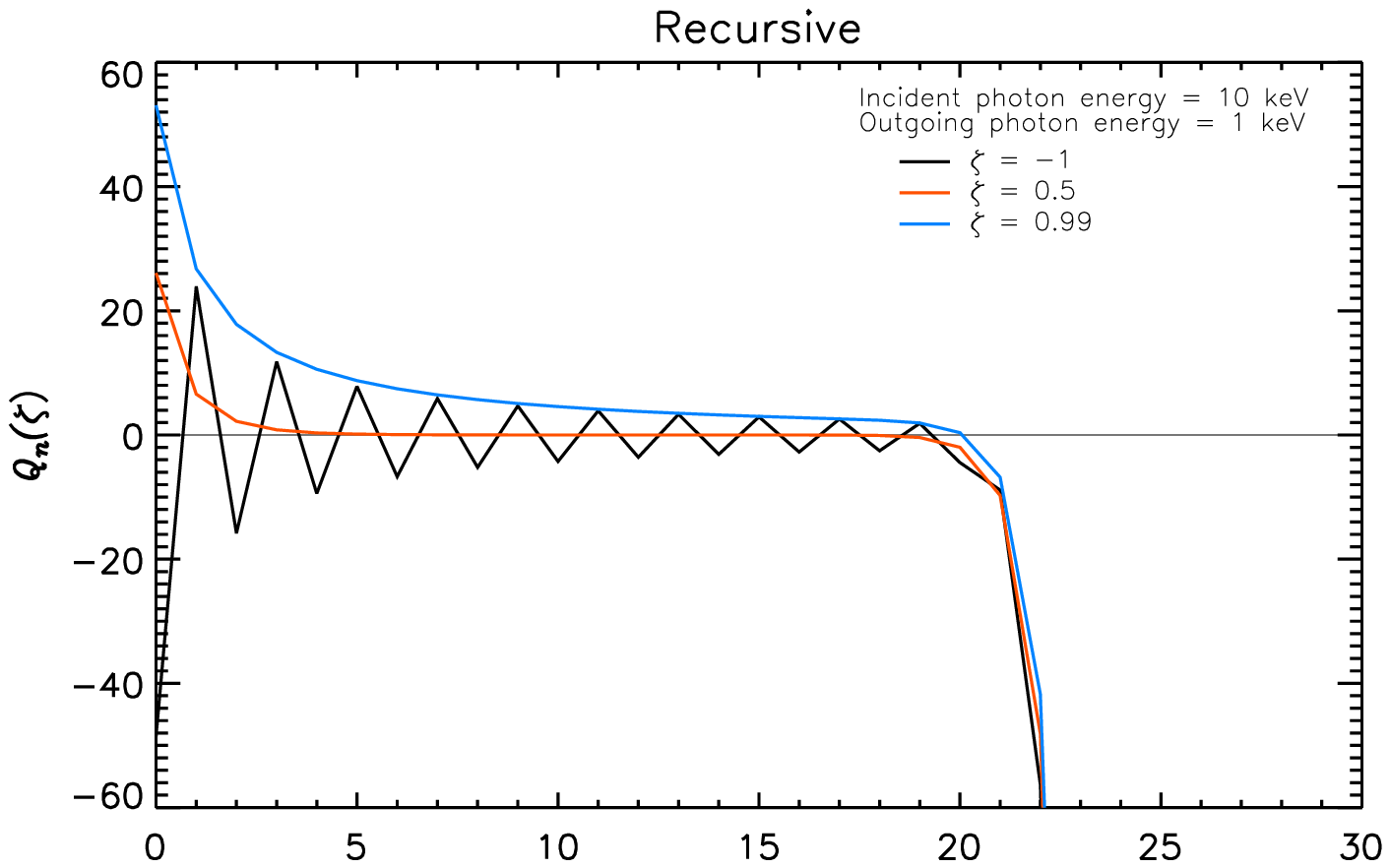}
  \includegraphics[width=0.49\textwidth]{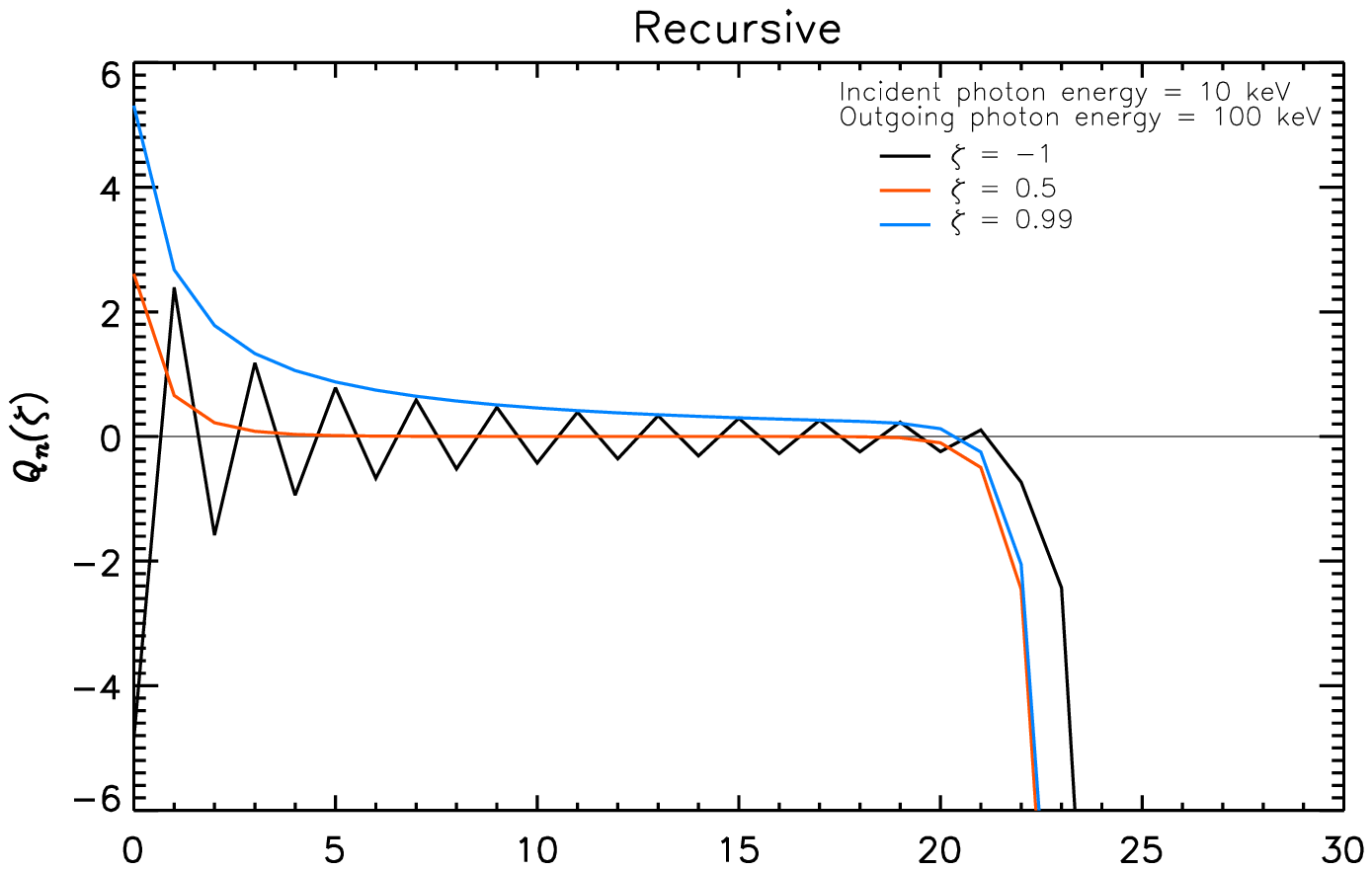}\vspace{-20mm}
  \\
  \includegraphics[width=0.49\textwidth]{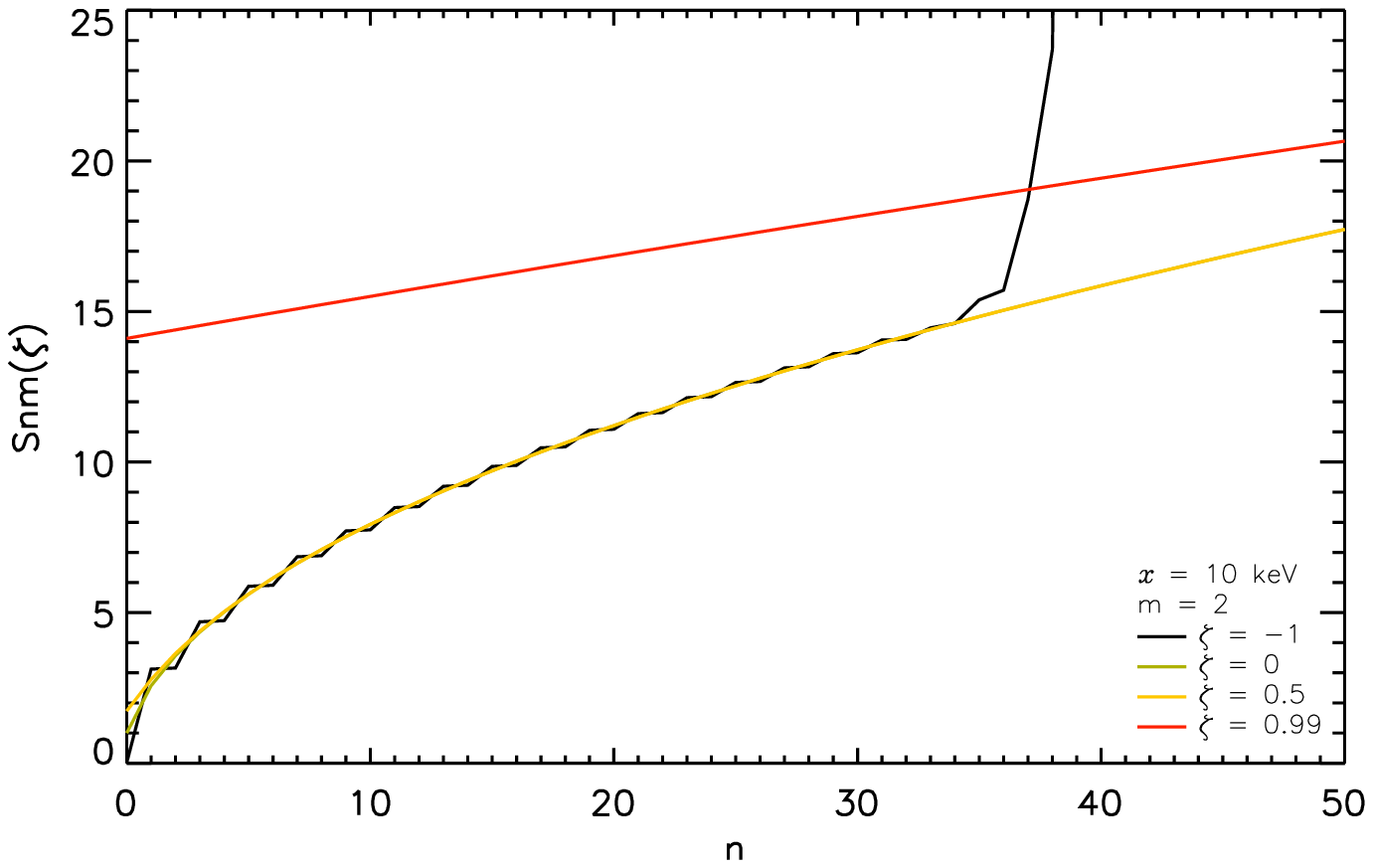}
  \includegraphics[width=0.49\textwidth]{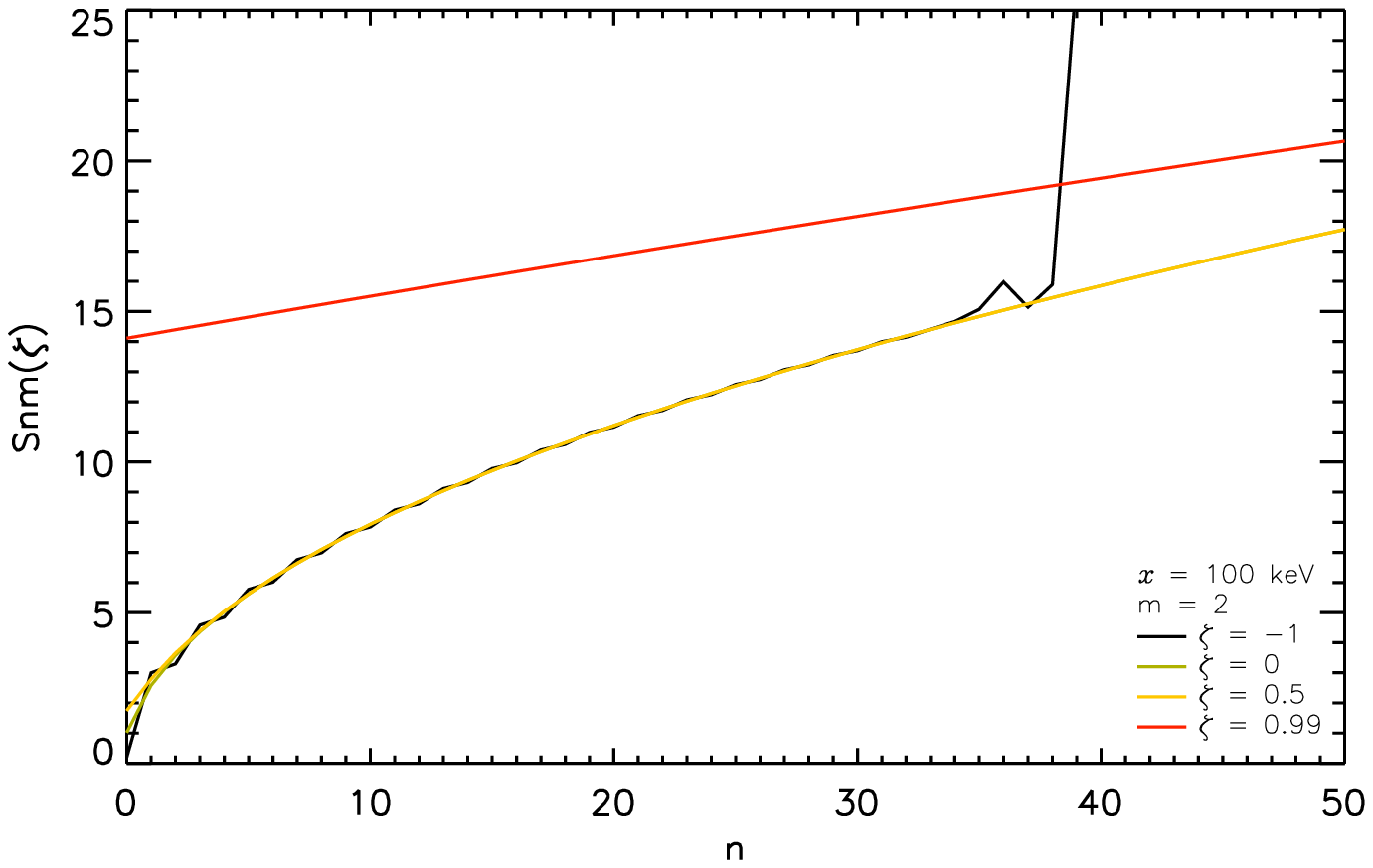}\vspace{-6mm}
\caption{Numerical \textsc{Fortran} evaluation of the moment integrals $Q_{n}$ and $S_{n,2}$, through recursion, for $x=10$ keV (left) and $x=100$ keV (right). For $Q_{n}$ numerical round-off errors occur beyond $n=20$. For $R_{n}$ and $S_{n,m}$ numerical round-off errors dominate beyond $n=30$. } 
\end{center}
\label{fig-2}
\end{figure}

\begin{figure}
\begin{center} 
  \includegraphics[width=0.49\textwidth]{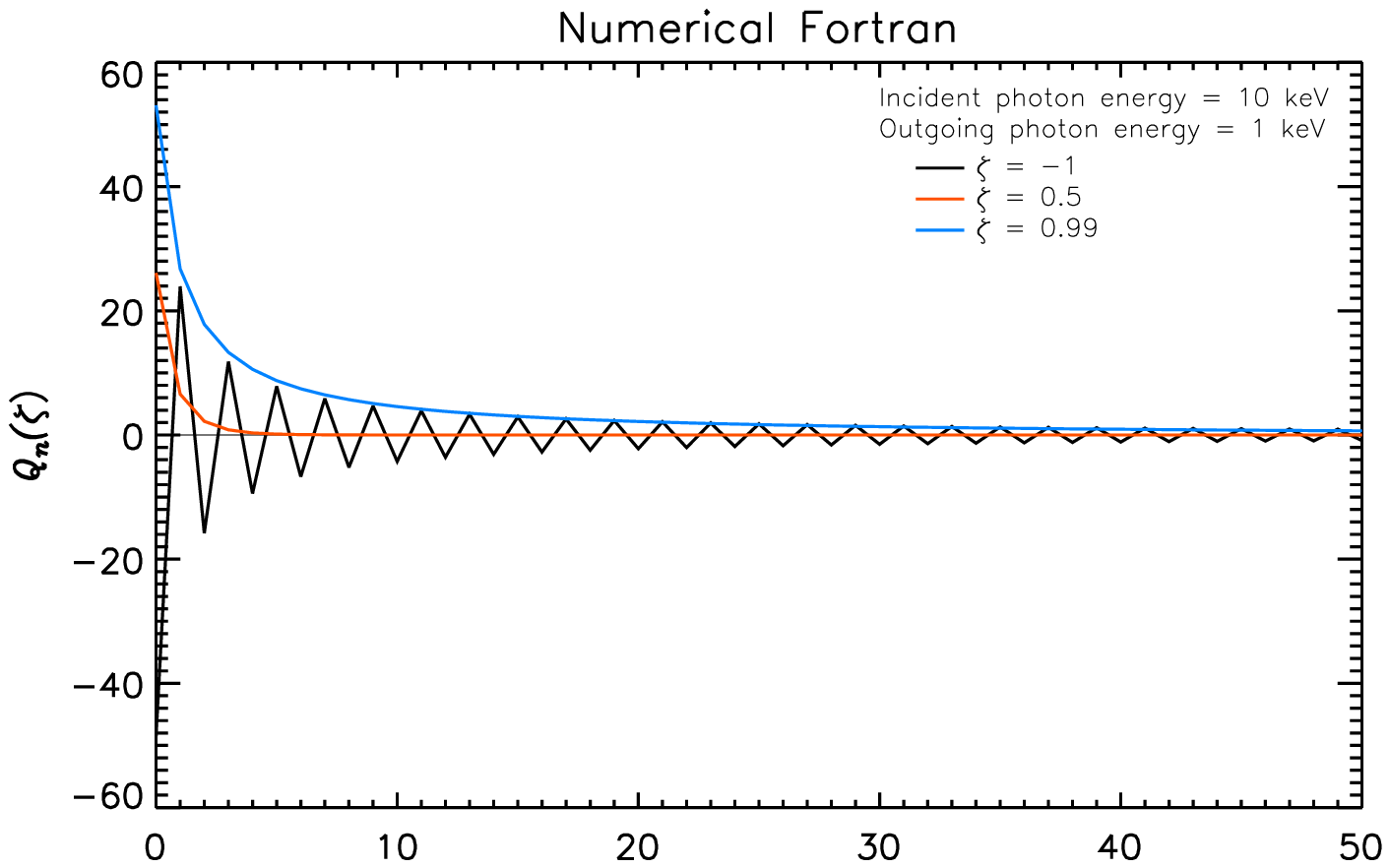}
  \includegraphics[width=0.49\textwidth]{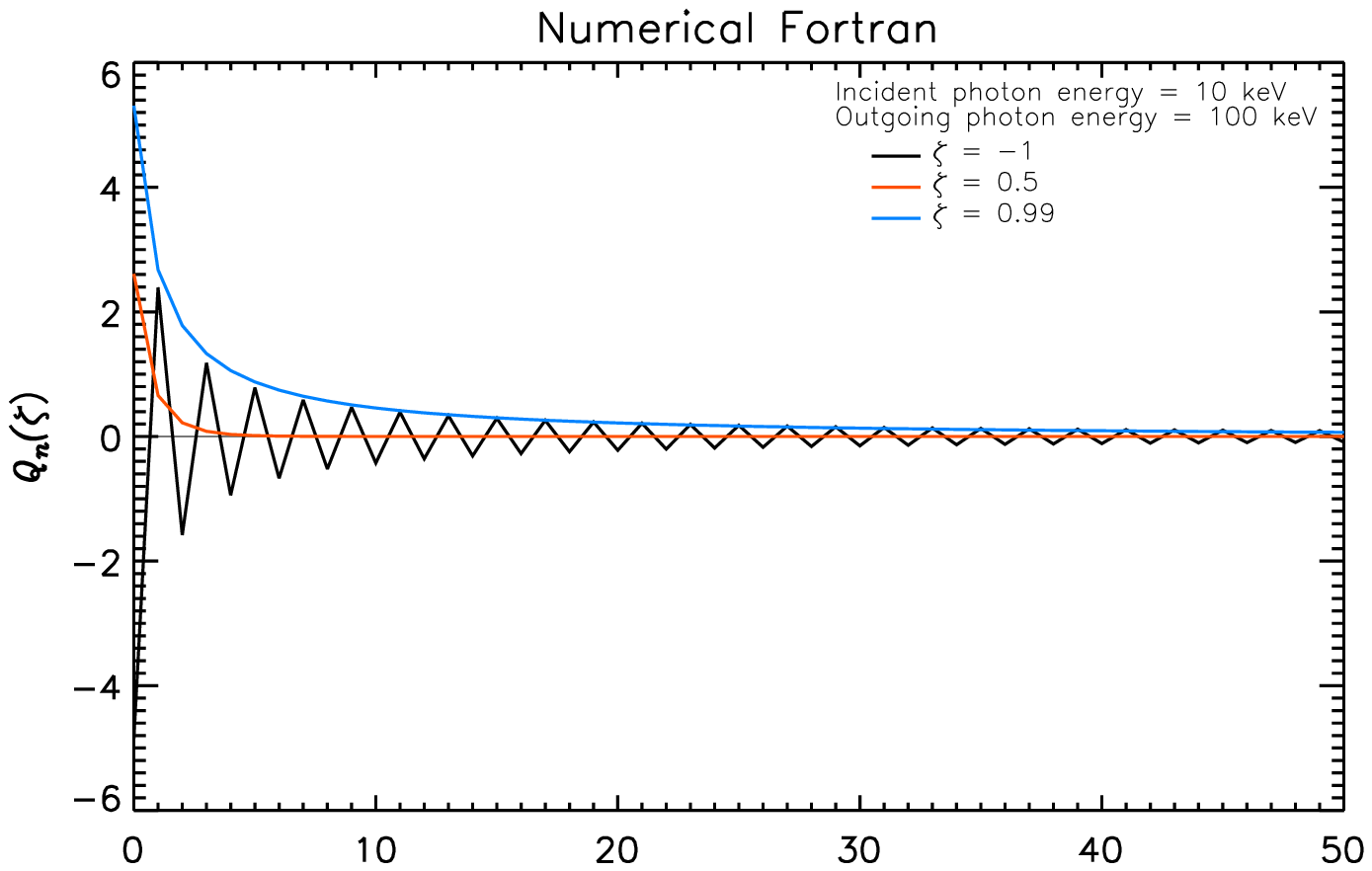}\vspace{-20mm}
  \\
  \includegraphics[width=0.49\textwidth]{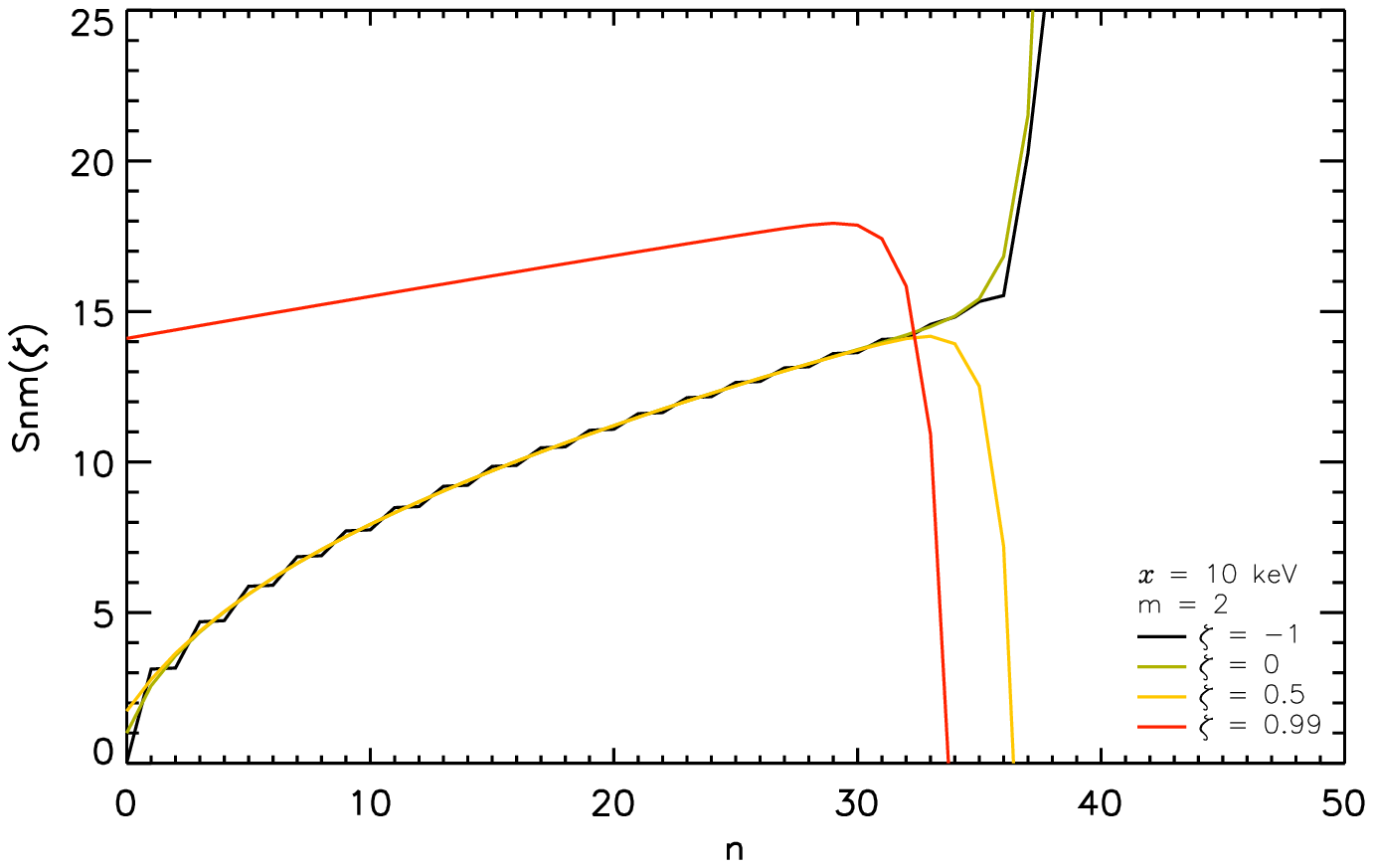}
  \includegraphics[width=0.49\textwidth]{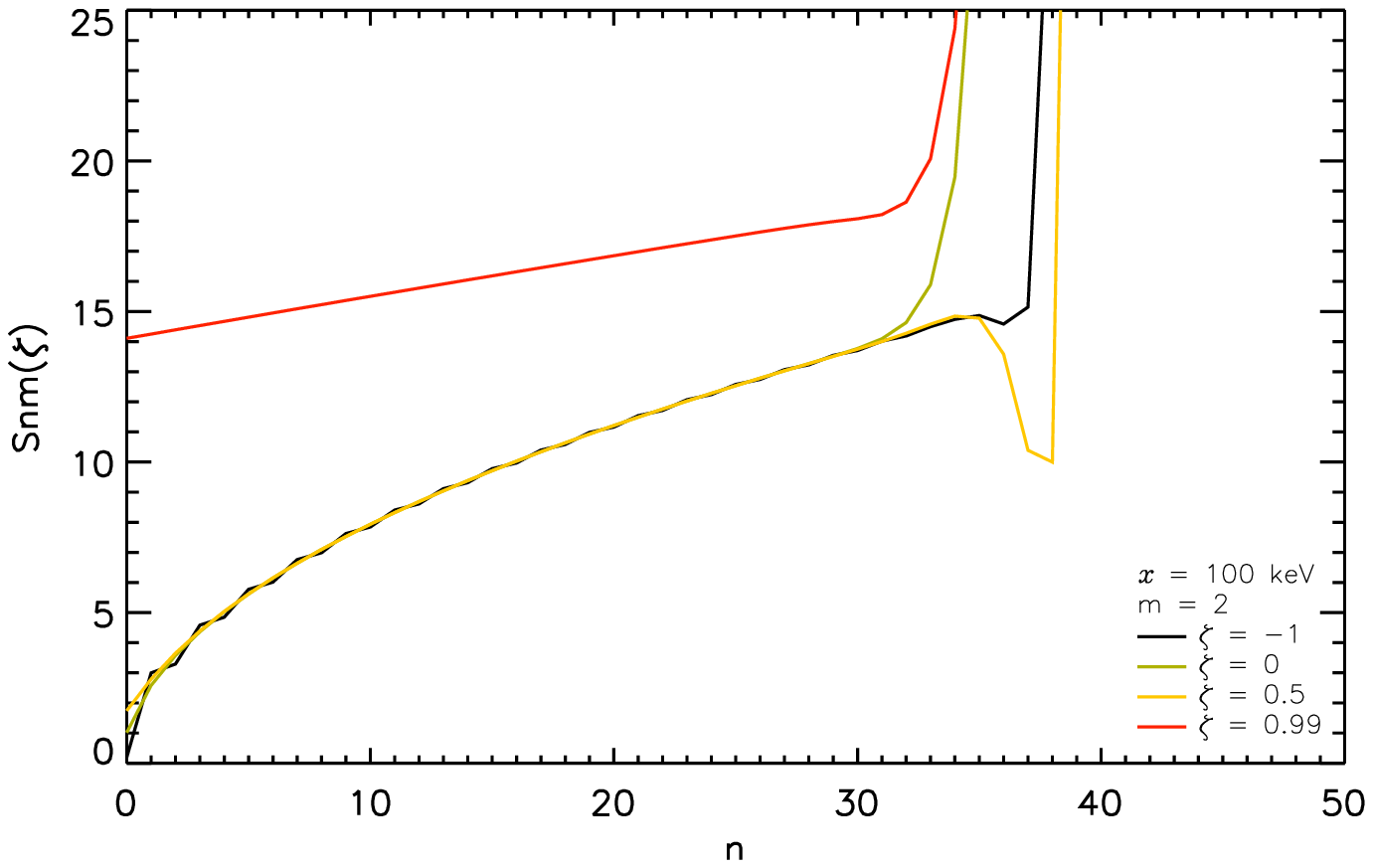}\vspace{-6mm}
\caption{Direct numerical evaluation of the moment integrals $Q_{n}$ and $S_{n,2}$, through the hypergeometric function method, for $x=10$ keV (left) and $x=100$ keV  (right). $Q_{n}$ is now numerically very stable, even beyond $n=100$ (not shown). However, for $R_{n}$ and $S_{n,m}$ there is no improvement compared to the recurrence relation method, and the results are in fact slightly worse for all scattering angles.} 
\end{center}
\label{fig-3}
\end{figure}

\begin{figure}
\begin{center} 
  \includegraphics[width=0.49\textwidth]{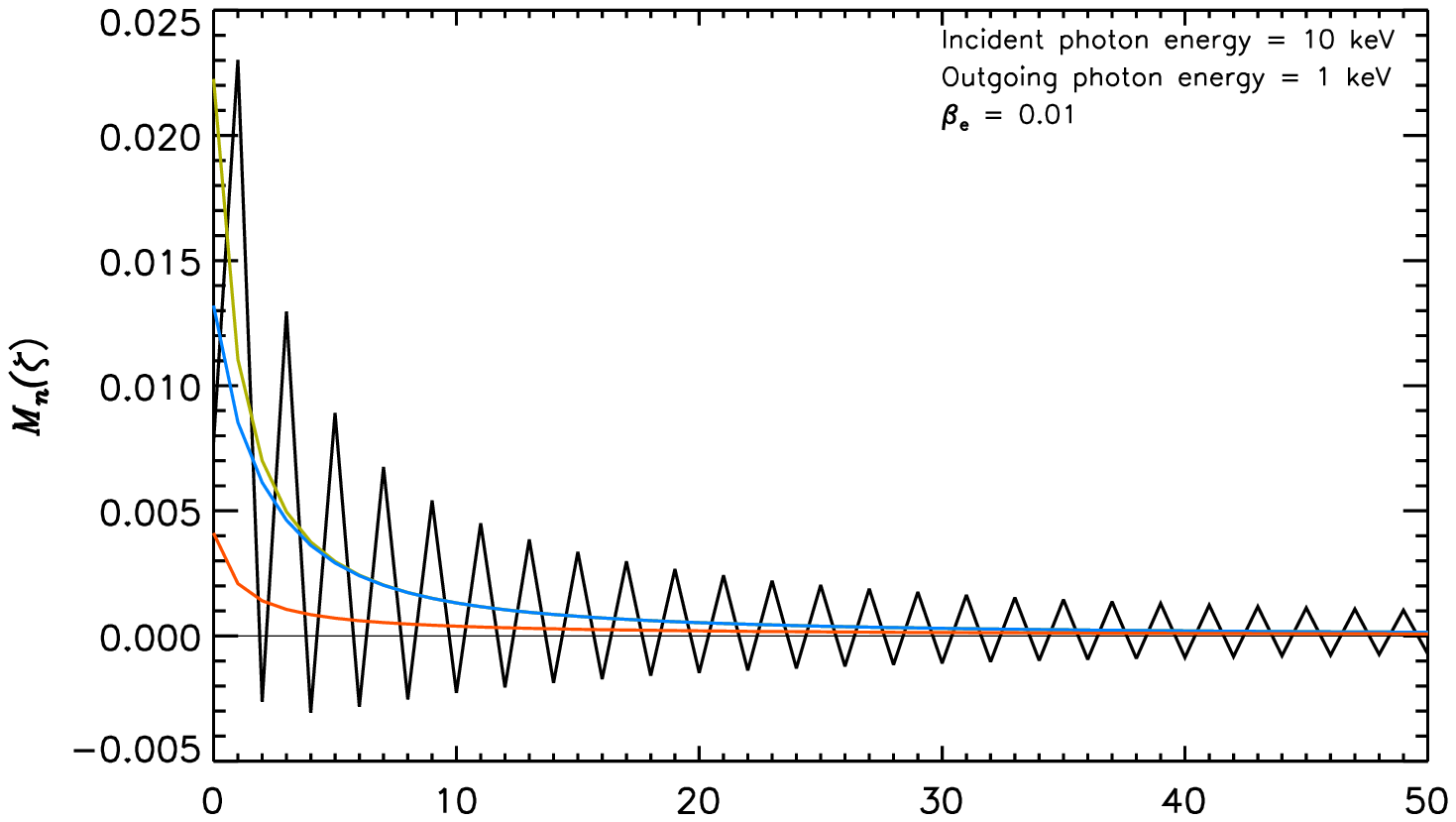}
  \includegraphics[width=0.49\textwidth]{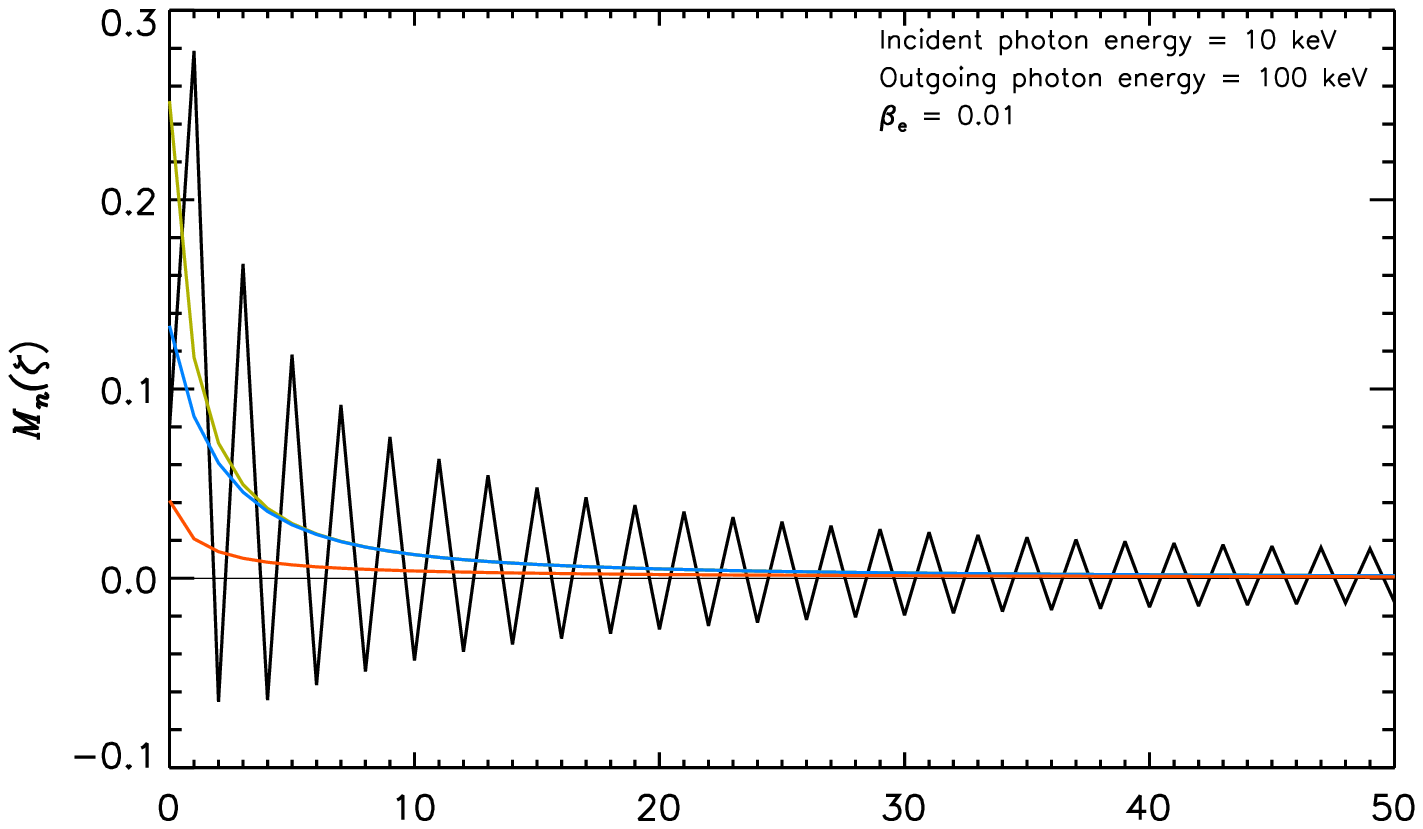}\vspace{-20mm}
  \\
  \includegraphics[width=0.49\textwidth]{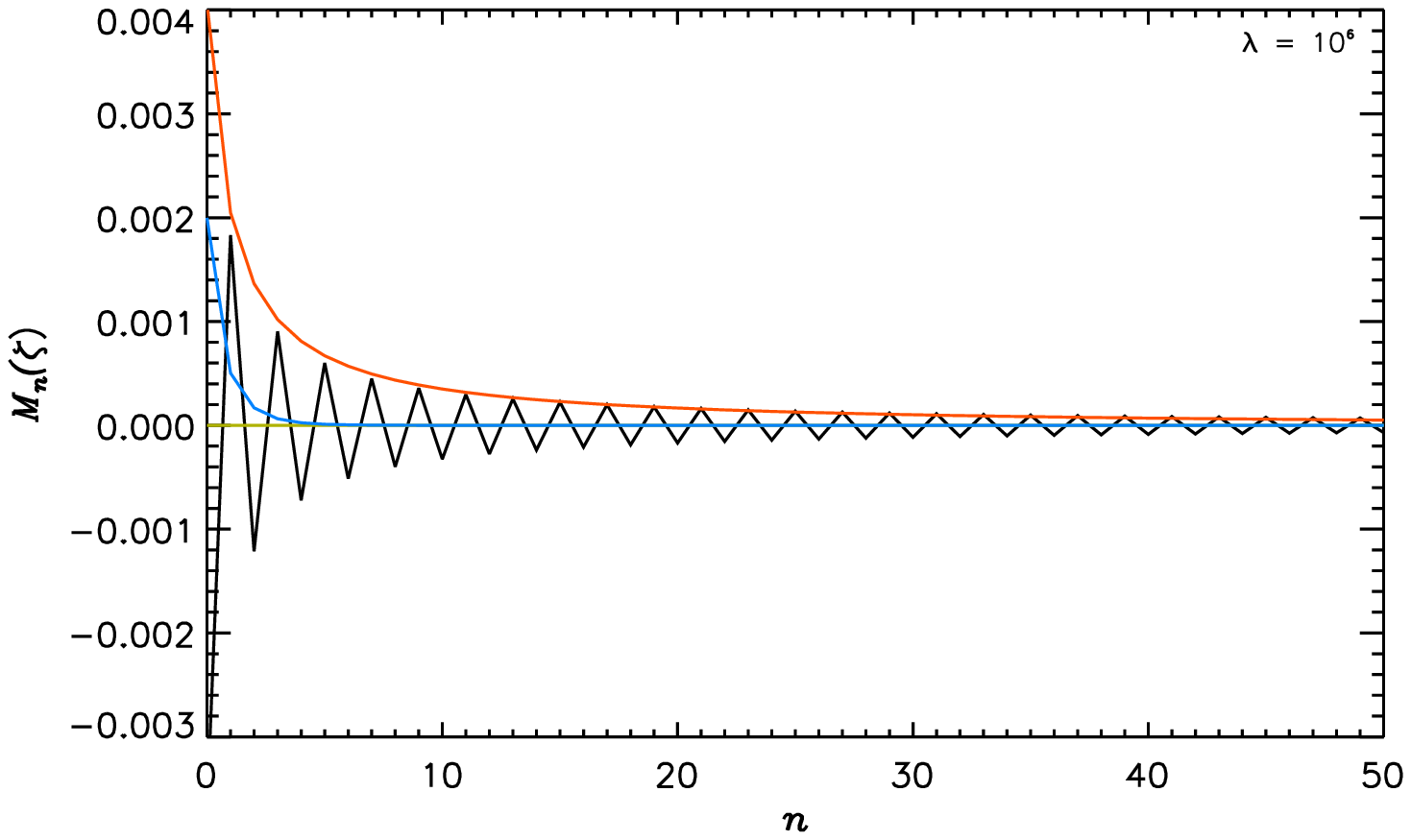}
  \includegraphics[width=0.49\textwidth]{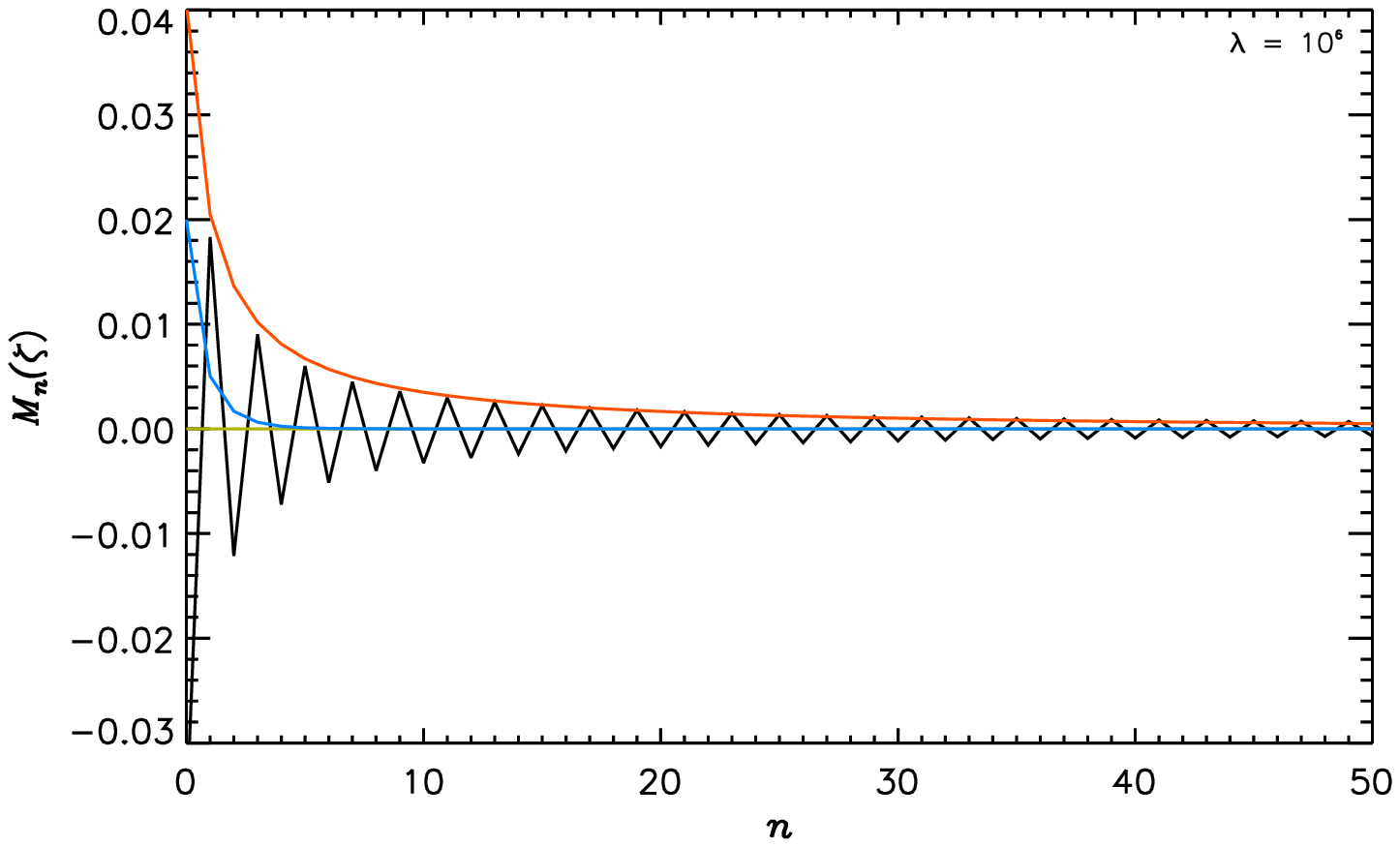}\vspace{-6mm}
\caption{Numerical evaluation of the moment integral $\mathcal{M}_{n}$ as a function of $n$. Same colour scheme as $S_{n,2}$ in Figures 2 and 3. In all plots the incident photon energy is $10$ keV. Left plots show Compton scattering resulting in an outgoing photon energy of $1$ keV, for (top to bottom) electron velocities of $\beta_{\mathrm{e}}=0.01$ and $\lambda=10^6$ ($\beta_{\mathrm{e}}\simeq 0.9999999999995$) respectively. Right plots show inverse Compton scattering for an outgoing photon of energy $100$ keV.} 
\end{center}
\label{fig-4}
\end{figure}

\begin{figure}
\begin{center} 
  \includegraphics[width=0.49\textwidth]{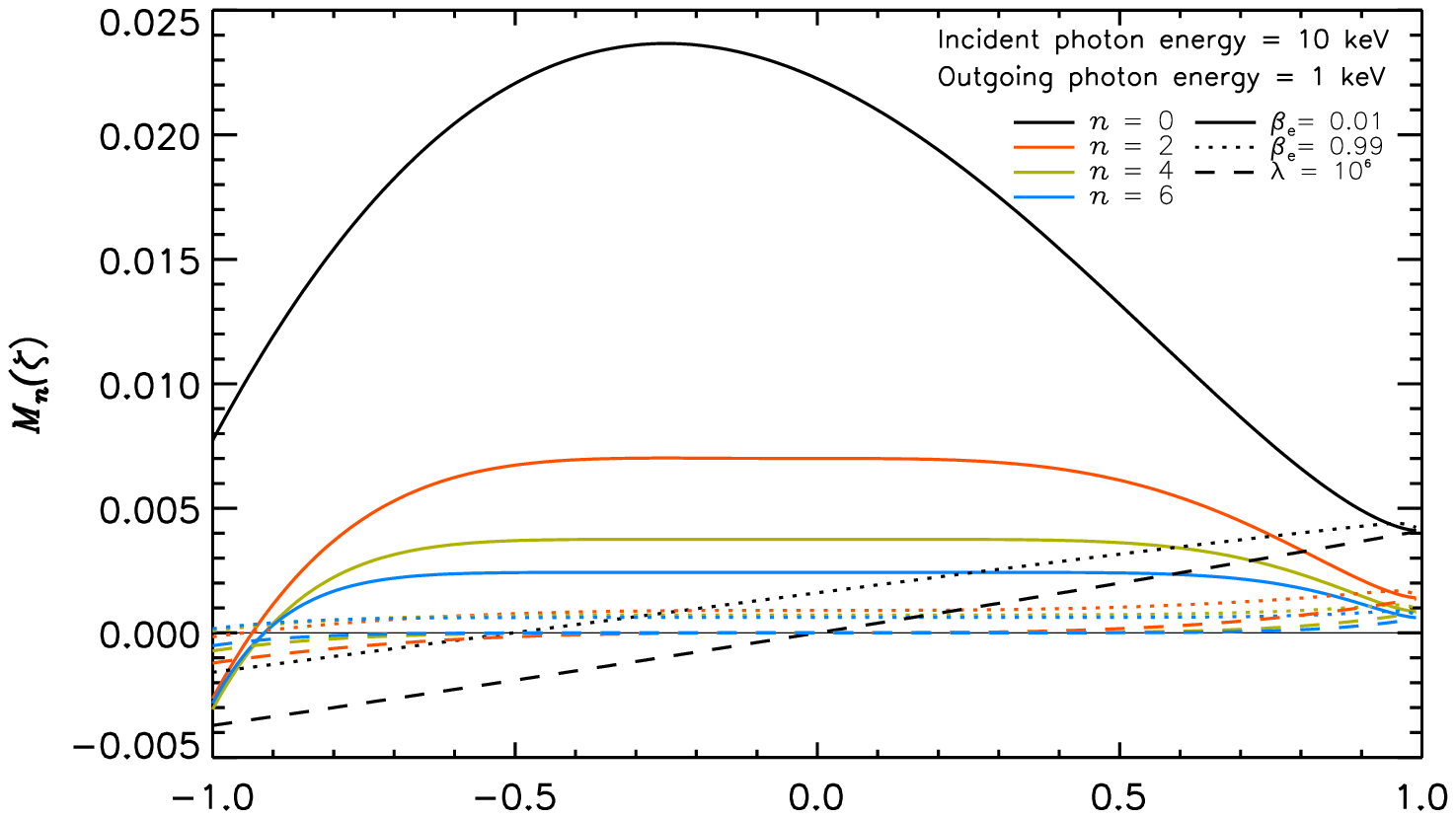}
  \includegraphics[width=0.49\textwidth]{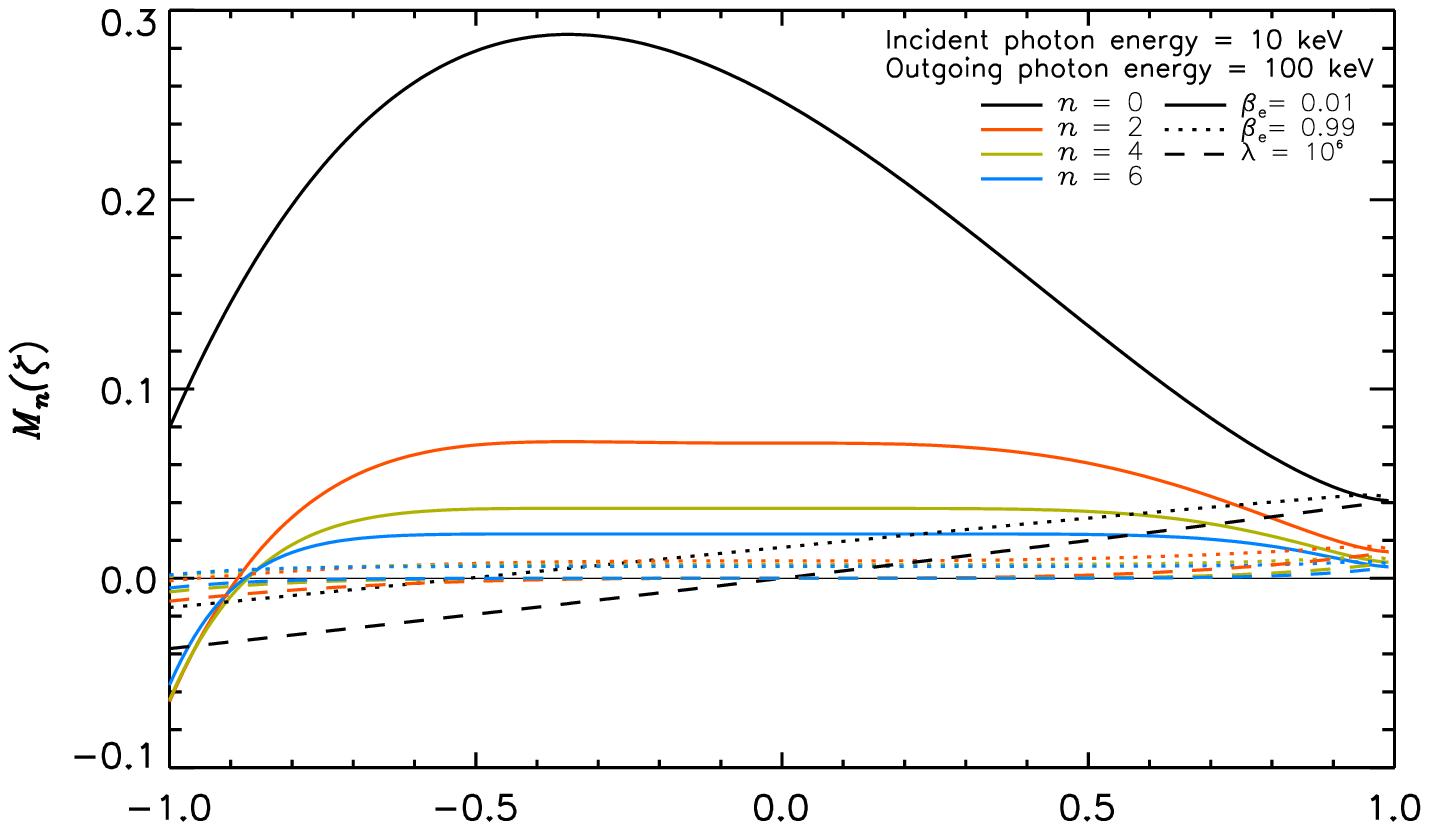}\vspace{-20mm}
  \\
  \includegraphics[width=0.49\textwidth]{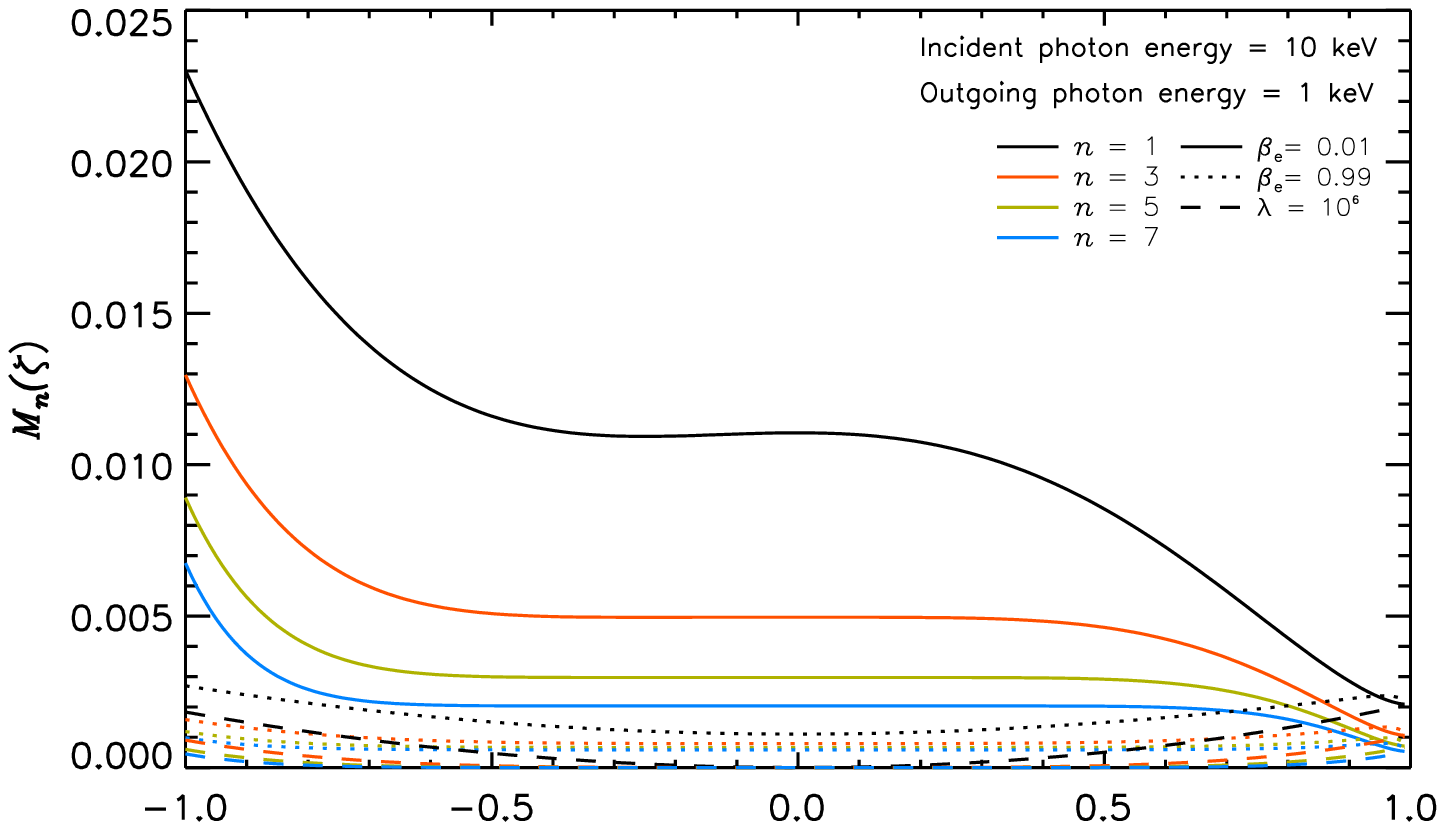}
  \includegraphics[width=0.49\textwidth]{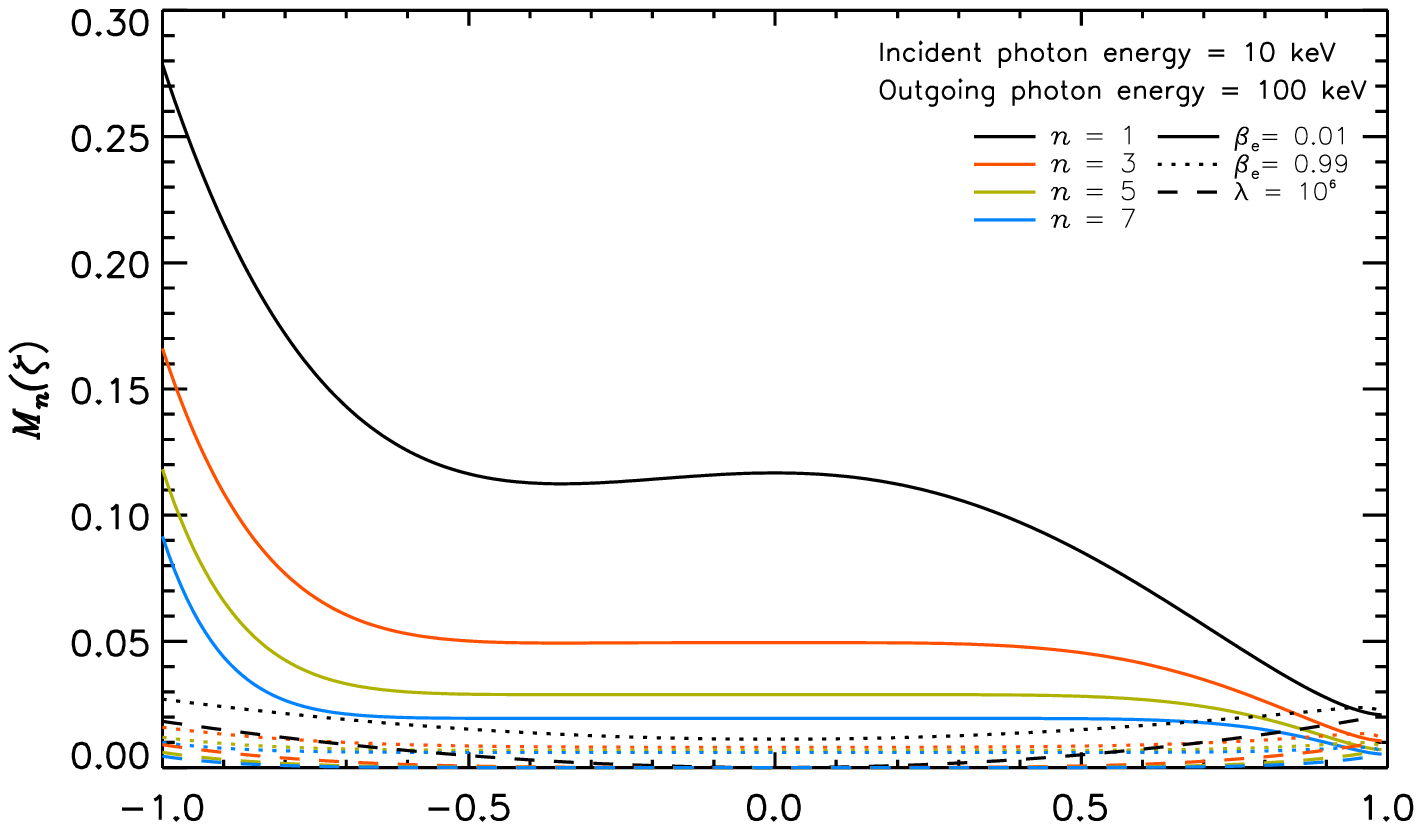}\vspace{-20mm}
  \\
  \includegraphics[width=0.49\textwidth]{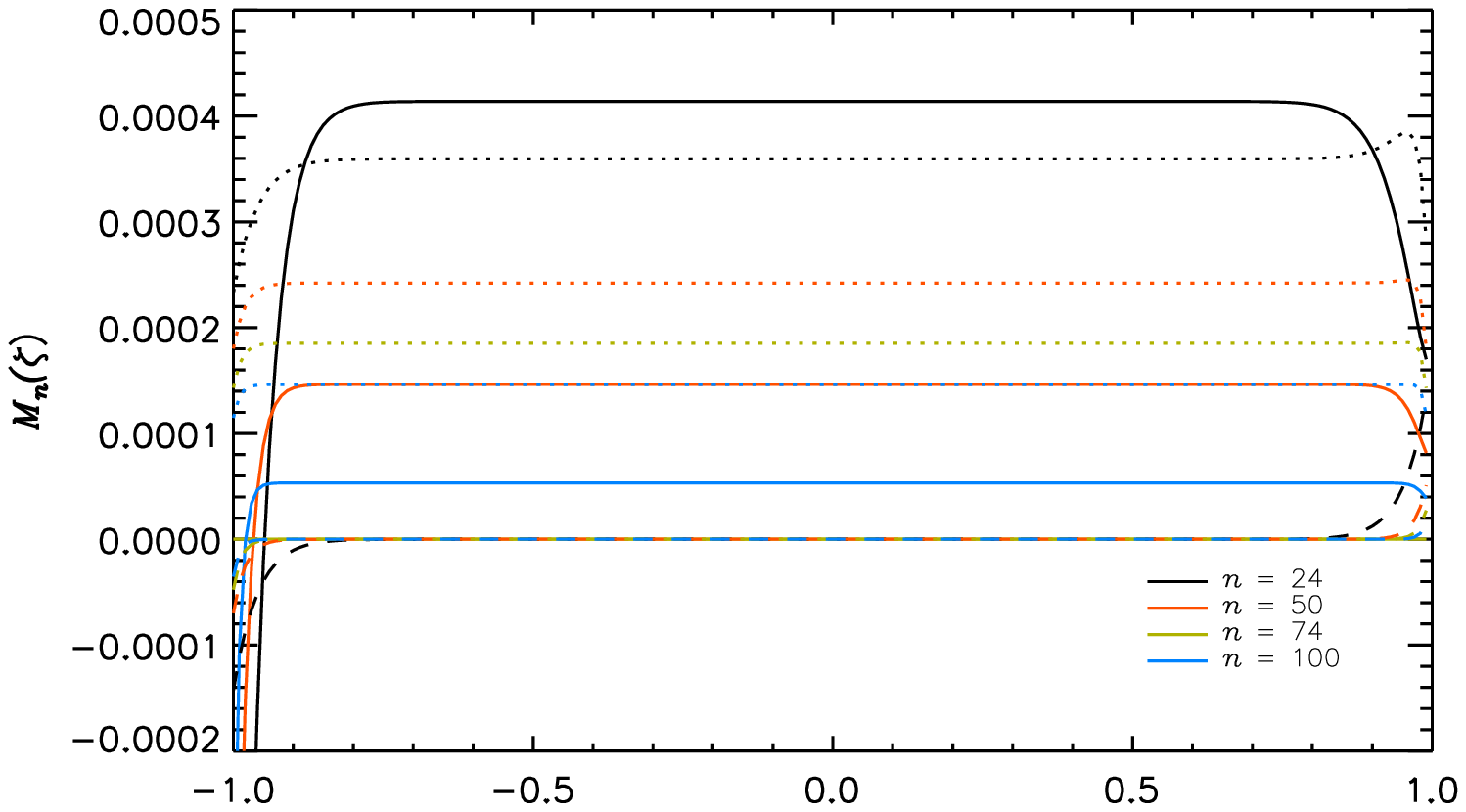}
  \includegraphics[width=0.49\textwidth]{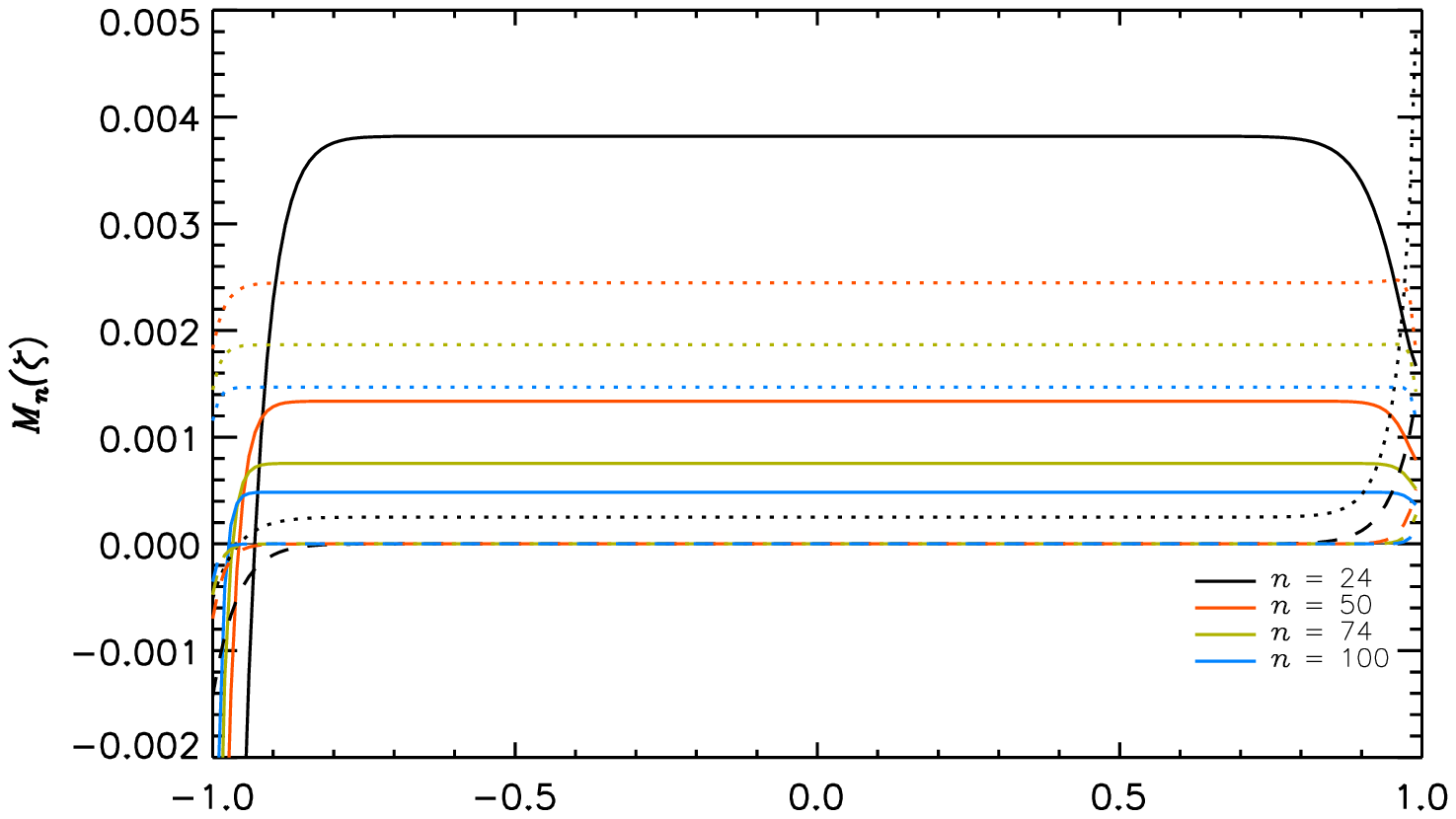}\vspace{-20mm}
  \\
  \includegraphics[width=0.49\textwidth]{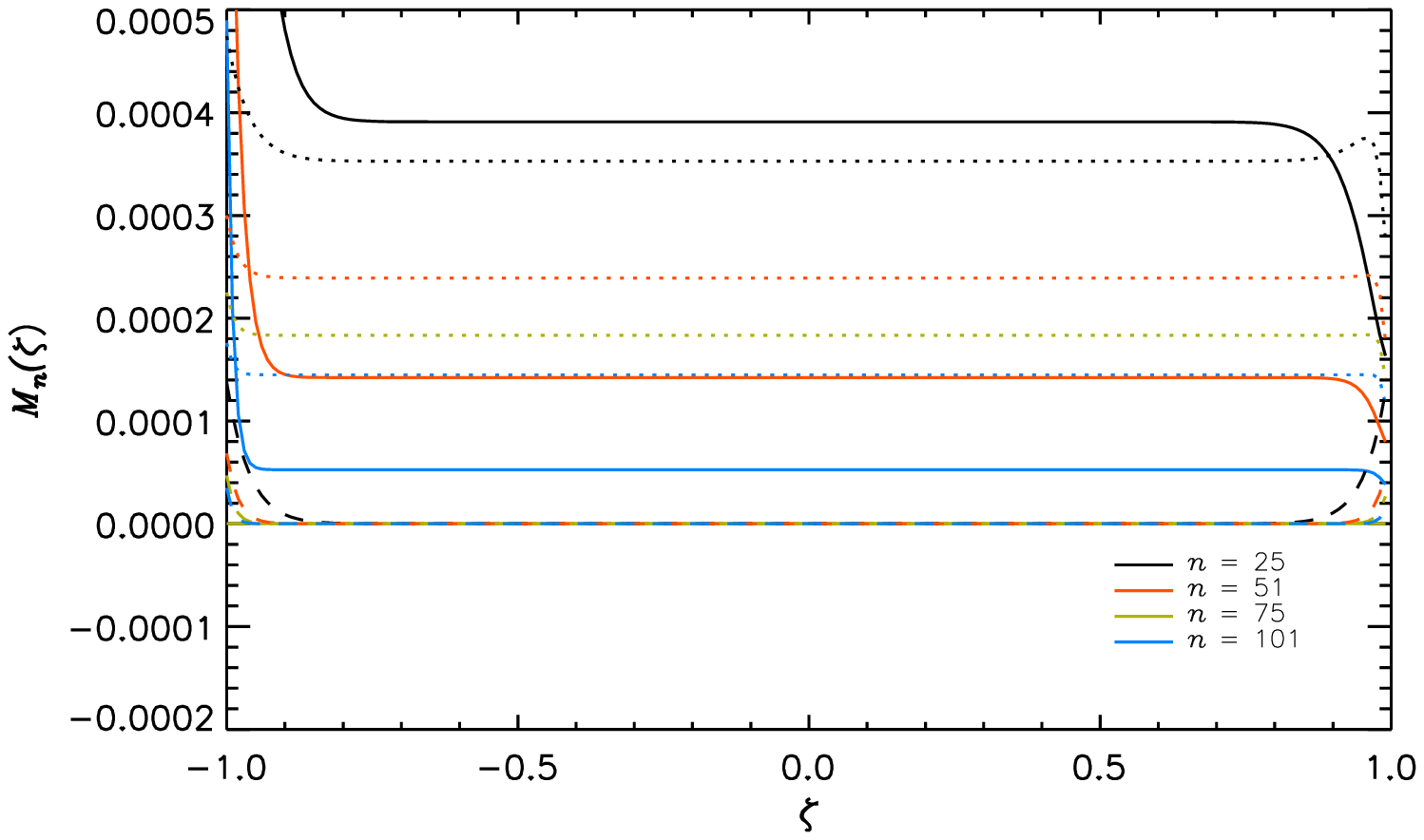}
  \includegraphics[width=0.49\textwidth]{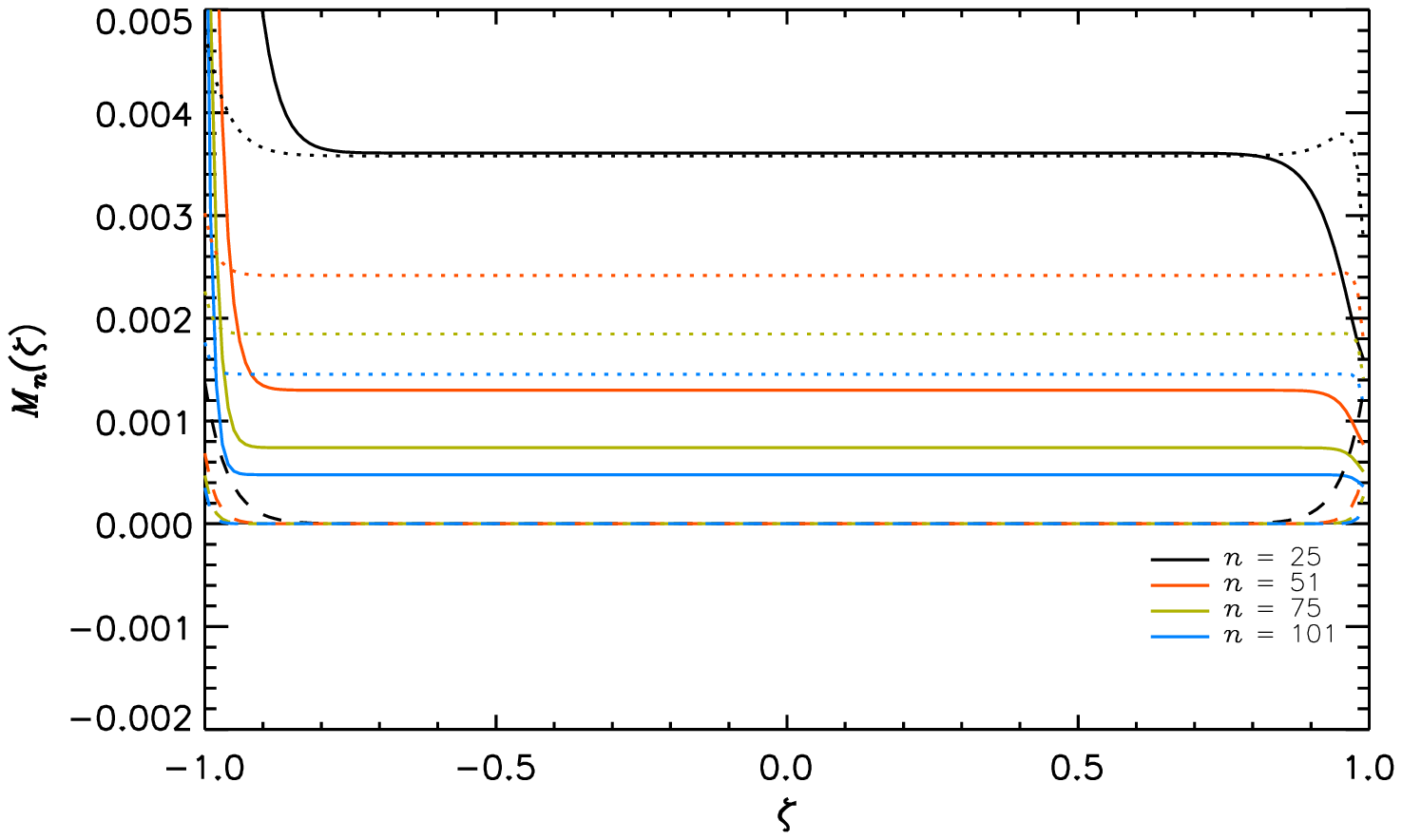} \vspace{-6mm}
\caption{Plots of the moment integral $\mathcal{M}_{n}$ as a function of $\zeta$ for an incident photon energy of $10$ keV. Plots on the left show an outgoing photon energy of $1$ keV, plots on the right an outgoing photon energy of $100$ keV (i.e. inverse Compton scattering). Left and right columns show, from top to bottom, $\mathcal{M}_{n}$ evaluated for $n=0$, 2, 4 and 6, $n=1$, 3, 5 and 7, $n=24$, 50, 74 and 100, and $n=25$, 51, 75 and 101, respectively. Solid, dotted, and dashed lines denote electron velocities of $\beta_{\mathrm{e}}=0.01$, $\beta_{\mathrm{e}}=0.99$ and $\lambda=10^6$ respectively. As $n$ increases, the angular moments become increasingly insensitive to a wider range of $\zeta$. The angular moments are strongly dependent on electron velocity.} 
\end{center}
\label{fig-5}
\end{figure}

\begin{figure}
\begin{center} 
  \includegraphics[width=0.9\textwidth]{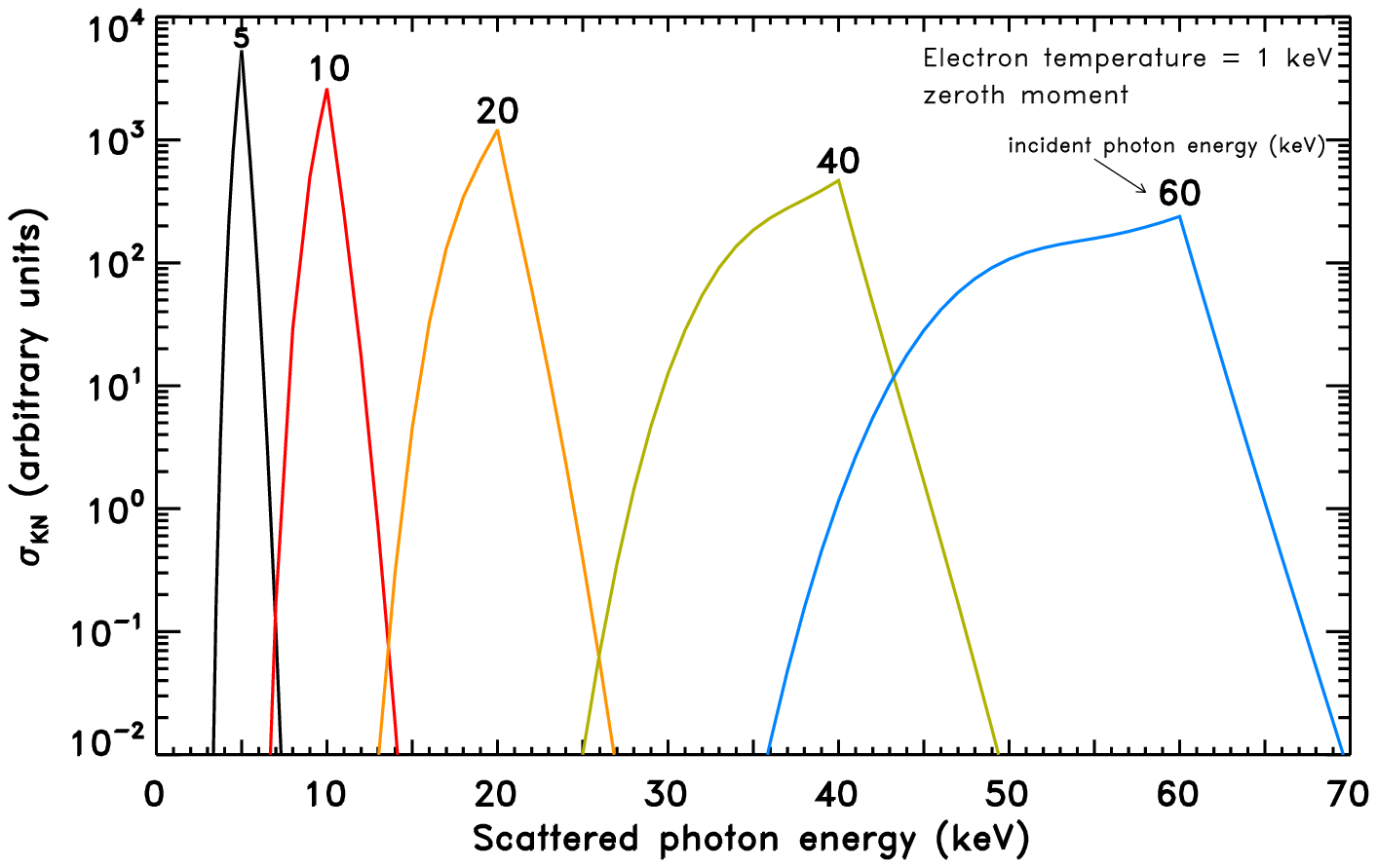} \vspace{-20mm}
  \\
  \includegraphics[width=0.9\textwidth]{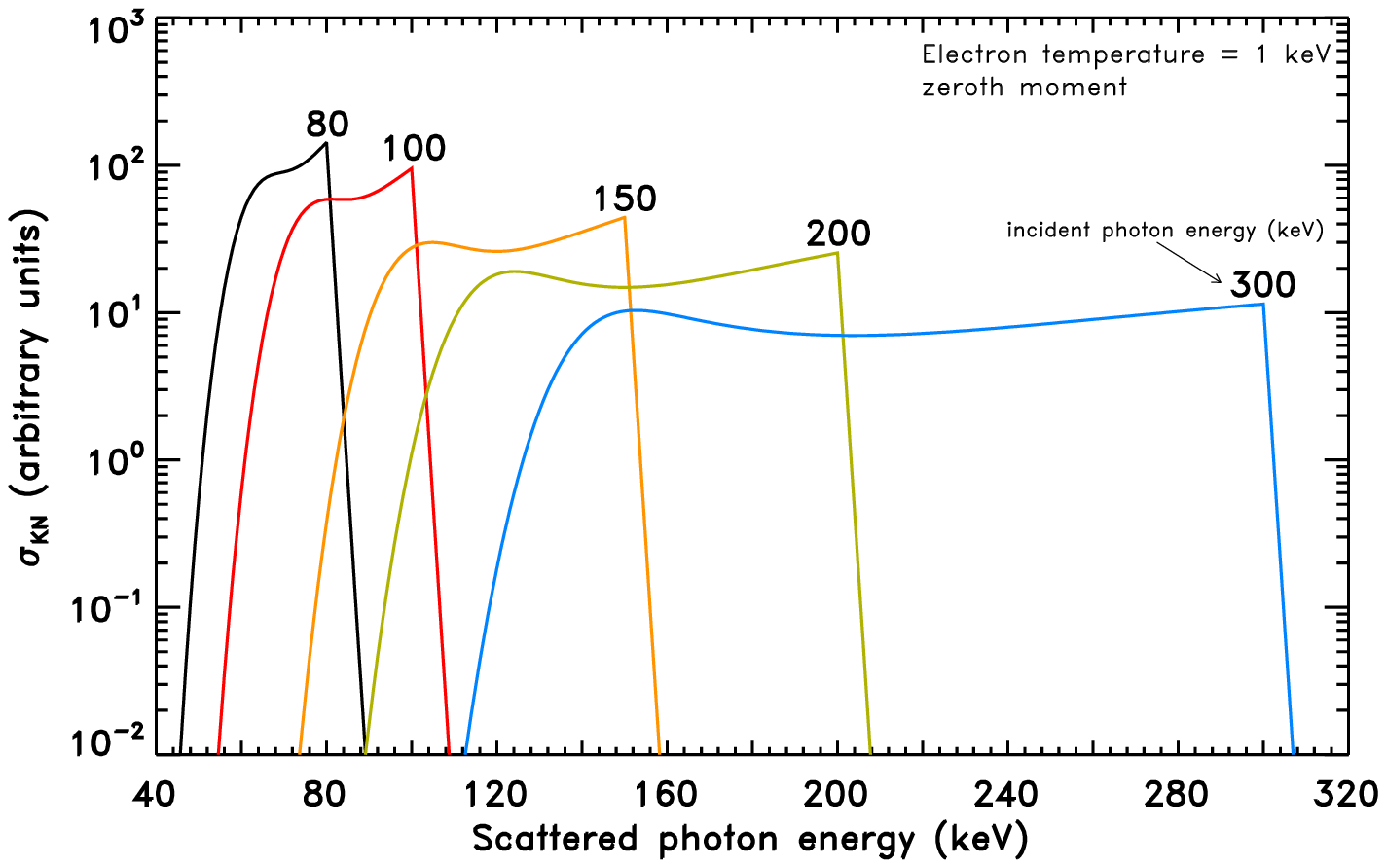}
\caption{Compton scattering kernel (as a function of scattered photon energy) evaluated for the zeroth moment for an electron temperature of $1$ keV. Top: kernel for incident photon energies of $5$ keV, $10$ keV, $20$ keV, $40$ keV and $60$ keV. Bottom: kernel for incident photon energies of $80$ keV, $100$ keV, $150$ keV, $200$ keV and $300$ keV.}
\end{center}
\label{fig-6}
\end{figure}

\begin{figure}
\begin{center}   
  \includegraphics[width=0.9\textwidth]{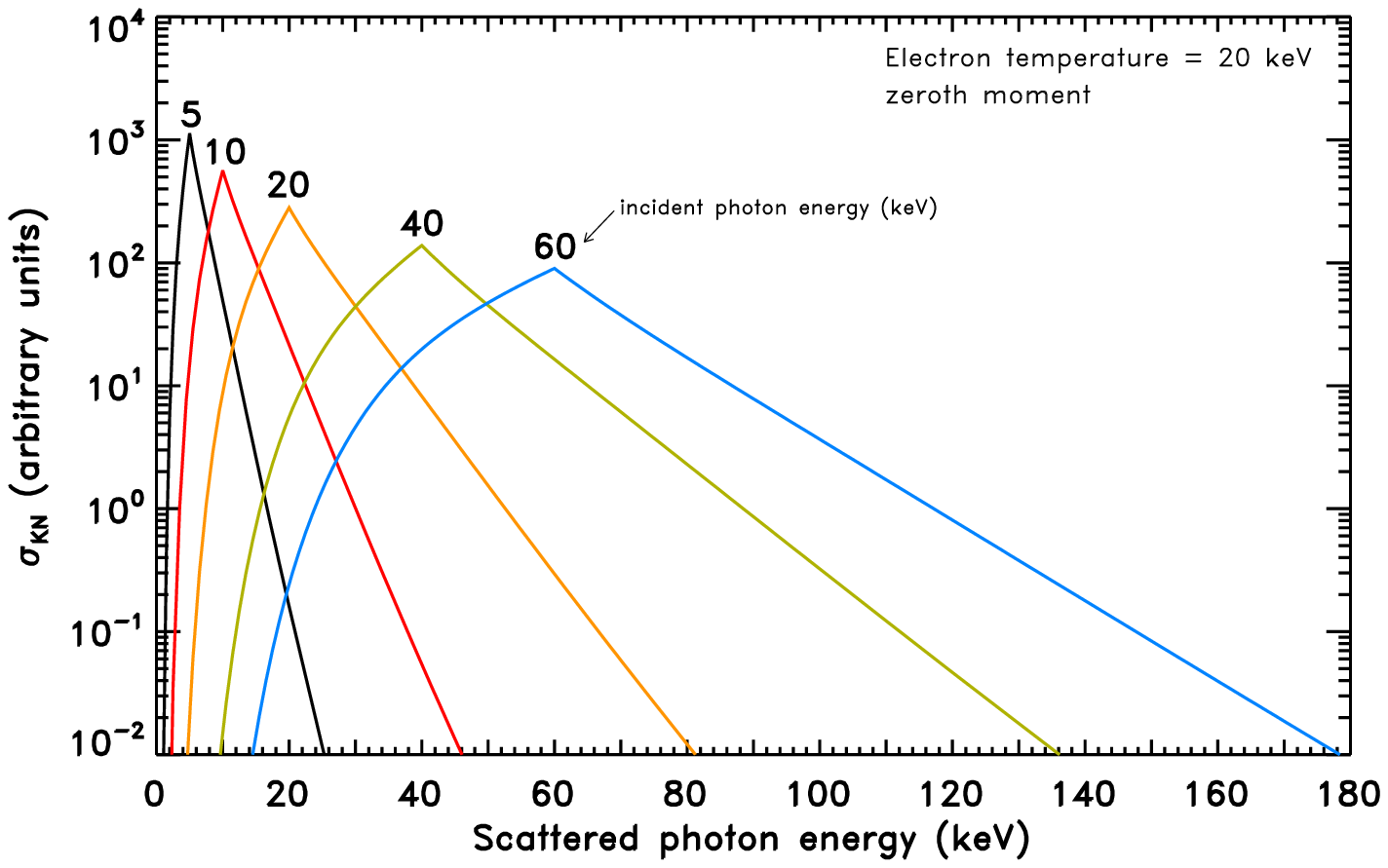} \vspace{-20mm}
  \\
  \includegraphics[width=0.9\textwidth]{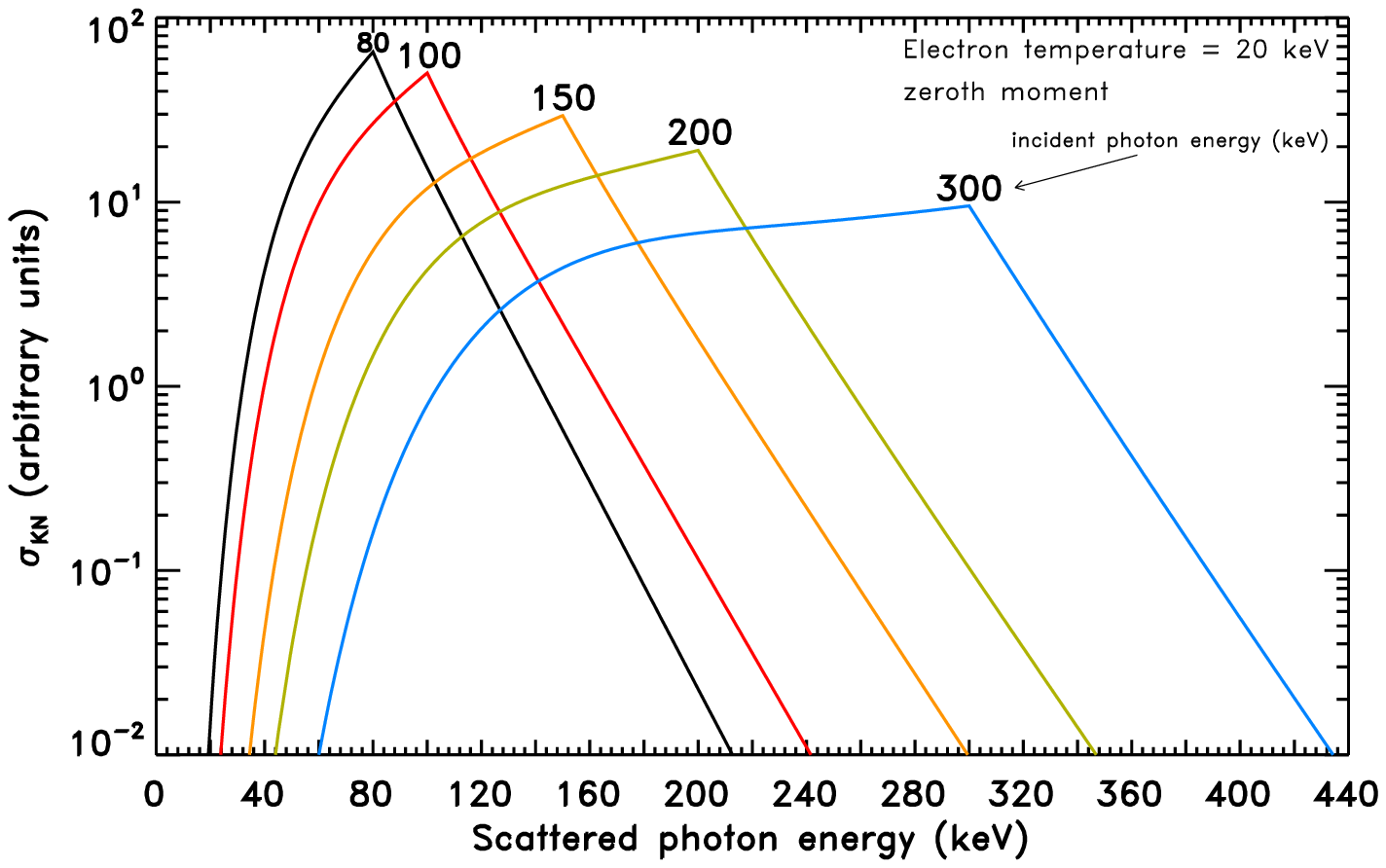}
\caption{Compton scattering kernel (as a function of scattered photon energy) evaluated for the zeroth moment for an electron temperature of $20$ keV. Top: kernel for incident photon energies of $5$ keV, $10$ keV, $20$ keV, $40$ keV and $60$ keV. Bottom: kernel for incident photon energies of $80$ keV, $100$ keV, $150$ keV, $200$ keV and $300$ keV.}
\end{center}
\label{fig-7}
\end{figure}

\begin{figure}
\begin{center}   
  \includegraphics[width=0.9\textwidth]{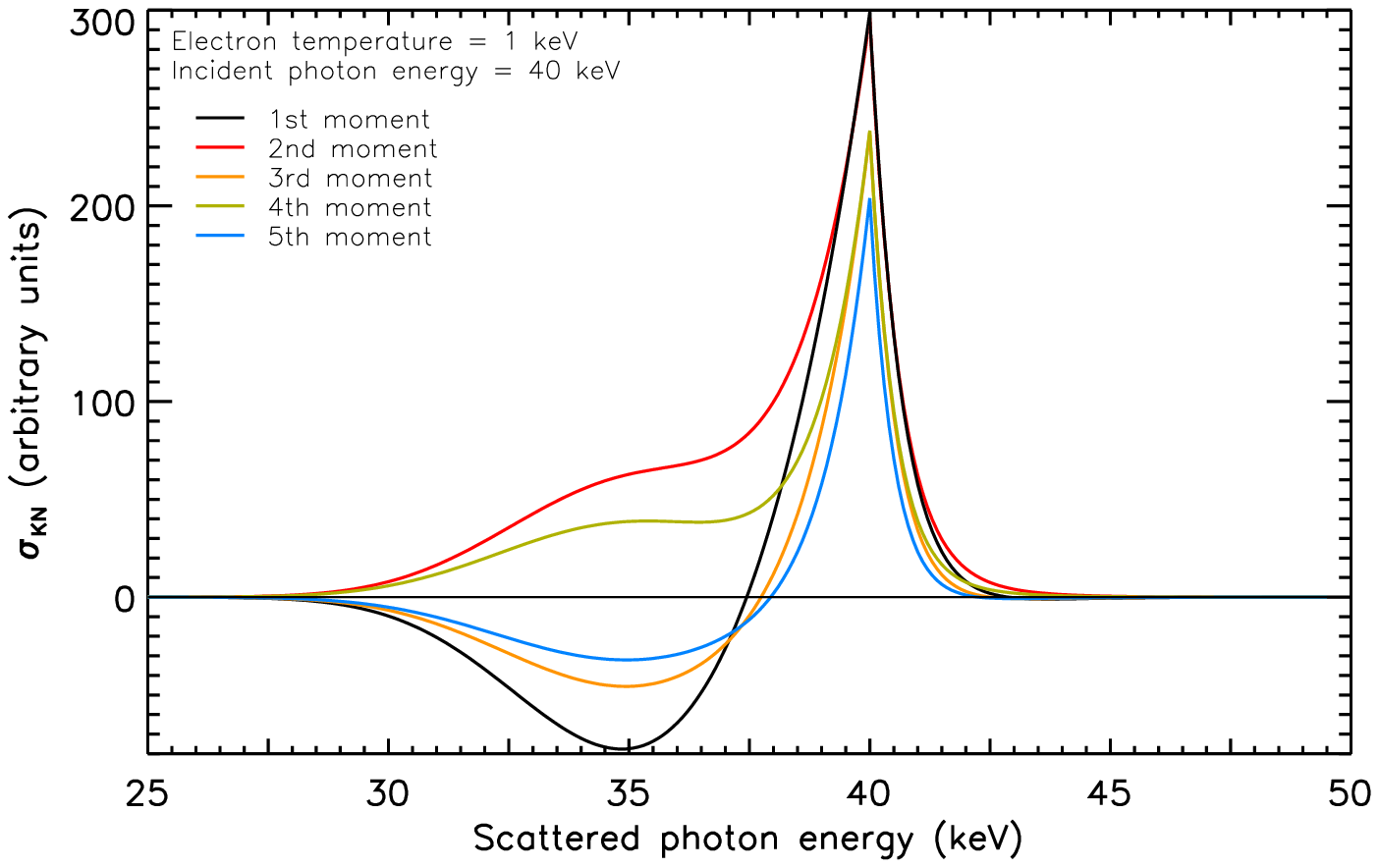} \vspace{-20mm}
  \\

  \includegraphics[width=0.9\textwidth]{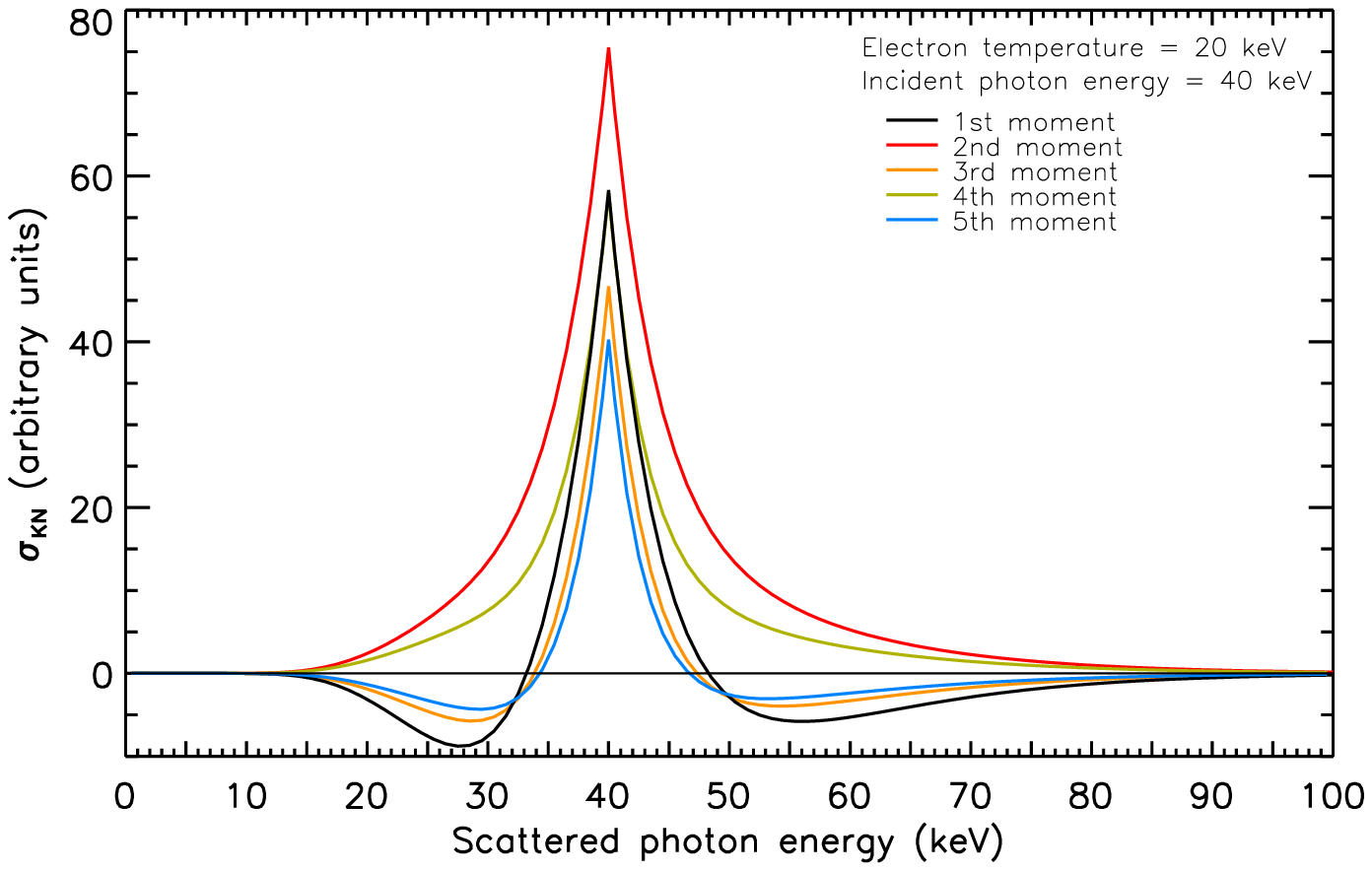}
\caption{Compton scattering kernel (as a function of scattered photon energy) evaluated for the $1$st, $2$nd, $3$rd, $4$th and $5$th moments, for an incident photon of energy $40$ keV. Top: moments of the Compton scattering kernel for electrons of temperature $1$ keV. Bottom: moments of the Compton scattering kernel for electrons of temperature $20$ keV.}
\end{center}
\label{fig-8}
\end{figure}

\begin{figure}
\begin{center}   
  \includegraphics[width=0.9\textwidth]{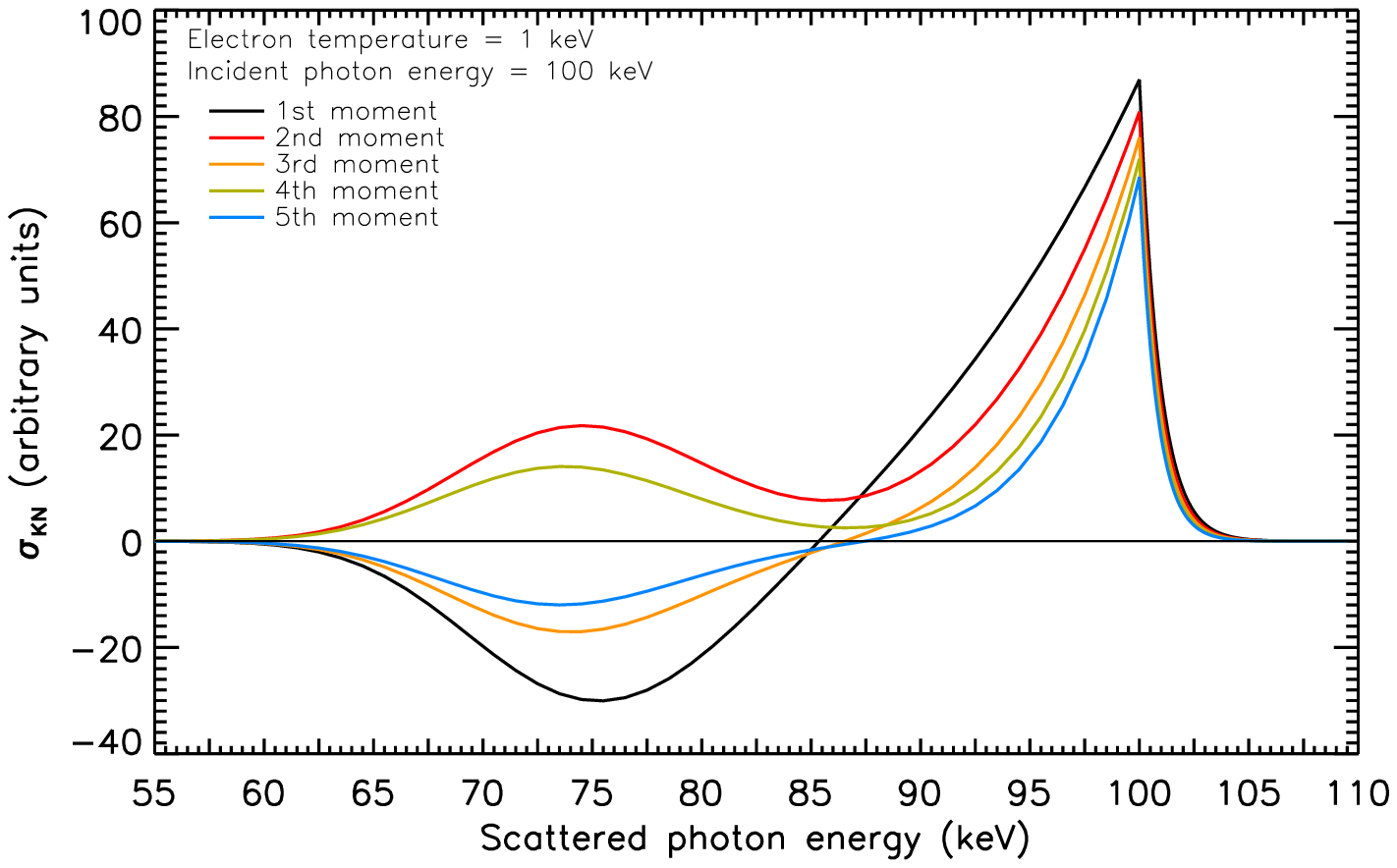} \vspace{-20mm}
  \\
  \includegraphics[width=0.9\textwidth]{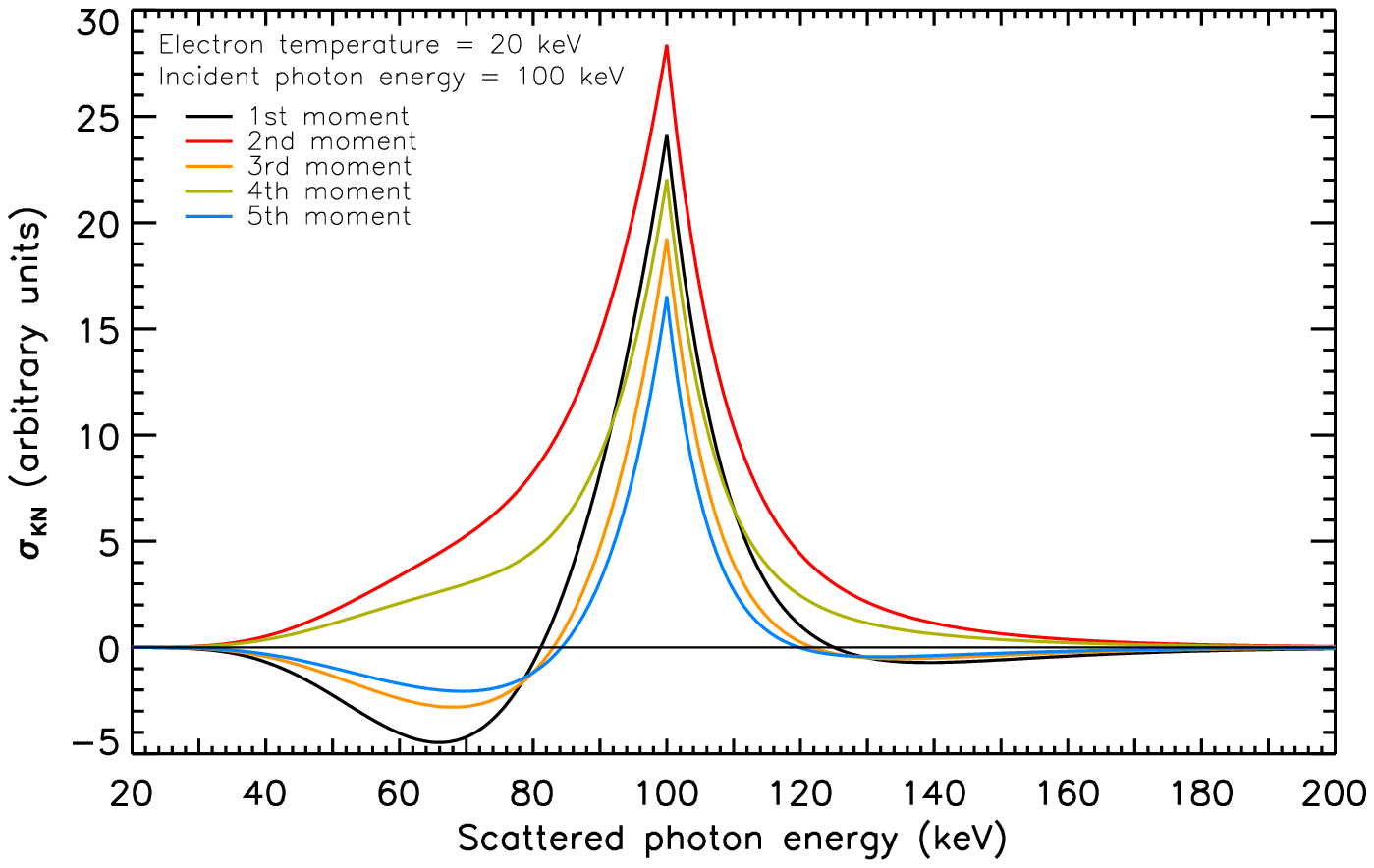}
\caption{Compton scattering kernel (as a function of scattered photon energy) evaluated for the $1$st, $2$nd, $3$rd, $4$th and $5$th moments, for an incident photon of energy $100$ keV. Top: moments of the Compton scattering kernel for electrons of temperature $1$ keV. Bottom: moments of the Compton scattering kernel for electrons of temperature $20$ keV.}
\end{center}
\label{fig-9}
\end{figure}

\begin{figure}
\begin{center}   
  \includegraphics[width=0.9\textwidth]{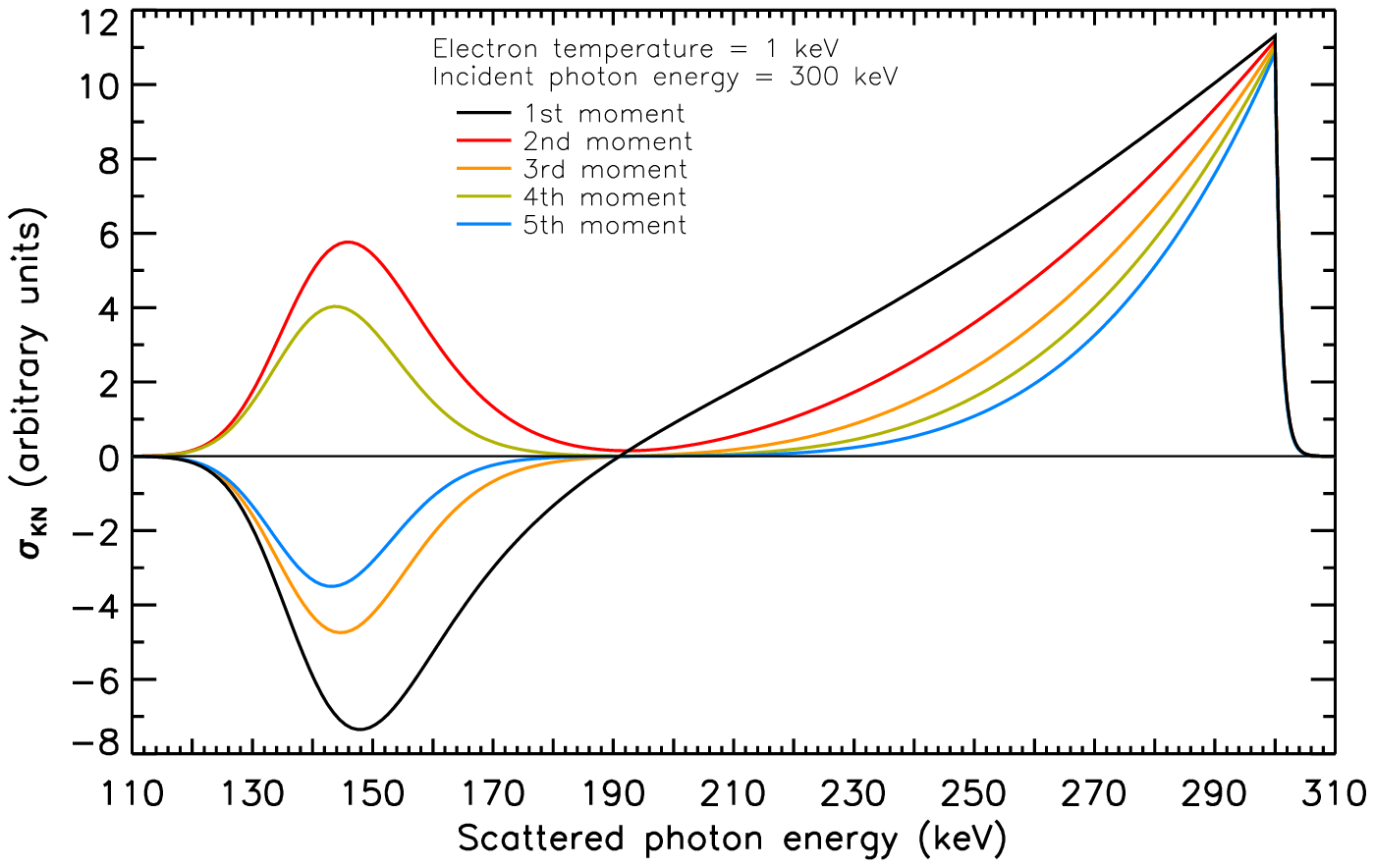} \vspace{-20mm}
  \\
  \includegraphics[width=0.9\textwidth]{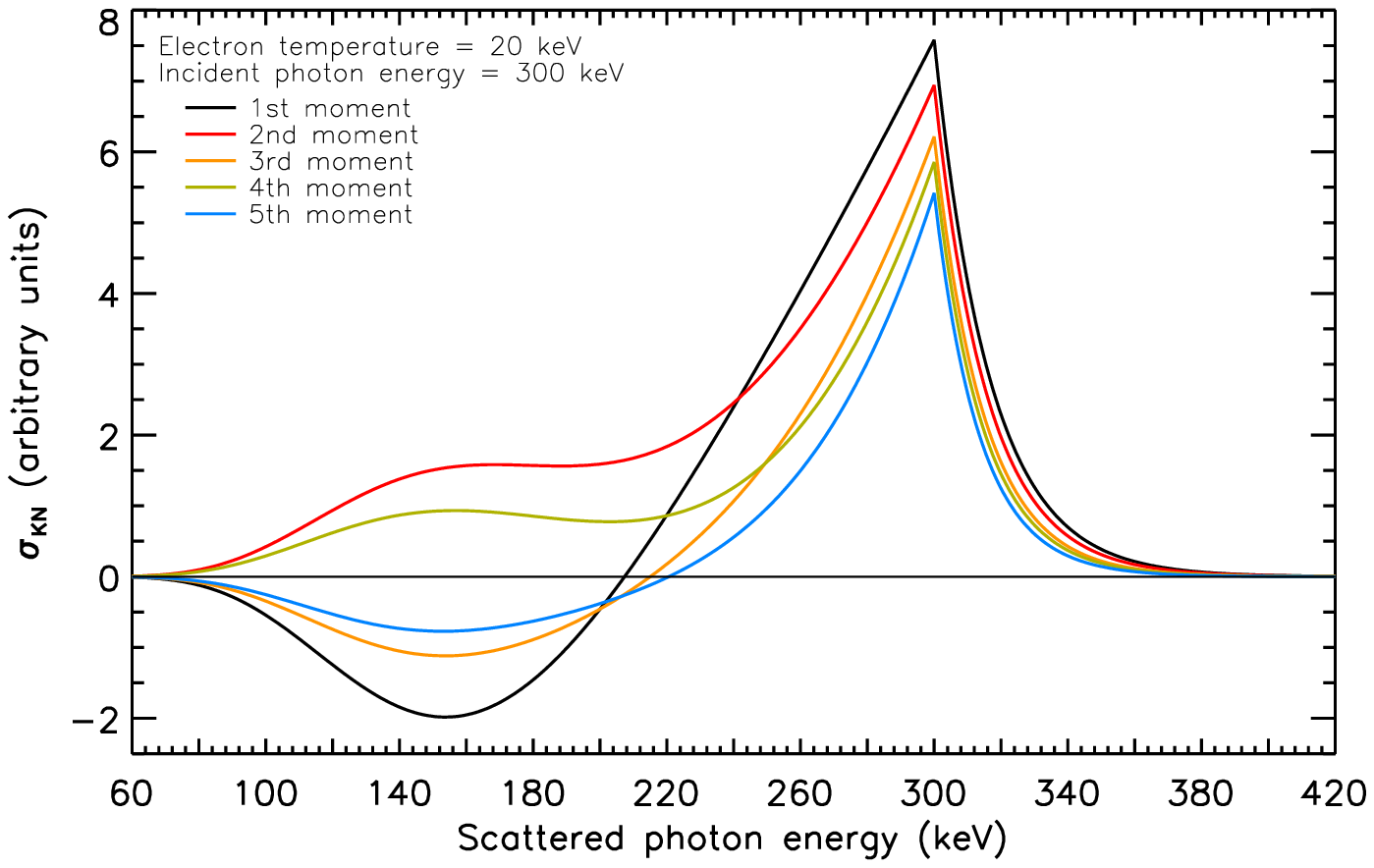}
\caption{Compton scattering kernel (as a function of scattered photon energy) evaluated for the $1$st, $2$nd, $3$rd, $4$th and $5$th moments, for an incident photon of energy $300$ keV. Top: moments of the Compton scattering kernel for electrons of temperature $1$ keV. Bottom: moments of the Compton scattering kernel for electrons of temperature $20$ keV.}
\end{center}
\label{fig-10}
\end{figure}

\clearpage

\appendix

\section{Appell Hypergeometric Function Method} \label{AppendixA}
The Appell $\mathrm{F}_{1}$ hypergeometric function is one of a set of four hypergeometric series of two variables \citep{Appell1880, Appell1926}. It is a very general class of special function, containing many other special functions as particular or limiting cases, including hypergeometric functions of one variable like the Gauss ${}_{2}\mathrm{F}_{1}$. The Appell $\mathrm{F}_{1}$ function is defined by the series expansion
\begin{equation}
\mathrm{F}_{1}\left(a;b_{1},b_{2};c;z_{1},z_{2}\right)=\sum_{k=0}^{\infty}\sum_{l=0}^{\infty}\frac{(a)_{k+l}(b_{1})_{k}(b_{2})_{l}}{(c)_{k+l}\ \! k! \ \!l!}z_{1}^{k}\ \! z_{2}^{l} \ ,
\end{equation}
This series is absolutely convergent for $|z_{1}|<1$, $|z_{2}|<1$. Cases outside of the unit disc of convergence can be calculated through analytic extension \citep{Olsson1964}, hence an algorithm can be constructed to evaluate the function numerically \citep[e.g.][]{Colavecchia2001,Colavecchia2004}.

Consider $R_{n}$ and $S_{n,m}$. $R_{n}$, after an appropriate substitution, may be expanded into a doubly-infinite series as
\begin{eqnarray}
\textrm{R}_{n}&=&\alpha^{-3/2}\int \mathrm{d}\zeta \ \! \zeta^{n}(1-\zeta)^{-1/2}\left(1+\frac{\beta}{\alpha}\zeta\right)^{-3/2} \nonumber \\
&=& -\frac{1}{2\sqrt{2}}\int \mathrm{d}u \ \! u^{-1/2}\left[ \sum_{k=0}^{\infty}\frac{(-n)_{k}}{k!}u^{k} \right] \left[ \sum_{l=0}^{\infty}\frac{(\frac{3}{2})_{l}}{l!}\left(\frac{\beta}{2}\right)^{l}u^{l}\right] \ ,
\end{eqnarray}
where $\alpha\equiv1+x^{2}$, $\beta\equiv 1-x^{2}$ and $u\equiv 1-\zeta$. Performing the integral over $u$ and using the identity $(2k+2l+1)=(3/2)_{k+l}/(1/2)_{k+l}$ the following closed-form expression for the moment integral $R_{n}$ is obtained:
\begin{equation}
R_{n}=-\frac{(1-\zeta)^{1/2}}{\sqrt{2}}\mathrm{F}_{1}\left[\frac{1}{2};-n,\frac{3}{2};\frac{3}{2};1-\zeta,\frac{1}{2}(1-x^{2})(1-\zeta) \right] \ . \label{Rn_Appell}
\end{equation}
By the same process, a closed-form expression for the moment integral $S_{n,m}$ also follows
\begin{equation}
\textrm{S}_{n,m}=-\frac{(1-\zeta)^{\frac{3}{2}-m}}{(\frac{3}{2}-m)\sqrt{2}}\mathrm{F}_{1}\left[\frac{3}{2}-m;-n,\frac{1}{2};\frac{5}{2}-m;1-\zeta,\frac{1}{2}(1-x^{2})(1-\zeta) \right] \ . \label{Snm_Appell}
\end{equation}
In the case $x^{2}=1$ these expressions simplify to Gauss hypergeometric functions of one variable through the following identity:
\begin{equation}
\mathrm{F}_{1}\left(a;b_{1},b_{2};c;x,0 \right)={}_{2}\mathrm{F}_{1}\left(a,b_{1};c;x\right) \ .
\end{equation}
 As expected from the integral expressions for the moment integrals in equations (\ref{Rn}) and (\ref{Snm}), equations (\ref{Rn_Appell}) and (\ref{Snm_Appell}) are identical in argument and differ only in their parameters $(a,b_{2},c)$.  For both of these expressions the parameter $b_{1}=-n$, and are hence absolutely convergent, since $(-n)_{k}=0$ for $k\ge n$. That is to say, by writing the Appell hypergeometric function as a single sum over Gauss hypergeometric functions \citep{Srivastava1985} the series always converges in $n+1$ terms.
 
 Although it may appear profitable to compute the scattering kernel in terms of Appell hypergeometric functions, since these simplify to finite sums of Gauss hypergeometric functions, it is not computationally cheaper and so the results in Section 9 are expressed in terms of the latter.

\section{Moment expansion in terms of more general polynomials} \label{AppendixB}
As has already been observed (see Fig. 4), a moment expansion in terms of $\zeta^{n}$, although convenient, is not strongly convergent for very small scattering angles. The expansion is inherently oscillatory in this instance, since even moments will always yield strictly positive results for the Compton scattering kernel, and odd moments are both positive and negative. The question naturally arises as to how the behaviour changes if a different moment formalism is chosen. This method can also be applied if the electron distribution is no longer isotropic, introducing a $\zeta$--dependence in the electron distribution function. Consider a generalised function of $\zeta$, $\mathcal{F}(\zeta)$, which can be represented as a Taylor series:
\begin{equation}
\mathcal{F}_{n}\left(\zeta\right)=\sum_{k=0}^{n}c\left(n,k\right)\zeta^{k}.
\end{equation}
Defining tilde variables as those which represent a moment expansion in terms of $\mathcal{F}_{n}(\zeta)$, it is readily shown that the generalised moment integrals may be written in terms of the usual $Q_{n}$, $R_{n}$ and $S_{n,m}$ as
\begin{eqnarray}
\widetilde{Q}_{n} &=& \int \mathrm{d}\zeta \ \frac{\mathcal{F}_{n}(\zeta)}{q} \nonumber \\
                       &=& \sum_{k=0}^{n}c \left(n,k\right)Q_{k} \ , \label{Qntilde} \\
\widetilde{R}_{n} &=& \int\frac{\mathrm{d}\zeta \ \mathcal{F}_{n}(\zeta)}{(1-\zeta)^{2}\left(x^{2}+\frac{1+\zeta}{1-\zeta}\right)^{3/2}} \nonumber \\
                       &=& \sum_{k=0}^{n}c \left(n,k\right)R_{k} \ , \\
\widetilde{S}_{n,m} &=& \int  \frac{\mathrm{d}\zeta\ \mathcal{F}_{n}(\zeta)}{(1-\zeta)^{m}\left(x^{2}+\frac{1+\zeta}{1-\zeta}\right)^{1/2}} \\
                           &=& \sum_{k=0}^{n}c \left(n,k\right)S_{k,m} \ , \label{Snmtilde}
\end{eqnarray}
from whence it follows
\begin{equation}
\widetilde{\mathcal{M}}_{n}=\sum_{k=0}^{n}c \left(n,k\right)\mathcal{M}_{k} \ .
\end{equation}
Therefore
\begin{eqnarray}
\widetilde{\sigma}_{\mathrm{KN}}(\gamma\rightarrow\gamma',\tau) &=& \int \mathrm{d}\zeta \! \ \mathcal{F}_{n}(\zeta) \ \! \sigma_{\mathrm{S}}(\gamma\rightarrow\gamma',\zeta,\tau) \nonumber \\
&=&\frac{ \mathcal{C}}{\gamma^{2}\ \! \tau \ \! \mathrm{K}_{2}(1/\tau)}\sum_{k=0}^{n}c\left(n,k\right)T(\gamma,\gamma',\tau) \ .
\end{eqnarray}

\label{lastpage}
\end{document}